\documentclass[aps,prb,reprint,floatfix]{revtex4-2}

\usepackage{amssymb}
\usepackage{graphicx}   % need for figures
\usepackage{color}      % use if color is used in text
\usepackage{hyperref}   % use for hypertext links, including those to external documents and URLs

\def\lnod{La$_2$NiO$_{4+\delta}$}
\def\lsno{La$_{2-x}$Sr$_x$NiO$_4$}
\def\lsco{La$_{2-x}$Sr$_x$CuO$_4$}
\def\lbco{La$_{2-x}$Ba$_x$CuO$_4$}

\def\lesco{La$_{1.8-x}$Eu$_{0.2}$Sr$_x$CuO$_4$}
\def\ybco{YBa$_2$Cu$_3$O$_{6+x}$}

\def\bscco{Bi$_2$Sr$_2$CaCu$_2$O$_{8+\delta}$}

\def\newr{\color{black}}

\begin{document}

\title{Topological Doping and Superconductivity in Cuprates:\\ An Experimental Perspective}

\author{John M. Tranquada}
\affiliation{Condensed Matter Physics and Materials Science Division, Brookhaven National Laboratory, Upton, New York 11973-5000, USA}

\date{\today} 

\begin{abstract}
Hole doping into a correlated antiferromagnet leads to topological stripe correlations, involving charge stripes that separate antiferromagnetic spin stripes of opposite phase.  Topological spin stripe order causes the spin degrees of freedom within the charge stripes to feel a geometric frustration with their environment.  In the case of cuprates, where the charge stripes have the character of a hole-doped two-leg spin ladder, with corresponding pairing correlations.  Anti-phase Josephson coupling across the spin stripes can lead to pair-density-wave order, in which broken translation symmetry of the superconducting wave function is accommodated by pairs with finite momentum.  This scenario has now been experimentally verified by recently reported measurements on \lbco\ with $x=1/8$.  While pair-density-wave order is not common as a cuprate ground state, it provides a basis for understanding the uniform $d$-wave  order that is more typical in superconducting cuprates.
\end{abstract}

\maketitle

\section{Introduction}

Charge order has now been observed in virtually all hole-doped cuprate superconductor families \cite{fran20,comi16,uchi21}.  In 214 cuprates such as \lsco\ (LSCO) and \lbco\ (LBCO), charge-stripe order is generally accompanied by spin-stripe order \cite{fuji04,huck11,fuji12a,wen19,miao21}, as originally observed in Nd-doped \lsco\ \cite{tran95a,ma21}; each of these orders breaks the translation symmetry of the square-lattice CuO$_2$ planes.  In a 1996 paper, Kivelson and Emery \cite{kive96} pointed out the topological character of the combined spin and charge stripe orders.  This corresponds to the fact that the period of the spin-stripe order is twice that of the charge-stripe order, as the antiferromagnetic phase flips by $\pi$ across each charge stripe, as illustrated in Fig.~1.

The topological character of stripes in cuprates is distinct from that of the topological insulators that have dominated attention more recently \cite{hasa10,qi11}.  In the latter case, the focus is on Bloch states in which spin-orbit effects play a special role.  In cuprates, in contrast, the effects of strong onsite Coulomb repulsion among Cu $3d$ electrons tend to make Bloch states of questionable relevance.  In a parent compound such as La$_2$CuO$_4$, one has a single unpaired Cu $3d_{x^2-y^2}$ electron on each Cu atom that acts as a local moment, with neighboring moments coupled antiferromagnetically by superexchange $J$, a local interaction.  While the electronic band gap has charge-transfer character due to O $2p$ states that lie between the lower and upper Hubbard bands associated with the Cu $3d_{x^2-y^2}$ orbital, it is the locally antiferromagnetic (AF) environment that limits the motion of doped holes.

It has taken quite some time to appreciate the significance of the topological order associated with spin stripes.  Experimentally, the same antiphase relationship of spin stripes seen in superconducting cuprates also occurs in the case of insulating behavior in \lsno\ \cite{tran13a} and in \lsco\ with $0.02\lesssim x \lesssim 0.05$ \cite{fuji02c} (where the stripes run diagonally with respect to the Ni-O or Cu-O bonds).  A theoretical analysis of interaction requirements for topological doping came to no firm conclusions \cite{prya99}.   Antiphase spin stripes have been obtained from many different approaches: from Hartree-Fock calculations on the Hubbard model \cite{zaan89}, from effective models that include long-range Coulomb interactions \cite{low94}, and from advanced variational and quantum Monte Carlo evaluations of the $t$-$J$ \cite{whit98a} or Hubbard model \cite{zhen17,huan18}.

\begin{figure}[t]
\begin{center}
\includegraphics[width=\columnwidth]{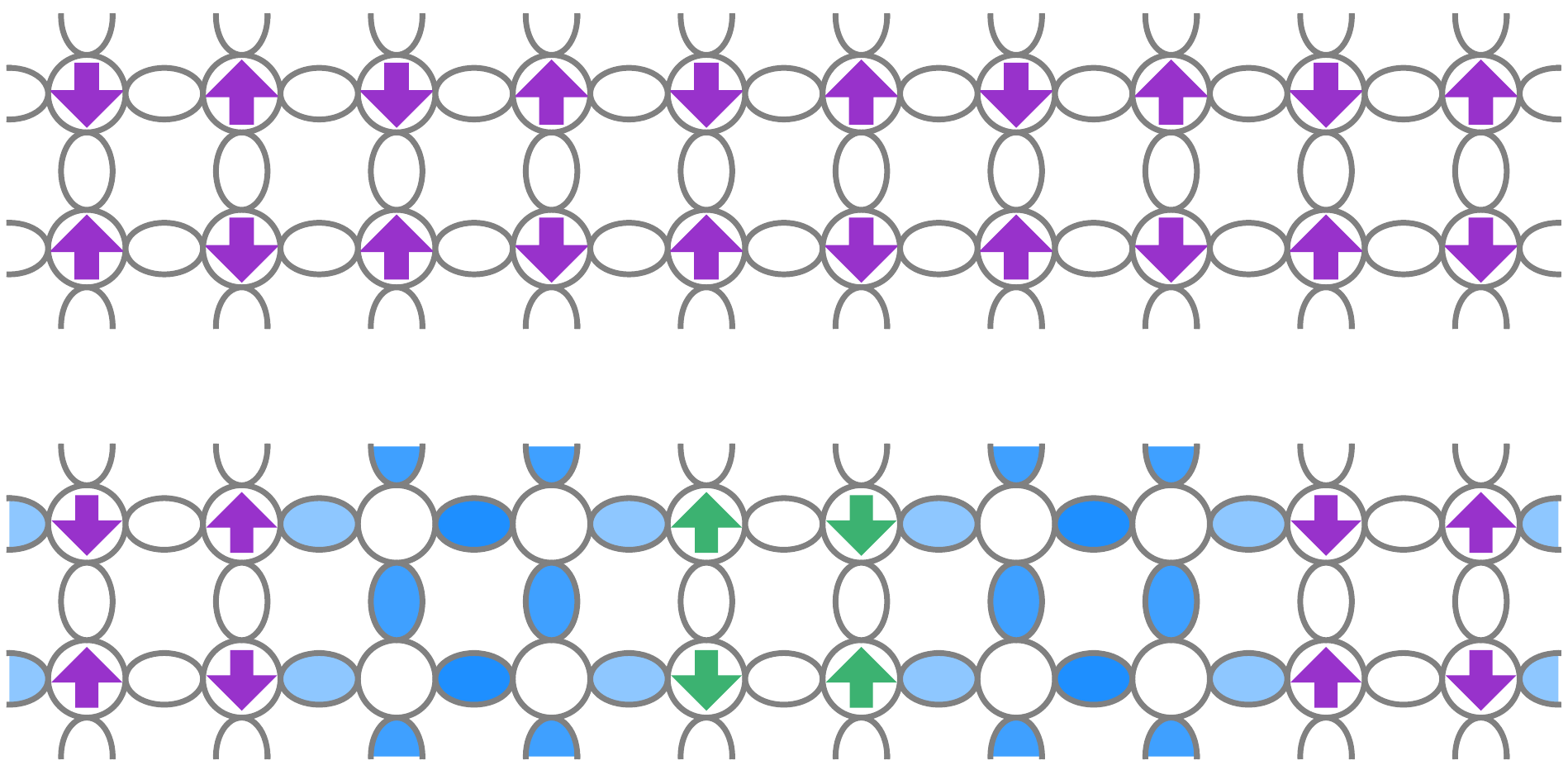}
\end{center}
\caption{Upper panel indicates the antiferromagnetic order of the undoped CuO$_2$ planes, with spin direction (arrows) indicated on Cu atoms (circles), separated by O atoms (ellipses).  Lower panel shows the spin configuration in the stripe-ordered phase at a doped-hole concentration of $p=1/8$, with doped hole density indicated by blue shading; antiphase spin stripe indicated in green.}
\end{figure}   

\begin{figure*}[t]
\begin{center}
\includegraphics[width=15 cm]{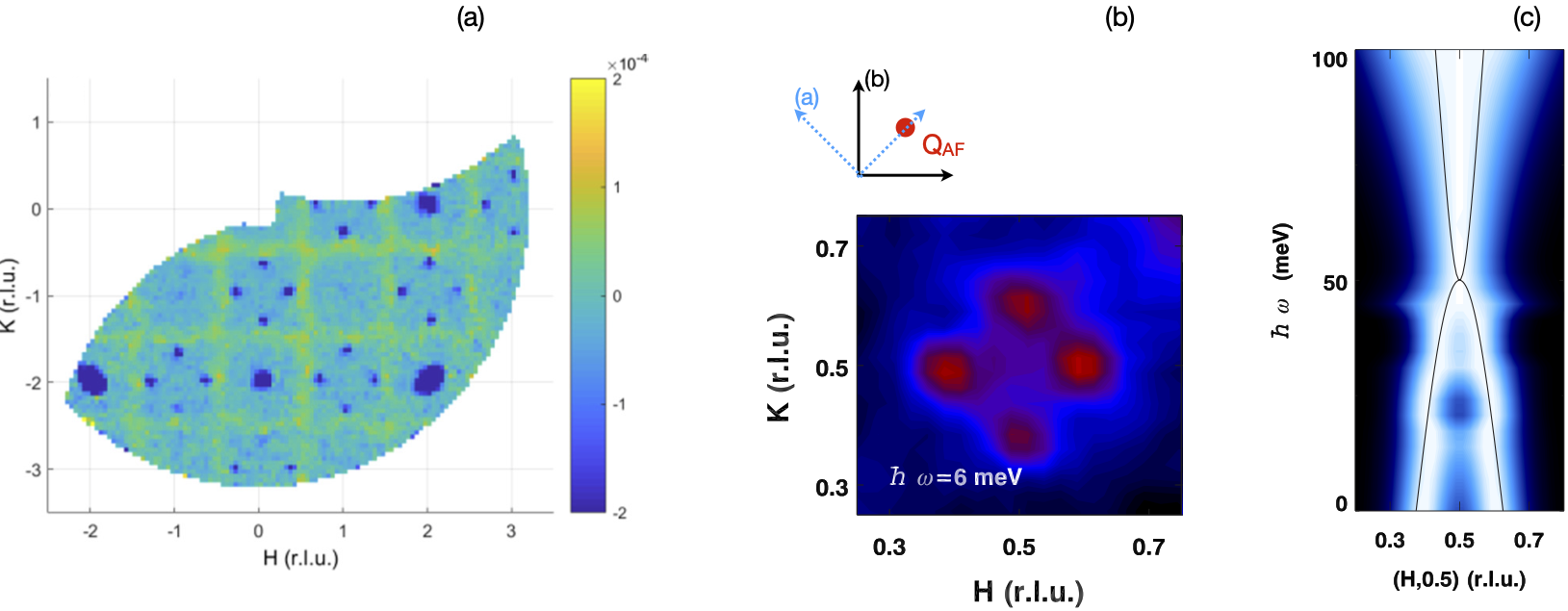}
\end{center}
\caption{\newr (a) Difference in neutron scattering intensity measured at 5 K and 70 K for $\hbar\omega=3\pm1$~meV in \lsno\ with $x=1/3$.  Dark blue points at positions of the type $(1\pm\frac13,0,0)$ and $(1,\pm\frac13,0)$ correspond to spin waves associated with the spin stripe order, where the AF wave vector, ${\bf Q}_{\rm AF}$, is $(1,0,0)$.  Yellow lines correspond to cuts through 2D planes of scattering from 1D spin correlations in charge stripes.  Note that twinning causes the measurement to include scattering from stripe domains rotated by 90$^\circ$.  Reprinted with permission from \cite{merr19}, \copyright(2019) by the American Physical Society.  (b) Neutron scattering intensity at $\hbar\omega=6$~meV and $T=10$~K for \lbco\ with $x=1/8$.  Here, ${\bf Q}_{\rm AF} = (0.5,0.5)$; inset shows relative orientations of axes in (a) and (b).  (c) Fitted dispersion and $Q$ widths of magnetic scattering in LBCO $x=1/8$ at 10~K along ${\bf Q}=(H,0.5)$.  Black line shows the hourglass dispersion often applied to such data.  (b) and (c) Reprinted with permission from \cite{xu07}, \copyright(2007) by the American Physical Society. }
\end{figure*}   

I have argued recently \cite{tran21a} that the key feature of topological doping is that the spin degrees of freedom within the charge stripes feel a geometric frustration of their interactions with the neighboring spin stripes.  This allows the charge stripes to develop quasi-one-dimensional spin correlations.  In the case of cuprates with bond-parallel stripes, the charge stripes may be viewed as hole-doped, two-leg, spin $S=1/2$ ladders, which are established to have strong superconducting correlations \cite{dago96,tsve11}.  This is a variation on the original proposal of superconducting charge stripes by Emery, Kivelson, and Zachar \cite{emer97}, who pointed out that a spin gap in a one-dimensional (1D) electron gas acts as a pairing amplitude; the difference is that they assumed that the spin gap would be transferred from the neighboring spin stripes, in which case one would never get superconductivity when spin-stripe order is present.  The advantage of the doped two-leg spin ladder is that it comes with its own spin gap.

To obtain superconducting order in the CuO$_2$ planes, it is necessary to establish phase coherence, via Josephson coupling, between neighboring charge stripes \cite{emer97}.  Because of the conflict between local AF order and hole motion, this needs to be antiphase superconducting order, resulting in a pair-density-wave (PDW) state \cite{berg09b,agte20}.  PDW order was initially proposed \cite{hime02,berg07} to explain the experimental observation of two-dimensional superconductivity in CuO$_2$ layers \cite{li07}, with frustration of the usual Josephson coupling between planes \cite{taji01}.

While the initial case for PDW order was circumstantial, direct phase-sensitive evidence of PDW order in LBCO $x=1/8$ has now been reported \cite{loza21b}.   This result is consistent with measurements of the Hall effect in high magnetic fields along the $c$-axis that suggest that the holes in the charge stripes remain paired even in the absence of superconducting order \cite{li19a}.  Hence, there is now a solid case that charge stripes in cuprates are essential to pairing.

Of course, the superconducting ground state of most cuprates is spatially-uniform $d$ wave, not PDW.  This is still compatible with pairing correlations developing within charge stripes, but it requires disordered spin stripes with an energy gap \cite{tran21a,li18}; uniform phase coherence can only be achieved at energies below the spin gap.  The antiphase spin stripes play a critical role for the superconducting order: they either need to be ordered to allow PDW phase order to be established, or gapped to enable spatially-uniform superconducting order.  As a consequence, uniform superconductivity will not coexist with a PDW ground state.  On the other hand, defects that require the superconducting order parameter to locally go to zero can favor local PDW order without spin order, as seen in studies by scanning tunneling spectroscopy \cite{edki19,du20}.

In the following, I fill in details that provide support for the story laid out above.

%The stripe order that develops has a topological character, with neighboring spin stripes being antiphase.  This was emphasized by Kivelson and Emery \cite{kive96}.  Charge stripes are not simple metallic systems; they also have Cu moments in them. 

%Current emphasis on topological order is associated with electronic wave functions that extend across a sample.  This is not so relevant for cuprates, because scattering from local moments limits the distance over which carriers can be coherent.  This is reflected in the bad-metal behavior.

%Instead, topological doping has relevance to magnetic correlations.  Geometric frustration for spins within the charge stripes.

\section{Stripe order and decoupling of spin excitations}

The holes doped into the CuO$_2$ planes tend to go into O $2p$ states \cite{chen91}.  As pointed out by Emery and Reiter \cite{emer88}, if one could localize a single hole, it would cause the neighboring Cu moments to be parallel, which frustrates the AF order of the undoped system.  In fact, it takes very few holes to kill the AF order.  In LSCO, commensurate AF order is gone by $p=0.02$, and even before that, one has phase separation at low temperature \cite{mats02}.  This transition occurs at a hole density that is 20 times smaller than the limit for percolation due to substitution of nonmagnetic ions, as verified in LSCO with nonmagnetic Zn and Mg substitution for Cu \cite{vajk02}.

Initially, the holes form diagonal stripes \cite{fuji02c} and the system is insulating.  This is similar to \lsno\ (LSNO) and \lnod\ \cite{tran94a,yosh00,ulbr12b}.  The case of LSNO with $x=1/3$ is of particular interest. {\newr Neutron scattering measurements of magnetic scattering are presented in Fig.~2(a); these can be understood in terms of the stripe order illustrated in Fig.~3.}  Note that, in contrast to the spin $S=1/2$ of Cu$^{2+}$, the Ni$^{2+}$ sites have $S=1$.  The Ni moments on the spin stripes order \cite{lee97} and exhibit well-defined spin waves \cite{boot03a,woo05}. Within the charge stripes, there is one hole per Ni site; a low-spin hybridization is expected to leave a net $S=1/2$ per Ni site along a charge stripe.  The interaction of each such moment with the neighboring spin stripes is geometrically frustrated.  It is still possible for the reduced Ni moments to couple antiferromagnetically along a charge stripe.  For such a decoupled 1D spin chain, one would expect to see no order, but spin excitations that disperse only along the stripe direction.  Just such 1D spin excitations were first identified by Boothroyd {\it et al.} \cite{boot03b}; the role of the decoupling of interactions due to such site-centered charge stripes between antiphase spin stripes was recognized and confirmed in \cite{merr19}.

\begin{figure}[t]
\begin{center}
\includegraphics[width=\columnwidth]{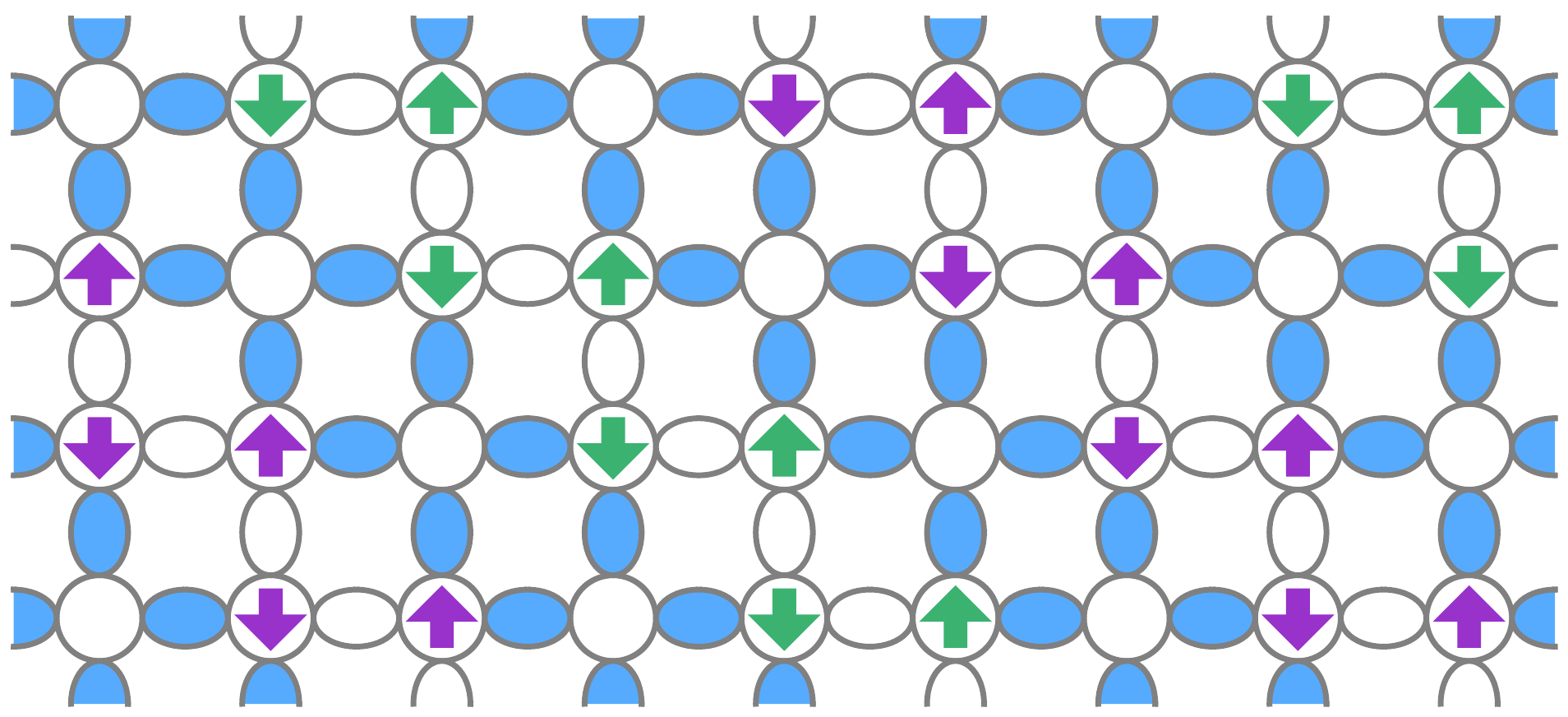}
\end{center}
\caption{Diagonal stripe order as observed in \lsno\ with $x=1/3$. Arrows indicate relative spin orientations on Ni sites (circles), with color change indicating antiphase domains.  Blue shading indicates distribution of doped holes on O sites (ellipses). }
\end{figure}   

In 214 cuprates, the stripe orientation rotates from diagonal to bond-parallel, and superconductivity appears, for $x\gtrsim0.05$ \cite{fuji02c,huck11,ma21}.  The stripe order is stabilized by coupling to lattice anisotropy, with the strongest stripe order correlated with a strong suppression of three-dimensional superconducting order at $x \approx 1/8$ \cite{tran95a,fuji04}.  The static spin order and the low-energy magnetic excitations correspond to the antiphase spin-stripe domains of Fig.~1{\newr; an example is shown in Fig. 2(b).}  The absence of any other low-energy magnetic excitations indicates that the spin degrees of freedom on the charge stripes are gapped.  {The size of the gap at ${\bf Q}_{\rm AF}$, apparent in Fig.~2(c),} is $\sim50$~meV, above which commensurate AF excitations appear  \cite{tran04}; the effective correlation length for the high-energy excitations is only about one lattice spacing \cite{xu07}.  (A two-component picture of the magnetic excitations has also been proposed in \cite{sato20}.)

We can reconcile the variations in the magnetic spectra through the model indicated schematically in Fig.~4.  If the charge stripes are centered on a row of bridging O atoms, then the charge stripes are effectively 2-leg spin ladders that are decoupled from the neighboring spin stripes due to frustration of the AF coupling \cite{tran21a}.  An undoped spin ladder is a spin liquid \cite{barn93}, with a spin gap that can be as large as $J/2$ \cite{dago96}.  The hole concentration in the 2-leg ladder picture of the charge stripes is 25\%.  With an effective $J$ of $\sim100$~meV \cite{tran04}, the holes will form pairs so as to avoid exciting the spins across the large spin gap.  As illustrated in Fig.~4, the spins can be viewed as forming a resonating-valence-bond (RVB) state of nearest-neighbor singlets.    Theoretical analysis indicates that the singlet-triplet excitation energy is essentially the pairing scale for the doped holes, and the pairs have $d$-wave-like character \cite{poil03,tsve11}.

\begin{figure}[b]
\begin{center}
\includegraphics[width=\columnwidth]{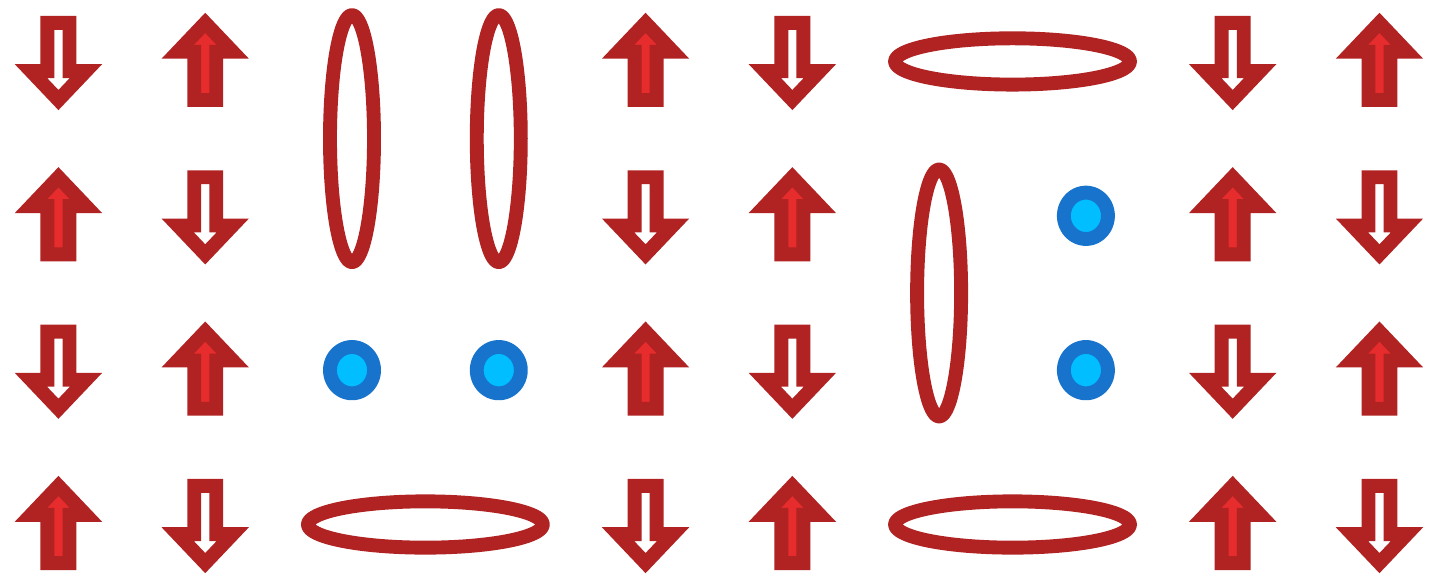}
\end{center}
\caption{Cartoon of cuprate spin stripe order at $p=1/8$, with resulting pairing correlations within the charge stripes, as proposed in \cite{tran21a}.  Here, only Cu sites are shown.  Arrows indicate ordered spins; blue circles are doped holes; ellipses are spin singlets on pairs of Cu sites. }
\end{figure}   

Note that the RVB state of the 2-leg ladder is a gapped spin liquid, in contrast to the gapless quantum spin liquid of Anderson's proposed RVB for the 2D square lattice \cite{ande87}. It is closer to the short-range RVB of Kivelson, Rokshar, and Sethna \cite{kive87}, in which a coupling to nearest-neighbor bond-length fluctuations (Peierls mechanism) stabliizes the singlet correlations.  In the stripe case, the charge segregation that enables the doped ladders is stabilized by soft phonons and a lattice distortion \cite{tran95a,rezn06,miao18,peng20,wang21a}.

The distinction between the spin stripes and the doped ladders breaks down for excitations above the pairing scale.  Such high-energy excitations can occur anywhere in the plane, and at such energies the holes are no longer confined to pairs within charge stripes.  The strong scattering between spins and holes leads to the short correlation length at high energies.

It is recognized that superconducting and charge-density-wave correlations compete with one another in 1D \cite{kive98}.  Recent calculations on 2-leg ladder models suggest that superconducting correlations survive in a 1-band Hubbard model \cite{dolf15} but not in a 3-band Hubbard model \cite{song21}.  In the experimental case of interest, we do not have individual ladders; while the spin components are decoupled by the magnetic topology, the holes in neighboring ladders will interact by long-range Coulomb repulsion and possibly other effects not considered in the calculations.  Furthermore, superconducting order requires Josephson coupling between the charge stripes \cite{emer97}.  So the important question is, what happens regarding superconductivity in experiments?

%In a charge-transfer insulator (cuprates and nickelates), antiferromagnetic order is present at low temperature and the low-energy excitations are magnetic; charge excitations are gapped.

%Hole doping leads to rapid destruction of AF order in cuprates; clearly, AF order and doped holes do not like to coexist.

%Correlated insulator with superexchange and local Cu moments

%Motion of doped holes flips spins, disrupting AF order; but $J$ is substantial, and at small hole concentrations, it is energetically favorable to maintain local AF correlations.

%\section{Pairing in charge stripes}

\section{PDW order}

In the case of optimal stripe order, LBCO with $x=1/8$, 2D superconducting correlations were observed to set in together with the spin-stripe order at $\sim40$~K \cite{li07,tran08}; 2D superconducting order was established through a Berezinskii-Kosterlitz-Thouless transition at 16~K, with 3D superconductivity developing only at $\sim5$~K.  Related behavior was observed in Nd-doped LSCO with $x=0.15$ \cite{ding08}, where the transition to the crystal structure that pins stripe order can be tuned with increasing Nd concentration \cite{buch94a}; measurements of $c$-axis optical conductivity demonstrated the loss of 3D superconductivity as the Nd concentration was tuned through the structural transition \cite{taji01}.

To explain the 2D superconductivity, a novel superconducting state was proposed: pair-density-wave order \cite{hime02,berg07}.  In the PDW state, the pair wave function oscillates from positive on one charge stripe to negative on the next, passing through zero in the spin stripes.  Because the stripe order is pinned to a lattice anisotropy that rotates by $90^\circ$ on passing from one layer to the next along the $c$ axis \cite{axe94}, the interlayer Josephson coupling associated with PDW order should be frustrated.  

The PDW order is characterized by a finite wave vector that matches that of the spin stripe order.  This finite-momentum of pairs is shared with the concept of superconductivity in a strong uniform magnetic field proposed by Fulde and Farrel \cite{fuld64} and Larkin and Ochinnikov \cite{lark64}; the difference is the absence of a net magnetic field.  (Experimental evidence for a field-induced FFLO state in a layered organic superconductor has been reported fairly recently \cite{kout16,sugi19}.)  It is also apt to note that there have been other proposals for pairing based on charge-density waves (CDWs) in cuprates.  In particular, Castellani, Di Castro, Grilli, and coworkers \cite{cast95,gril09,capr17} have proposed that dynamical CDWs underly the superconductivity of cuprates.  The fluctuating CDWs would provide a pairing interaction between extended quasiparticles as in the general case of bosonic fluctuations near a quantum critical point \cite{wang16}.  From this perspective, static CDW order would tend to compete with superconductivity.

The evidence of 2D superconductivity in LBCO, together with the PDW proposal, supported the alternative concept of intertwined order \cite{frad15}.  Here the idea is that the interactions that drive pairing and spin order actually work together, but benefit from spatial segregation.  This approach builds on theoretical evidence that static spatial inhomogeneity can enhance pairing \cite{carl03,tsai08,jian21b}.  The concept of pairing within charge stripes has also had to evolve.  The initial proposal for pairing in charge stripes relied on interacting with a spin gap in the neighboring spin stripes \cite{emer97}, which is not consistent with the presence of spin stripe order.  The idea of charge stripes as doped 2-leg spin ladders resolves this problem \cite{tran21a}.

While the proposed PDW state can explain the 2D superconductivity in LBCO, the story would be more compelling with direct evidence for PDW order in LBCO.  Phase-sensitive evidence has now been reported \cite{loza21b}.  Yang \cite{yang13} had predicted that one could (at least partially) restore the interlayer Josephson coupling by application of an in-plane magnetic field, and that the maximum effect would occur with the field at $45^\circ$ to the in-plane Cu-O bonds.  This angular dependence of the superconducting critical current density is now confirmed by experiment \cite{loza21b}.  Hence, PDW order coexisting with spin stripe order is experimentally verified.  The conclusion that the charge stripes are the source of pairing seems unavoidable.

Further evidence of the last conclusion comes from transport measurements in a large magnetic field applied perpendicular to the planes.  Such measurements on LBCO $x=1/8$ revealed, beyond a reentrant 2D superconducting phase at a field of 20 T, an ultra-quantum metal phase, with a very large sheet resistance (twice the quantum of resistance for pairs) that appears to saturate at low temperature \cite{li19a}.  The Hall resistance in this phase, as in the 3D and 2D superconducting phases, is zero within the error bars.  (Similar results have been obtained for La$_{1.7}$Eu$_{0.2}$Sr$_{0.10}$CuO$_4$ and La$_{1.48}$Nd$_{0.4}$Sr$_{0.12}$CuO$_4$ \cite{shi21}.)  A possible interpretation of the Hall resistance at high field is that the doped holes remain paired, even with the loss of PDW phase coherence between neighboring charge stripes.  Theoretically, if one takes the disorder into account, this could be a Bose-metal phase \cite{ren20}.

%pairing correlations even at high field perpendicular to the planes \cite{li19a}.
%Also found evidence for pairing in charge stripes from high-field measurements \cite{li19a}.
%Also seen in LNSCO and LESCO (Dragana Popovic)

Incoherent stripe correlations are also present in the normal state of LSCO \cite{miao21,wen19}.  Intriguingly, a study of shot noise in tunnel junctions involving LSCO films has found evidence for pairs in the the normal state of underdoped samples \cite{zhou19}.  That result is at least compatible with the concept of pair correlations in the charge stripes even at $T>T_c$.
%pairs in normal state---incoherent pairs in stripes; cite Bozovic noise study \cite{zhou19}

%Observation of 2D SC

%Explained by proposed PDW order \cite{hime02,berg07}.

%Now PDW order has been directly confirmed \cite{loza21b}.

%Stripes pinned by LTT phase.  Induced by Nd substitution for La in \lsco.  For $x=0.15$, superconductivity along the $c$ axis disappears when the structure becomes LTT and stripes are pinned \cite{taji01}.  For LBCO $x=1/8$, 2D SC directly observed to onset with spin-stripe order \cite{li07}.  

%\section{Decoupled magnetic excitations}

%LSNO

%LBCO

%similar dispersions in other cuprates

%\section{Pairing scheme}

%doped spin ladders

%differences from other schemes---not associated with QCP

\section{PDW vs. uniform $d$-wave superconductivity}

Dynamic topological doping, in the form of incommensurate spin excitations, is a common feature of underdoped cuprates \cite{enok13}.  Spectroscopically, the differences between cuprates with PDW order \cite{vall06,he09,home12} and those with uniform $d$-wave superconductivity \cite{dama03,baso05} are small, while charge stripes, static or dynamic, are common \cite{fran20,comi16,uchi21}.  Hence, it seems quite reasonable to propose that charge stripes are the common pairing centers.

The difference between the PDW and uniform superconducting states is associated with the presence or absence of static spin-stripe order.  In the PDW state, static spin-stripe order is essential for the pair correlations to develop (anti-)phase coherence between neighboring charge stripes; purely fluctuating spin stripes oppose superconducting phase order.  We should take a moment to acknowledge that it is surprising that we can have such spin order at all.  There is already a large tendency toward spin fluctuations in the undoped CuO$_2$ planes \cite{chak88}, while 1D spin chains have no static order.  There must be some degree of spin anisotropy present in order for the spin stripes to order.  Besides being an open question, this represents a challenge for simulations using the Hubbard or $t$-$J$ models, as they lack any term that would tend to induce spin order.  As a consequence, attempts to identify PDW order in numerical simulations have generally been unsuccessful \cite{whit15}.

While the spin stripes are good for isolating the doped spin ladders that yield pairing, they stand in the way of spatially-uniform superconducting order.  If they can be gapped, then it should be possible to develop a uniform superconducting phase among electronic states at energies below the spin gap.  Indeed, an analysis of the available experimental results on cuprate families indicates that the energy gap for incommensurate spin excitations is an upper limit for the superconducting gap associated with long-range coherence \cite{li18}.  Note that the local pairing scale within the charge stripes will be larger than the coherent gap of the uniform order.  This is consistent with observations by angle-resolved photoemission \cite{lee07,yosh16,droz18} and Raman scattering \cite{munn11,sacu13} of antinodal gaps that are much larger than the scale of the coherent gap \cite{tran21a}.  On the theory side, recent density-matrix-renormalization-group calculations of the Hubbard model (on a lattice of width 4 or 6 Cu sites, with boundaries joined to form a cylinder) find that a modulation of the hopping between neighboring sites in one direction (around the circumference of the cylinder), as one might expect for charge stripes without spin order, enhances the superconducting correlations \cite{jian21b}.

The close relationship between the PDW and uniform superconducting orders is illustrated by a study of the phase transitions in LBCO $x=0.115$ as a function of uniaxial strain \cite{gugu20}.  In the absence of strain, bulk susceptibility measurements suggest an onset of 2D superconducting correlations, along with spin-stripe order, near 40~K; however, the spin-stripe order is weaker than at $x=1/8$, so that 3D superconducting order develops below $\sim12$~K \cite{huck11}.  Application of significant in-plane stress causes the bulk superconducting $T_c$ to rise to 32~K, while muon-spin-rotation spectra indicated a reduction of the magnetically-ordered volume fraction by more than 50\%, consistent with a decrease in the volume of spin-stripe order and associated PDW order \cite{gugu20}.  While the dominant character of the superconducting state changes under strain, the onset temperature of superconducting coherence changes relatively little.

Another connection is seen through the impact of proton irradiation on LBCO $x=1/8$ \cite{lero19}.  Protons create narrow tracks of structural defects, often used as pinning centers for magnetic vortices.  In LBCO, moderate proton irradiation resulted in an increase in the bulk $T_c$, from 4~K to 6~K, while also reducing the correlation length of the charge stripes.  It is difficult to see how the structural disorder induced by the bombardment would directly enhance pairing.  Instead, the induced disorder must modify the coherent coupling among the correlated pairs already present.

%Strain can cause transition from PDW order to uniform SC, as indicated by $\mu$SR study on LBCO $x=0.115$ \cite{gugu20}.  Shift in $T_c$ from 2D to 3D is not big, and volume fraction with local spin order decreases substantially.  Confirms relevance of spin order.

%mention Jiang and SAK paper here.

%the same but different

%spectroscopically, the SC states look similar, at least in-plane

%LBCO, don't see QP peaks in ARPES, but dispersion similar and large antinodal gap; I have argued that the near-nodal dispersion in LBCO is a consequence of the spin-stripe order.

%In Bi2212, QP peaks go away in the normal state

%topological spin excitations limit $T_c$ of uniform phase

%IC spin excitations are common to underdoped cuprates \cite{enok13}

%importance of spin order
%coupling of static spin order and SC phase coupling

Lee \cite{lee14} has proposed that PDW order is the dominant order in cuprates and that it explains the pseudogap behavior.  While the proposal is interesting, there are a number of problematic issues with it.  For one thing, the PDW order in LBCO $x=1/8$ sets in at a temperature far below the $T^*$ crossover temperature associated with pseudogap phenomena.  While there are dynamic charge and spin stripes at higher temperatures \cite{miao17,fuji04}, and there could be pairing correlations within those dynamic charge stripes, there is no evidence of coherence of pairs between neighboring charge stripes, which would be essential for a reasonable definition of PDW correlations, let alone PDW order.  For another thing, PDW order as defined in \cite{berg09b} is not generic to most cuprate families.  For example, in \ybco, charge-density-wave order develops together with a gap in the incommensurate spin excitations \cite{huck14}.  As discussed above, the spin gap is compatible with the uniform $d$-wave superconductivity that orders at lower temperatures.

%PAL's suggestion of PDW as dominant state \cite{lee14}

%direct evidence for PDW order in LBCO

Another distinction between different cuprates concerns the magnitude of the wave vector ${\bf Q}_{\rm co}$ for the charge order and its variation with doping.  In 214 cuprates, $Q_{\rm co}$ grows linearly with hole density $p$ up to $p\approx 1/8$, where it saturates at $\approx1/8$ reciprocal lattice unit (rlu) \cite{birg06,huck11}.  Common contrasting behavior is typified by \ybco, where $Q_{\rm co}$ starts at $\sim0.34$ rlu for $p\approx 0.08$, and then decreases by about $\sim10$\%\ with doping \cite{huck14,blan14}.  These distinct doping dependences raise questions about the relationship between the orders in different compounds.

A new study of the doping and temperature dependence of the charge-stripe order in \lesco\ brings new insight to this issue \cite{lee21c}.  At low temperature, where both charge- and spin-stripe orders are observed, $Q_{\rm co}$ follows the behavior identified in other 214 systems.  With rising temperature, however, $Q_{\rm co}$ tends to grow in the disordered regime, especially at smaller hole density.  A physically-inspired Landau-Ginzburg model, when fitted to the temperature-dependent $Q_{\rm co}$ measurements, provides an extrapolation that, at $T\sim400$~K, shows behavior of $Q_{\rm co}$ vs.\ $p$ very similar to that found in YBCO \cite{huck14,blan14}.  Hence, it is plausible that the charge orders found in various cuprate families have a common origin.

\section{PDW around defects}

Scanning tunneling microscopy (STM) experiments have provided evidence for a local coexistence of PDW and uniform superconducting orders around defects in the superconducting order, such as magnetic vortex cores \cite{edki19,hoff02}.  The detected signature corresponds to an induced charge modulation that results from the superimposed, locally-coexisting orders \cite{berg09b}.  The main system studied by STM has been \bscco, which tends to have a large spin gap \cite{xu09}.  Local, short-range PDW correlations have also been detected through spatial modulation of the superconducting gap \cite{du20}.

The absence of spin-stripe order removes the conflict between PDW and uniform superconducting orders discussed above.  At the same time, the PDW order detected by STM appears to be induced by local defects, which is distinct from the PDW ground state detected in LBCO, where spin-stripe order appears to be an intrinsic component.  A defect such as a magnetic vortex or a Zn atom substituted for Cu causes the superconducting order to go to zero \cite{nach96}; that may make PDW order energetically favored in the local environment \cite{tran21b}. 

%STM of vortices  (PAL and cheap vortices)

This line of reasoning provides an interesting connection to the LBCO system.  A crystal with $x=0.095$ shows a bulk $T_c = 32$~K in zero magnetic field, along with weakened stripe order relative to $x=1/8$ \cite{huck11}.  Application of a $c$-axis magnetic field causes an enhancement of stripe order and a decoupling of the superconducting planes \cite{wen12b,steg13}, presumably due to dominance of PDW order.  The regions of uniform superconductivity should act as pinning centers for the magnetic vortices, since one can get an energy gain from inducing PDW order there.

If Zn defects act like magnetic vortices in terms of locally favoring PDW order, then enough Zn should, like the magnetic field, cause a decoupling of superconducting planes.  Indeed, this effect was confirmed in a crystal of LBCO $x=0.095$ with 1\%\ Zn \cite{loza21a}.  Similar behavior has been observed in \lsco\ with $x=0.13$ and 1\%\ Fe \cite{huan21}.  One difference between Zn defects and magnetic vortices is that Zn is known to induce pinning of spin-stripe order \cite{hiro01}; however, it may lead to a reduction in the spin-stripe-ordering temperature when introduced to a system that already has strong spin-stripe order \cite{gugu17}.

%Zn known to induce stripe order, but with reduced ordering temperature - Guguchia.

%compare irradiation of LBCO raising $T_c$; modification of lattice distortion?  Allowing coupling along $c$-axis?  Similar to in-plane field?

%LBCO $x=0.095$ must have some uniform SC at zero field, but $c$-axis magnetic field kills enough of that order to decouple the planes \cite{wen12b,steg13}.  The field also enhances the spin stripe order.  Given that PDW order seems to be favored around defects such as vortex cores \cite{edki19,hoff02} and Zn dopants \cite{loza21a}, it seems likely that the initial vortex cores are centered in regions with uniform $d$-wave.  That provides pinning that stabilizes the vortex lattice at high field.

%compare YBCO

\section{Relations to other superconductors}

Topological doping is important in cuprates because it establishes regions of reduced dimensionality where pairing can develop in the presence of repulsive interactions.   There is a natural connection with systems in which the lattice is formed from a coupling of lower-dimensional components.  One example is alkali-doped C$_{60}$ \cite{gani10}, where the dopants provide the charge carriers while the the C$_{60}$ molecules provide the interactions.  Chakravarty and Kivelson \cite{chak01c} proposed a model in which electrons could minimize repulsive interactions by hopping onto C$_{60}$ molecules in pairs.  In fact, they made a direct comparison to pairing in a doped 2-leg spin ladder.  Another obvious parallel is with organic superconductors \cite{arda12}, where superconductivity tends to occur in proximity to spin-density-wave order \cite{kang10,kawa19}.

%Connection with systems made up of lower-D components that are coupled together.
%Comparison with doped C$_{60}$.  Pairing on substructures that get coupled together.  Organic SC

The situation is different if we compare with other layered superconductors.  In electron-doped cuprates such as Nd$_{2-x}$Ce$_x$CuO$_4$, the carriers and the Cu moments do not spatially segregate.  As a result, commensurate antiferromagnetic order survives to a higher carrier concentration, and superconductivity appears only when static order disappears \cite{moto07}.  There is a good deal of commensurate inelastic magnetic spectral weight at low energy.  If this were a good thing for superconductivity, then one might expect to get a very high $T_c$; instead, the highest $T_c$ is lower than that in essentially all hole-doped families of cuprate superconductors.  Neutron scattering studies on electron-doped superconductors show that the low-energy antiferromagnetic excitations become gapped on an energy scale comparable to the superconducting gap \cite{yama03,zhao07,yu10}.

\section{Acknowledgment}Work at Brookhaven is supported by the Office of Basic Energy Sciences, Materials Sciences and Engineering Division, U.S. Department of Energy (DOE) under Contract No.\ DE-SC0012704.

%\conflictsofinterest{The author declares no conflict of interest.}

%\end{paracol}

%\reftitle{References}

%\externalbibliography{yes}
%\bibliographystyle{mdpi}
\bibliography{LNO,theory,topological}

\begin{thebibliography}{122}%
\makeatletter
\providecommand \@ifxundefined [1]{%
 \@ifx{#1\undefined}
}%
\providecommand \@ifnum [1]{%
 \ifnum #1\expandafter \@firstoftwo
 \else \expandafter \@secondoftwo
 \fi
}%
\providecommand \@ifx [1]{%
 \ifx #1\expandafter \@firstoftwo
 \else \expandafter \@secondoftwo
 \fi
}%
\providecommand \natexlab [1]{#1}%
\providecommand \enquote  [1]{``#1''}%
\providecommand \bibnamefont  [1]{#1}%
\providecommand \bibfnamefont [1]{#1}%
\providecommand \citenamefont [1]{#1}%
\providecommand \href@noop [0]{\@secondoftwo}%
\providecommand \href [0]{\begingroup \@sanitize@url \@href}%
\providecommand \@href[1]{\@@startlink{#1}\@@href}%
\providecommand \@@href[1]{\endgroup#1\@@endlink}%
\providecommand \@sanitize@url [0]{\catcode `\\12\catcode `\$12\catcode
  `\&12\catcode `\#12\catcode `\^12\catcode `\_12\catcode `\%12\relax}%
\providecommand \@@startlink[1]{}%
\providecommand \@@endlink[0]{}%
\providecommand \url  [0]{\begingroup\@sanitize@url \@url }%
\providecommand \@url [1]{\endgroup\@href {#1}{\urlprefix }}%
\providecommand \urlprefix  [0]{URL }%
\providecommand \Eprint [0]{\href }%
\providecommand \doibase [0]{https://doi.org/}%
\providecommand \selectlanguage [0]{\@gobble}%
\providecommand \bibinfo  [0]{\@secondoftwo}%
\providecommand \bibfield  [0]{\@secondoftwo}%
\providecommand \translation [1]{[#1]}%
\providecommand \BibitemOpen [0]{}%
\providecommand \bibitemStop [0]{}%
\providecommand \bibitemNoStop [0]{.\EOS\space}%
\providecommand \EOS [0]{\spacefactor3000\relax}%
\providecommand \BibitemShut  [1]{\csname bibitem#1\endcsname}%
\let\auto@bib@innerbib\@empty
%</preamble>
\bibitem [{\citenamefont {Frano}\ \emph {et~al.}(2020)\citenamefont {Frano},
  \citenamefont {Blanco-Canosa}, \citenamefont {Keimer},\ and\ \citenamefont
  {Birgeneau}}]{fran20}%
  \BibitemOpen
  \bibfield  {author} {\bibinfo {author} {\bibfnamefont {A.}~\bibnamefont
  {Frano}}, \bibinfo {author} {\bibfnamefont {S.}~\bibnamefont
  {Blanco-Canosa}}, \bibinfo {author} {\bibfnamefont {B.}~\bibnamefont
  {Keimer}},\ and\ \bibinfo {author} {\bibfnamefont {R.~J.}\ \bibnamefont
  {Birgeneau}},\ }\bibfield  {title} {\bibinfo {title} {{Charge ordering in
  superconducting copper oxides}},\ }\href
  {https://doi.org/10.1088/1361-648x/ab6140} {\bibfield  {journal} {\bibinfo
  {journal} {J. Phys. Condens. Matter}\ }\textbf {\bibinfo {volume} {32}},\
  \bibinfo {pages} {374005} (\bibinfo {year} {2020})}\BibitemShut {NoStop}%
\bibitem [{\citenamefont {Comin}\ and\ \citenamefont
  {Damascelli}(2016)}]{comi16}%
  \BibitemOpen
  \bibfield  {author} {\bibinfo {author} {\bibfnamefont {R.}~\bibnamefont
  {Comin}}\ and\ \bibinfo {author} {\bibfnamefont {A.}~\bibnamefont
  {Damascelli}},\ }\bibfield  {title} {\bibinfo {title} {{Resonant X-Ray
  Scattering Studies of Charge Order in Cuprates}},\ }\href
  {https://doi.org/10.1146/annurev-conmatphys-031115-011401} {\bibfield
  {journal} {\bibinfo  {journal} {Annu. Rev. Condens. Matter Phys.}\ }\textbf
  {\bibinfo {volume} {7}},\ \bibinfo {pages} {369} (\bibinfo {year}
  {2016})}\BibitemShut {NoStop}%
\bibitem [{\citenamefont {Uchida}(2021)}]{uchi21}%
  \BibitemOpen
  \bibfield  {author} {\bibinfo {author} {\bibfnamefont {S.-i.}\ \bibnamefont
  {Uchida}},\ }\bibfield  {title} {\bibinfo {title} {{Ubiquitous Charge Order
  Correlations in High-Temperature Superconducting Cuprates}},\ }\href
  {https://doi.org/10.7566/JPSJ.90.111001} {\bibfield  {journal} {\bibinfo
  {journal} {J. Phys. Soc. Jpn.}\ }\textbf {\bibinfo {volume} {90}},\ \bibinfo
  {pages} {111001} (\bibinfo {year} {2021})}\BibitemShut {NoStop}%
\bibitem [{\citenamefont {Fujita}\ \emph {et~al.}(2004)\citenamefont {Fujita},
  \citenamefont {Goka}, \citenamefont {Yamada}, \citenamefont {Tranquada},\
  and\ \citenamefont {Regnault}}]{fuji04}%
  \BibitemOpen
  \bibfield  {author} {\bibinfo {author} {\bibfnamefont {M.}~\bibnamefont
  {Fujita}}, \bibinfo {author} {\bibfnamefont {H.}~\bibnamefont {Goka}},
  \bibinfo {author} {\bibfnamefont {K.}~\bibnamefont {Yamada}}, \bibinfo
  {author} {\bibfnamefont {J.~M.}\ \bibnamefont {Tranquada}},\ and\ \bibinfo
  {author} {\bibfnamefont {L.~P.}\ \bibnamefont {Regnault}},\ }\bibfield
  {title} {\bibinfo {title} {{Stripe order, depinning, and fluctuations in
  ${\mathrm{La}}_{1.875}{\mathrm{Ba}}_{0.125}{\mathrm{CuO}}_{4}$ and
  ${\mathrm{La}}_{1.875}{\mathrm{Ba}}_{0.075}{\mathrm{Sr}}_{0.050}{\mathrm{CuO}}_{4}$}},\
  }\href {https://doi.org/10.1103/PhysRevB.70.104517} {\bibfield  {journal}
  {\bibinfo  {journal} {Phys. Rev. B}\ }\textbf {\bibinfo {volume} {70}},\
  \bibinfo {pages} {104517} (\bibinfo {year} {2004})}\BibitemShut {NoStop}%
\bibitem [{\citenamefont {H\"ucker}\ \emph {et~al.}(2011)\citenamefont
  {H\"ucker}, \citenamefont {v.~Zimmermann}, \citenamefont {Gu}, \citenamefont
  {Xu}, \citenamefont {Wen}, \citenamefont {Xu}, \citenamefont {Kang},
  \citenamefont {Zheludev},\ and\ \citenamefont {Tranquada}}]{huck11}%
  \BibitemOpen
  \bibfield  {author} {\bibinfo {author} {\bibfnamefont {M.}~\bibnamefont
  {H\"ucker}}, \bibinfo {author} {\bibfnamefont {M.}~\bibnamefont
  {v.~Zimmermann}}, \bibinfo {author} {\bibfnamefont {G.~D.}\ \bibnamefont
  {Gu}}, \bibinfo {author} {\bibfnamefont {Z.~J.}\ \bibnamefont {Xu}}, \bibinfo
  {author} {\bibfnamefont {J.~S.}\ \bibnamefont {Wen}}, \bibinfo {author}
  {\bibfnamefont {G.}~\bibnamefont {Xu}}, \bibinfo {author} {\bibfnamefont
  {H.~J.}\ \bibnamefont {Kang}}, \bibinfo {author} {\bibfnamefont
  {A.}~\bibnamefont {Zheludev}},\ and\ \bibinfo {author} {\bibfnamefont
  {J.~M.}\ \bibnamefont {Tranquada}},\ }\bibfield  {title} {\bibinfo {title}
  {{Stripe order in superconducting La$_{2-x}$Ba$_{x}$CuO$_{4}$
  ($0.095\le{}x\le{}0.155$)}},\ }\href
  {https://doi.org/10.1103/PhysRevB.83.104506} {\bibfield  {journal} {\bibinfo
  {journal} {Phys. Rev. B}\ }\textbf {\bibinfo {volume} {83}},\ \bibinfo
  {pages} {104506} (\bibinfo {year} {2011})}\BibitemShut {NoStop}%
\bibitem [{\citenamefont {Fujita}\ \emph {et~al.}(2012)\citenamefont {Fujita},
  \citenamefont {Hiraka}, \citenamefont {Matsuda}, \citenamefont {Matsuura},
  \citenamefont {M.~Tranquada}, \citenamefont {Wakimoto}, \citenamefont {Xu},\
  and\ \citenamefont {Yamada}}]{fuji12a}%
  \BibitemOpen
  \bibfield  {author} {\bibinfo {author} {\bibfnamefont {M.}~\bibnamefont
  {Fujita}}, \bibinfo {author} {\bibfnamefont {H.}~\bibnamefont {Hiraka}},
  \bibinfo {author} {\bibfnamefont {M.}~\bibnamefont {Matsuda}}, \bibinfo
  {author} {\bibfnamefont {M.}~\bibnamefont {Matsuura}}, \bibinfo {author}
  {\bibfnamefont {J.}~\bibnamefont {M.~Tranquada}}, \bibinfo {author}
  {\bibfnamefont {S.}~\bibnamefont {Wakimoto}}, \bibinfo {author}
  {\bibfnamefont {G.}~\bibnamefont {Xu}},\ and\ \bibinfo {author}
  {\bibfnamefont {K.}~\bibnamefont {Yamada}},\ }\bibfield  {title} {\bibinfo
  {title} {{Progress in Neutron Scattering Studies of Spin Excitations in
  High-$T_c$ Cuprates}},\ }\href {https://doi.org/10.1143/JPSJ.81.011007}
  {\bibfield  {journal} {\bibinfo  {journal} {J. Phys. Soc. Jpn.}\ }\textbf
  {\bibinfo {volume} {81}},\ \bibinfo {pages} {011007} (\bibinfo {year}
  {2012})}\BibitemShut {NoStop}%
\bibitem [{\citenamefont {Wen}\ \emph {et~al.}(2019)\citenamefont {Wen},
  \citenamefont {Huang}, \citenamefont {Lee}, \citenamefont {Jang},
  \citenamefont {Knight}, \citenamefont {Lee}, \citenamefont {Fujita},
  \citenamefont {Suzuki}, \citenamefont {Asano}, \citenamefont {Kivelson},
  \citenamefont {Kao},\ and\ \citenamefont {Lee}}]{wen19}%
  \BibitemOpen
  \bibfield  {author} {\bibinfo {author} {\bibfnamefont {J.~J.}\ \bibnamefont
  {Wen}}, \bibinfo {author} {\bibfnamefont {H.}~\bibnamefont {Huang}}, \bibinfo
  {author} {\bibfnamefont {S.~J.}\ \bibnamefont {Lee}}, \bibinfo {author}
  {\bibfnamefont {H.}~\bibnamefont {Jang}}, \bibinfo {author} {\bibfnamefont
  {J.}~\bibnamefont {Knight}}, \bibinfo {author} {\bibfnamefont {Y.~S.}\
  \bibnamefont {Lee}}, \bibinfo {author} {\bibfnamefont {M.}~\bibnamefont
  {Fujita}}, \bibinfo {author} {\bibfnamefont {K.~M.}\ \bibnamefont {Suzuki}},
  \bibinfo {author} {\bibfnamefont {S.}~\bibnamefont {Asano}}, \bibinfo
  {author} {\bibfnamefont {S.~A.}\ \bibnamefont {Kivelson}}, \bibinfo {author}
  {\bibfnamefont {C.~C.}\ \bibnamefont {Kao}},\ and\ \bibinfo {author}
  {\bibfnamefont {J.~S.}\ \bibnamefont {Lee}},\ }\bibfield  {title} {\bibinfo
  {title} {{Observation of two types of charge-density-wave orders in
  superconducting La$_{2-x}$Sr$_x$CuO$_4$}},\ }\href
  {https://doi.org/10.1038/s41467-019-11167-z} {\bibfield  {journal} {\bibinfo
  {journal} {Nat. Commun.}\ }\textbf {\bibinfo {volume} {10}},\ \bibinfo
  {pages} {3269} (\bibinfo {year} {2019})}\BibitemShut {NoStop}%
\bibitem [{\citenamefont {Miao}\ \emph {et~al.}(2021)\citenamefont {Miao},
  \citenamefont {Fabbris}, \citenamefont {Koch}, \citenamefont {Mazzone},
  \citenamefont {Nelson}, \citenamefont {Acevedo-Esteves}, \citenamefont {Gu},
  \citenamefont {Li}, \citenamefont {Yilimaz}, \citenamefont {Kaznatcheev},
  \citenamefont {Vescovo}, \citenamefont {Oda}, \citenamefont {Kurosawa},
  \citenamefont {Momono}, \citenamefont {Assefa}, \citenamefont {Robinson},
  \citenamefont {Bozin}, \citenamefont {Tranquada}, \citenamefont {Johnson},\
  and\ \citenamefont {Dean}}]{miao21}%
  \BibitemOpen
  \bibfield  {author} {\bibinfo {author} {\bibfnamefont {H.}~\bibnamefont
  {Miao}}, \bibinfo {author} {\bibfnamefont {G.}~\bibnamefont {Fabbris}},
  \bibinfo {author} {\bibfnamefont {R.~J.}\ \bibnamefont {Koch}}, \bibinfo
  {author} {\bibfnamefont {D.~G.}\ \bibnamefont {Mazzone}}, \bibinfo {author}
  {\bibfnamefont {C.~S.}\ \bibnamefont {Nelson}}, \bibinfo {author}
  {\bibfnamefont {R.}~\bibnamefont {Acevedo-Esteves}}, \bibinfo {author}
  {\bibfnamefont {G.~D.}\ \bibnamefont {Gu}}, \bibinfo {author} {\bibfnamefont
  {Y.}~\bibnamefont {Li}}, \bibinfo {author} {\bibfnamefont {T.}~\bibnamefont
  {Yilimaz}}, \bibinfo {author} {\bibfnamefont {K.}~\bibnamefont
  {Kaznatcheev}}, \bibinfo {author} {\bibfnamefont {E.}~\bibnamefont
  {Vescovo}}, \bibinfo {author} {\bibfnamefont {M.}~\bibnamefont {Oda}},
  \bibinfo {author} {\bibfnamefont {T.}~\bibnamefont {Kurosawa}}, \bibinfo
  {author} {\bibfnamefont {N.}~\bibnamefont {Momono}}, \bibinfo {author}
  {\bibfnamefont {T.}~\bibnamefont {Assefa}}, \bibinfo {author} {\bibfnamefont
  {I.~K.}\ \bibnamefont {Robinson}}, \bibinfo {author} {\bibfnamefont {E.~S.}\
  \bibnamefont {Bozin}}, \bibinfo {author} {\bibfnamefont {J.~M.}\ \bibnamefont
  {Tranquada}}, \bibinfo {author} {\bibfnamefont {P.~D.}\ \bibnamefont
  {Johnson}},\ and\ \bibinfo {author} {\bibfnamefont {M.~P.~M.}\ \bibnamefont
  {Dean}},\ }\bibfield  {title} {\bibinfo {title} {{Charge density waves in
  cuprate superconductors beyond the critical doping}},\ }\href
  {https://doi.org/10.1038/s41535-021-00327-4} {\bibfield  {journal} {\bibinfo
  {journal} {npj Quantum Mater.}\ }\textbf {\bibinfo {volume} {6}},\ \bibinfo
  {pages} {31} (\bibinfo {year} {2021})}\BibitemShut {NoStop}%
\bibitem [{\citenamefont {Tranquada}\ \emph {et~al.}(1995)\citenamefont
  {Tranquada}, \citenamefont {Sternlieb}, \citenamefont {Axe}, \citenamefont
  {Nakamura},\ and\ \citenamefont {Uchida}}]{tran95a}%
  \BibitemOpen
  \bibfield  {author} {\bibinfo {author} {\bibfnamefont {J.~M.}\ \bibnamefont
  {Tranquada}}, \bibinfo {author} {\bibfnamefont {B.~J.}\ \bibnamefont
  {Sternlieb}}, \bibinfo {author} {\bibfnamefont {J.~D.}\ \bibnamefont {Axe}},
  \bibinfo {author} {\bibfnamefont {Y.}~\bibnamefont {Nakamura}},\ and\
  \bibinfo {author} {\bibfnamefont {S.}~\bibnamefont {Uchida}},\ }\bibfield
  {title} {\bibinfo {title} {{Evidence for stripe correlations of spins and
  holes in copper oxide superconductors}},\ }\href
  {https://doi.org/10.1038/375561a0} {\bibfield  {journal} {\bibinfo  {journal}
  {Nature}\ }\textbf {\bibinfo {volume} {375}},\ \bibinfo {pages} {561}
  (\bibinfo {year} {1995})}\BibitemShut {NoStop}%
\bibitem [{\citenamefont {Ma}\ \emph {et~al.}(2021)\citenamefont {Ma},
  \citenamefont {Rule}, \citenamefont {Cronkwright}, \citenamefont {Dragomir},
  \citenamefont {Mitchell}, \citenamefont {Smith}, \citenamefont {Chi},
  \citenamefont {Kolesnikov}, \citenamefont {Stone},\ and\ \citenamefont
  {Gaulin}}]{ma21}%
  \BibitemOpen
  \bibfield  {author} {\bibinfo {author} {\bibfnamefont {Q.}~\bibnamefont
  {Ma}}, \bibinfo {author} {\bibfnamefont {K.~C.}\ \bibnamefont {Rule}},
  \bibinfo {author} {\bibfnamefont {Z.~W.}\ \bibnamefont {Cronkwright}},
  \bibinfo {author} {\bibfnamefont {M.}~\bibnamefont {Dragomir}}, \bibinfo
  {author} {\bibfnamefont {G.}~\bibnamefont {Mitchell}}, \bibinfo {author}
  {\bibfnamefont {E.~M.}\ \bibnamefont {Smith}}, \bibinfo {author}
  {\bibfnamefont {S.}~\bibnamefont {Chi}}, \bibinfo {author} {\bibfnamefont
  {A.~I.}\ \bibnamefont {Kolesnikov}}, \bibinfo {author} {\bibfnamefont
  {M.~B.}\ \bibnamefont {Stone}},\ and\ \bibinfo {author} {\bibfnamefont
  {B.~D.}\ \bibnamefont {Gaulin}},\ }\bibfield  {title} {\bibinfo {title}
  {{Parallel spin stripes and their coexistence with superconducting ground
  states at optimal and high doping in
  ${\mathrm{La}}_{1.6\ensuremath{-}x}{\mathrm{Nd}}_{0.4}{\mathrm{Sr}}_{x}{\mathrm{CuO}}_{4}$}},\
  }\href {https://doi.org/10.1103/PhysRevResearch.3.023151} {\bibfield
  {journal} {\bibinfo  {journal} {Phys. Rev. Research}\ }\textbf {\bibinfo
  {volume} {3}},\ \bibinfo {pages} {023151} (\bibinfo {year}
  {2021})}\BibitemShut {NoStop}%
\bibitem [{\citenamefont {Kivelson}\ and\ \citenamefont
  {Emery}(1996)}]{kive96}%
  \BibitemOpen
  \bibfield  {author} {\bibinfo {author} {\bibfnamefont {S.}~\bibnamefont
  {Kivelson}}\ and\ \bibinfo {author} {\bibfnamefont {V.}~\bibnamefont
  {Emery}},\ }\bibfield  {title} {\bibinfo {title} {Topological doping of
  correlated insulators},\ }\href
  {https://doi.org/https://doi.org/10.1016/S0379-6779(96)03696-X} {\bibfield
  {journal} {\bibinfo  {journal} {Synth. Met.}\ }\textbf {\bibinfo {volume}
  {80}},\ \bibinfo {pages} {151} (\bibinfo {year} {1996})}\BibitemShut
  {NoStop}%
\bibitem [{\citenamefont {Hasan}\ and\ \citenamefont {Kane}(2010)}]{hasa10}%
  \BibitemOpen
  \bibfield  {author} {\bibinfo {author} {\bibfnamefont {M.~Z.}\ \bibnamefont
  {Hasan}}\ and\ \bibinfo {author} {\bibfnamefont {C.~L.}\ \bibnamefont
  {Kane}},\ }\bibfield  {title} {\bibinfo {title} {{Colloquium: Topological
  Insulators}},\ }\href {https://doi.org/10.1103/RevModPhys.82.3045} {\bibfield
   {journal} {\bibinfo  {journal} {Rev. Mod. Phys.}\ }\textbf {\bibinfo
  {volume} {82}},\ \bibinfo {pages} {3045} (\bibinfo {year}
  {2010})}\BibitemShut {NoStop}%
\bibitem [{\citenamefont {Qi}\ and\ \citenamefont {Zhang}(2011)}]{qi11}%
  \BibitemOpen
  \bibfield  {author} {\bibinfo {author} {\bibfnamefont {X.-L.}\ \bibnamefont
  {Qi}}\ and\ \bibinfo {author} {\bibfnamefont {S.-C.}\ \bibnamefont {Zhang}},\
  }\bibfield  {title} {\bibinfo {title} {{Topological insulators and
  superconductors}},\ }\href {https://doi.org/10.1103/RevModPhys.83.1057}
  {\bibfield  {journal} {\bibinfo  {journal} {Rev. Mod. Phys.}\ }\textbf
  {\bibinfo {volume} {83}},\ \bibinfo {pages} {1057} (\bibinfo {year}
  {2011})}\BibitemShut {NoStop}%
\bibitem [{\citenamefont {Tranquada}(2013)}]{tran13a}%
  \BibitemOpen
  \bibfield  {author} {\bibinfo {author} {\bibfnamefont {J.~M.}\ \bibnamefont
  {Tranquada}},\ }\bibfield  {title} {\bibinfo {title} {{Spins, stripes, and
  superconductivity in hole-doped cuprates}},\ }\href
  {https://doi.org/10.1063/1.4818402} {\bibfield  {journal} {\bibinfo
  {journal} {AIP Conf. Proc.}\ }\textbf {\bibinfo {volume} {1550}},\ \bibinfo
  {pages} {114} (\bibinfo {year} {2013})}\BibitemShut {NoStop}%
\bibitem [{\citenamefont {Fujita}\ \emph {et~al.}(2002)\citenamefont {Fujita},
  \citenamefont {Yamada}, \citenamefont {Hiraka}, \citenamefont {Gehring},
  \citenamefont {Lee}, \citenamefont {Wakimoto},\ and\ \citenamefont
  {Shirane}}]{fuji02c}%
  \BibitemOpen
  \bibfield  {author} {\bibinfo {author} {\bibfnamefont {M.}~\bibnamefont
  {Fujita}}, \bibinfo {author} {\bibfnamefont {K.}~\bibnamefont {Yamada}},
  \bibinfo {author} {\bibfnamefont {H.}~\bibnamefont {Hiraka}}, \bibinfo
  {author} {\bibfnamefont {P.~M.}\ \bibnamefont {Gehring}}, \bibinfo {author}
  {\bibfnamefont {S.~H.}\ \bibnamefont {Lee}}, \bibinfo {author} {\bibfnamefont
  {S.}~\bibnamefont {Wakimoto}},\ and\ \bibinfo {author} {\bibfnamefont
  {G.}~\bibnamefont {Shirane}},\ }\bibfield  {title} {\bibinfo {title} {{Static
  magnetic correlations near the insulating-superconducting phase boundary in
  ${\mathrm{La}}_{2-x}{\mathrm{Sr}}_{x}{\mathrm{CuO}}_{4}$}},\ }\href
  {https://doi.org/10.1103/PhysRevB.65.064505} {\bibfield  {journal} {\bibinfo
  {journal} {Phys. Rev. B}\ }\textbf {\bibinfo {volume} {65}},\ \bibinfo
  {pages} {064505} (\bibinfo {year} {2002})}\BibitemShut {NoStop}%
\bibitem [{\citenamefont {Pryadko}\ \emph {et~al.}(1999)\citenamefont
  {Pryadko}, \citenamefont {Kivelson}, \citenamefont {Emery}, \citenamefont
  {Bazaliy},\ and\ \citenamefont {Demler}}]{prya99}%
  \BibitemOpen
  \bibfield  {author} {\bibinfo {author} {\bibfnamefont {L.~P.}\ \bibnamefont
  {Pryadko}}, \bibinfo {author} {\bibfnamefont {S.~A.}\ \bibnamefont
  {Kivelson}}, \bibinfo {author} {\bibfnamefont {V.~J.}\ \bibnamefont {Emery}},
  \bibinfo {author} {\bibfnamefont {Y.~B.}\ \bibnamefont {Bazaliy}},\ and\
  \bibinfo {author} {\bibfnamefont {E.~A.}\ \bibnamefont {Demler}},\ }\bibfield
   {title} {\bibinfo {title} {{Topological doping and the stability of stripe
  phases}},\ }\href {https://doi.org/10.1103/PhysRevB.60.7541} {\bibfield
  {journal} {\bibinfo  {journal} {Phys. Rev. B}\ }\textbf {\bibinfo {volume}
  {60}},\ \bibinfo {pages} {7541} (\bibinfo {year} {1999})}\BibitemShut
  {NoStop}%
\bibitem [{\citenamefont {Zaanen}\ and\ \citenamefont
  {Gunnarsson}(1989)}]{zaan89}%
  \BibitemOpen
  \bibfield  {author} {\bibinfo {author} {\bibfnamefont {J.}~\bibnamefont
  {Zaanen}}\ and\ \bibinfo {author} {\bibfnamefont {O.}~\bibnamefont
  {Gunnarsson}},\ }\bibfield  {title} {\bibinfo {title} {{Charged magnetic
  domain lines and the magnetism of high-$T_c$ oxides}},\ }\href@noop {}
  {\bibfield  {journal} {\bibinfo  {journal} {Phys. Rev. B}\ }\textbf {\bibinfo
  {volume} {40}},\ \bibinfo {pages} {7391} (\bibinfo {year}
  {1989})}\BibitemShut {NoStop}%
\bibitem [{\citenamefont {L\"ow}\ \emph {et~al.}(1994)\citenamefont {L\"ow},
  \citenamefont {Emery}, \citenamefont {Fabricius},\ and\ \citenamefont
  {Kivelson}}]{low94}%
  \BibitemOpen
  \bibfield  {author} {\bibinfo {author} {\bibfnamefont {U.}~\bibnamefont
  {L\"ow}}, \bibinfo {author} {\bibfnamefont {V.~J.}\ \bibnamefont {Emery}},
  \bibinfo {author} {\bibfnamefont {K.}~\bibnamefont {Fabricius}},\ and\
  \bibinfo {author} {\bibfnamefont {S.~A.}\ \bibnamefont {Kivelson}},\
  }\bibfield  {title} {\bibinfo {title} {{Study of an Ising model with
  competing long- and short-range interactions}},\ }\href
  {https://doi.org/10.1103/PhysRevLett.72.1918} {\bibfield  {journal} {\bibinfo
   {journal} {Phys. Rev. Lett.}\ }\textbf {\bibinfo {volume} {72}},\ \bibinfo
  {pages} {1918} (\bibinfo {year} {1994})}\BibitemShut {NoStop}%
\bibitem [{\citenamefont {White}\ and\ \citenamefont
  {Scalapino}(1998)}]{whit98a}%
  \BibitemOpen
  \bibfield  {author} {\bibinfo {author} {\bibfnamefont {S.~R.}\ \bibnamefont
  {White}}\ and\ \bibinfo {author} {\bibfnamefont {D.~J.}\ \bibnamefont
  {Scalapino}},\ }\bibfield  {title} {\bibinfo {title} {{Density Matrix
  Renormalization Group Study of the Striped Phase in the 2D
  $\mathit{t}-\mathit{J}$ Model}},\ }\href@noop {} {\bibfield  {journal}
  {\bibinfo  {journal} {Phys. Rev. Lett.}\ }\textbf {\bibinfo {volume} {80}},\
  \bibinfo {pages} {1272} (\bibinfo {year} {1998})}\BibitemShut {NoStop}%
\bibitem [{\citenamefont {Zheng}\ \emph {et~al.}(2017)\citenamefont {Zheng},
  \citenamefont {Chung}, \citenamefont {Corboz}, \citenamefont {Ehlers},
  \citenamefont {Qin}, \citenamefont {Noack}, \citenamefont {Shi},
  \citenamefont {White}, \citenamefont {Zhang},\ and\ \citenamefont
  {Chan}}]{zhen17}%
  \BibitemOpen
  \bibfield  {author} {\bibinfo {author} {\bibfnamefont {B.-X.}\ \bibnamefont
  {Zheng}}, \bibinfo {author} {\bibfnamefont {C.-M.}\ \bibnamefont {Chung}},
  \bibinfo {author} {\bibfnamefont {P.}~\bibnamefont {Corboz}}, \bibinfo
  {author} {\bibfnamefont {G.}~\bibnamefont {Ehlers}}, \bibinfo {author}
  {\bibfnamefont {M.-P.}\ \bibnamefont {Qin}}, \bibinfo {author} {\bibfnamefont
  {R.~M.}\ \bibnamefont {Noack}}, \bibinfo {author} {\bibfnamefont
  {H.}~\bibnamefont {Shi}}, \bibinfo {author} {\bibfnamefont {S.~R.}\
  \bibnamefont {White}}, \bibinfo {author} {\bibfnamefont {S.}~\bibnamefont
  {Zhang}},\ and\ \bibinfo {author} {\bibfnamefont {G.~K.-L.}\ \bibnamefont
  {Chan}},\ }\bibfield  {title} {\bibinfo {title} {{Stripe order in the
  underdoped region of the two-dimensional Hubbard model}},\ }\href
  {https://doi.org/10.1126/science.aam7127} {\bibfield  {journal} {\bibinfo
  {journal} {Science}\ }\textbf {\bibinfo {volume} {358}},\ \bibinfo {pages}
  {1155} (\bibinfo {year} {2017})}\BibitemShut {NoStop}%
\bibitem [{\citenamefont {Huang}\ \emph {et~al.}(2018)\citenamefont {Huang},
  \citenamefont {Mendl}, \citenamefont {Jiang}, \citenamefont {Moritz},\ and\
  \citenamefont {Devereaux}}]{huan18}%
  \BibitemOpen
  \bibfield  {author} {\bibinfo {author} {\bibfnamefont {E.~W.}\ \bibnamefont
  {Huang}}, \bibinfo {author} {\bibfnamefont {C.~B.}\ \bibnamefont {Mendl}},
  \bibinfo {author} {\bibfnamefont {H.-C.}\ \bibnamefont {Jiang}}, \bibinfo
  {author} {\bibfnamefont {B.}~\bibnamefont {Moritz}},\ and\ \bibinfo {author}
  {\bibfnamefont {T.~P.}\ \bibnamefont {Devereaux}},\ }\bibfield  {title}
  {\bibinfo {title} {{Stripe order from the perspective of the Hubbard
  model}},\ }\href {https://doi.org/10.1038/s41535-018-0097-0} {\bibfield
  {journal} {\bibinfo  {journal} {npj Quantum Mater.}\ }\textbf {\bibinfo
  {volume} {3}},\ \bibinfo {pages} {22} (\bibinfo {year} {2018})}\BibitemShut
  {NoStop}%
\bibitem [{\citenamefont {Tranquada}(2020)}]{tran21a}%
  \BibitemOpen
  \bibfield  {author} {\bibinfo {author} {\bibfnamefont {J.~M.}\ \bibnamefont
  {Tranquada}},\ }\bibfield  {title} {\bibinfo {title} {{Cuprate
  superconductors as viewed through a striped lens}},\ }\href
  {https://doi.org/10.1080/00018732.2021.1935698} {\bibfield  {journal}
  {\bibinfo  {journal} {Adv. Phys.}\ }\textbf {\bibinfo {volume} {69}},\
  \bibinfo {pages} {437} (\bibinfo {year} {2020})}\BibitemShut {NoStop}%
\bibitem [{\citenamefont {Dagotto}\ and\ \citenamefont {Rice}(1996)}]{dago96}%
  \BibitemOpen
  \bibfield  {author} {\bibinfo {author} {\bibfnamefont {E.}~\bibnamefont
  {Dagotto}}\ and\ \bibinfo {author} {\bibfnamefont {T.~M.}\ \bibnamefont
  {Rice}},\ }\bibfield  {title} {\bibinfo {title} {{Surprises on the Way from
  One- to Two-Dimensional Quantum Magnets: The Ladder Materials}},\ }\href
  {https://doi.org/10.1126/science.271.5249.618} {\bibfield  {journal}
  {\bibinfo  {journal} {Science}\ }\textbf {\bibinfo {volume} {271}},\ \bibinfo
  {pages} {618} (\bibinfo {year} {1996})}\BibitemShut {NoStop}%
\bibitem [{\citenamefont {Tsvelik}(2011)}]{tsve11}%
  \BibitemOpen
  \bibfield  {author} {\bibinfo {author} {\bibfnamefont {A.~M.}\ \bibnamefont
  {Tsvelik}},\ }\bibfield  {title} {\bibinfo {title} {{Field theory for a
  fermionic ladder with generic intrachain interactions}},\ }\href
  {https://doi.org/10.1103/PhysRevB.83.104405} {\bibfield  {journal} {\bibinfo
  {journal} {Phys. Rev. B}\ }\textbf {\bibinfo {volume} {83}},\ \bibinfo
  {pages} {104405} (\bibinfo {year} {2011})}\BibitemShut {NoStop}%
\bibitem [{\citenamefont {Emery}\ \emph {et~al.}(1997)\citenamefont {Emery},
  \citenamefont {Kivelson},\ and\ \citenamefont {Zachar}}]{emer97}%
  \BibitemOpen
  \bibfield  {author} {\bibinfo {author} {\bibfnamefont {V.~J.}\ \bibnamefont
  {Emery}}, \bibinfo {author} {\bibfnamefont {S.~A.}\ \bibnamefont
  {Kivelson}},\ and\ \bibinfo {author} {\bibfnamefont {O.}~\bibnamefont
  {Zachar}},\ }\bibfield  {title} {\bibinfo {title} {{Spin-gap proximity effect
  mechanism of high-temperature superconductivity}},\ }\href@noop {} {\bibfield
   {journal} {\bibinfo  {journal} {Phys. Rev. B}\ }\textbf {\bibinfo {volume}
  {56}},\ \bibinfo {pages} {6120} (\bibinfo {year} {1997})}\BibitemShut
  {NoStop}%
\bibitem [{\citenamefont {Berg}\ \emph {et~al.}(2009)\citenamefont {Berg},
  \citenamefont {Fradkin}, \citenamefont {Kivelson},\ and\ \citenamefont
  {Tranquada}}]{berg09b}%
  \BibitemOpen
  \bibfield  {author} {\bibinfo {author} {\bibfnamefont {E.}~\bibnamefont
  {Berg}}, \bibinfo {author} {\bibfnamefont {E.}~\bibnamefont {Fradkin}},
  \bibinfo {author} {\bibfnamefont {S.~A.}\ \bibnamefont {Kivelson}},\ and\
  \bibinfo {author} {\bibfnamefont {J.~M.}\ \bibnamefont {Tranquada}},\
  }\bibfield  {title} {\bibinfo {title} {{Striped superconductors: how spin,
  charge and superconducting orders intertwine in the cuprates}},\ }\href
  {http://stacks.iop.org/1367-2630/11/i=11/a=115004} {\bibfield  {journal}
  {\bibinfo  {journal} {New J. Phys.}\ }\textbf {\bibinfo {volume} {11}},\
  \bibinfo {pages} {115004} (\bibinfo {year} {2009})}\BibitemShut {NoStop}%
\bibitem [{\citenamefont {Agterberg}\ \emph {et~al.}(2020)\citenamefont
  {Agterberg}, \citenamefont {Davis}, \citenamefont {Edkins}, \citenamefont
  {Fradkin}, \citenamefont {Van~Harlingen}, \citenamefont {Kivelson},
  \citenamefont {Lee}, \citenamefont {Radzihovsky}, \citenamefont {Tranquada},\
  and\ \citenamefont {Wang}}]{agte20}%
  \BibitemOpen
  \bibfield  {author} {\bibinfo {author} {\bibfnamefont {D.~F.}\ \bibnamefont
  {Agterberg}}, \bibinfo {author} {\bibfnamefont {J.~S.}\ \bibnamefont
  {Davis}}, \bibinfo {author} {\bibfnamefont {S.~D.}\ \bibnamefont {Edkins}},
  \bibinfo {author} {\bibfnamefont {E.}~\bibnamefont {Fradkin}}, \bibinfo
  {author} {\bibfnamefont {D.~J.}\ \bibnamefont {Van~Harlingen}}, \bibinfo
  {author} {\bibfnamefont {S.~A.}\ \bibnamefont {Kivelson}}, \bibinfo {author}
  {\bibfnamefont {P.~A.}\ \bibnamefont {Lee}}, \bibinfo {author} {\bibfnamefont
  {L.}~\bibnamefont {Radzihovsky}}, \bibinfo {author} {\bibfnamefont {J.~M.}\
  \bibnamefont {Tranquada}},\ and\ \bibinfo {author} {\bibfnamefont
  {Y.}~\bibnamefont {Wang}},\ }\bibfield  {title} {\bibinfo {title} {{The
  Physics of Pair-Density Waves: Cuprate Superconductors and Beyond}},\ }\href
  {https://doi.org/10.1146/annurev-conmatphys-031119-050711} {\bibfield
  {journal} {\bibinfo  {journal} {Annu. Rev. Condens. Matter Phys.}\ }\textbf
  {\bibinfo {volume} {11}},\ \bibinfo {pages} {231} (\bibinfo {year}
  {2020})}\BibitemShut {NoStop}%
\bibitem [{\citenamefont {Himeda}\ \emph {et~al.}(2002)\citenamefont {Himeda},
  \citenamefont {Kato},\ and\ \citenamefont {Ogata}}]{hime02}%
  \BibitemOpen
  \bibfield  {author} {\bibinfo {author} {\bibfnamefont {A.}~\bibnamefont
  {Himeda}}, \bibinfo {author} {\bibfnamefont {T.}~\bibnamefont {Kato}},\ and\
  \bibinfo {author} {\bibfnamefont {M.}~\bibnamefont {Ogata}},\ }\bibfield
  {title} {\bibinfo {title} {{Stripe States with Spatially Oscillating $d$-Wave
  Superconductivity in the Two-Dimensional $t-t'{}-J$ Model}},\ }\href@noop {}
  {\bibfield  {journal} {\bibinfo  {journal} {Phys. Rev. Lett.}\ }\textbf
  {\bibinfo {volume} {88}},\ \bibinfo {pages} {117001} (\bibinfo {year}
  {2002})}\BibitemShut {NoStop}%
\bibitem [{\citenamefont {Berg}\ \emph {et~al.}(2007)\citenamefont {Berg},
  \citenamefont {Fradkin}, \citenamefont {Kim}, \citenamefont {Kivelson},
  \citenamefont {Oganesyan}, \citenamefont {Tranquada},\ and\ \citenamefont
  {Zhang}}]{berg07}%
  \BibitemOpen
  \bibfield  {author} {\bibinfo {author} {\bibfnamefont {E.}~\bibnamefont
  {Berg}}, \bibinfo {author} {\bibfnamefont {E.}~\bibnamefont {Fradkin}},
  \bibinfo {author} {\bibfnamefont {E.-A.}\ \bibnamefont {Kim}}, \bibinfo
  {author} {\bibfnamefont {S.~A.}\ \bibnamefont {Kivelson}}, \bibinfo {author}
  {\bibfnamefont {V.}~\bibnamefont {Oganesyan}}, \bibinfo {author}
  {\bibfnamefont {J.~M.}\ \bibnamefont {Tranquada}},\ and\ \bibinfo {author}
  {\bibfnamefont {S.~C.}\ \bibnamefont {Zhang}},\ }\bibfield  {title} {\bibinfo
  {title} {{Dynamical Layer Decoupling in a Stripe-Ordered High-${T}_{c}$
  Superconductor}},\ }\href {https://doi.org/10.1103/PhysRevLett.99.127003}
  {\bibfield  {journal} {\bibinfo  {journal} {Phys. Rev. Lett.}\ }\textbf
  {\bibinfo {volume} {99}},\ \bibinfo {pages} {127003} (\bibinfo {year}
  {2007})}\BibitemShut {NoStop}%
\bibitem [{\citenamefont {Li}\ \emph {et~al.}(2007)\citenamefont {Li},
  \citenamefont {{H\"ucker}}, \citenamefont {Gu}, \citenamefont {Tsvelik},\
  and\ \citenamefont {Tranquada}}]{li07}%
  \BibitemOpen
  \bibfield  {author} {\bibinfo {author} {\bibfnamefont {Q.}~\bibnamefont
  {Li}}, \bibinfo {author} {\bibfnamefont {M.}~\bibnamefont {{H\"ucker}}},
  \bibinfo {author} {\bibfnamefont {G.~D.}\ \bibnamefont {Gu}}, \bibinfo
  {author} {\bibfnamefont {A.~M.}\ \bibnamefont {Tsvelik}},\ and\ \bibinfo
  {author} {\bibfnamefont {J.~M.}\ \bibnamefont {Tranquada}},\ }\bibfield
  {title} {\bibinfo {title} {{Two-Dimensional Superconducting Fluctuations in
  Stripe-Ordered La$_{1.875}$Ba$_{0.125}$CuO$_4$}},\ }\href
  {https://doi.org/10.1103/PhysRevLett.99.067001} {\bibfield  {journal}
  {\bibinfo  {journal} {Phys. Rev. Lett.}\ }\textbf {\bibinfo {volume} {99}},\
  \bibinfo {eid} {067001} (\bibinfo {year} {2007})}\BibitemShut {NoStop}%
\bibitem [{\citenamefont {Tajima}\ \emph {et~al.}(2001)\citenamefont {Tajima},
  \citenamefont {Noda}, \citenamefont {Eisaki},\ and\ \citenamefont
  {Uchida}}]{taji01}%
  \BibitemOpen
  \bibfield  {author} {\bibinfo {author} {\bibfnamefont {S.}~\bibnamefont
  {Tajima}}, \bibinfo {author} {\bibfnamefont {T.}~\bibnamefont {Noda}},
  \bibinfo {author} {\bibfnamefont {H.}~\bibnamefont {Eisaki}},\ and\ \bibinfo
  {author} {\bibfnamefont {S.}~\bibnamefont {Uchida}},\ }\bibfield  {title}
  {\bibinfo {title} {{$c$-Axis Optical Response in the Static Stripe Ordered
  Phase of the Cuprates}},\ }\href@noop {} {\bibfield  {journal} {\bibinfo
  {journal} {Phys. Rev. Lett.}\ }\textbf {\bibinfo {volume} {86}},\ \bibinfo
  {pages} {500} (\bibinfo {year} {2001})}\BibitemShut {NoStop}%
\bibitem [{\citenamefont {Lozano}\ \emph
  {et~al.}(2021{\natexlab{a}})\citenamefont {Lozano}, \citenamefont {Ren},
  \citenamefont {Gu}, \citenamefont {Tsvelik}, \citenamefont {Tranquada},\ and\
  \citenamefont {Li}}]{loza21b}%
  \BibitemOpen
  \bibfield  {author} {\bibinfo {author} {\bibfnamefont {P.~M.}\ \bibnamefont
  {Lozano}}, \bibinfo {author} {\bibfnamefont {T.}~\bibnamefont {Ren}},
  \bibinfo {author} {\bibfnamefont {G.~D.}\ \bibnamefont {Gu}}, \bibinfo
  {author} {\bibfnamefont {A.~M.}\ \bibnamefont {Tsvelik}}, \bibinfo {author}
  {\bibfnamefont {J.~M.}\ \bibnamefont {Tranquada}},\ and\ \bibinfo {author}
  {\bibfnamefont {Q.}~\bibnamefont {Li}},\ }\href@noop {} {\bibinfo {title}
  {{Phase-sensitive evidence of pair-density-wave order in a cuprate}}},\
  \bibinfo {howpublished} {arXiv:2110.05513} (\bibinfo {year}
  {2021}{\natexlab{a}})\BibitemShut {NoStop}%
\bibitem [{\citenamefont {Li}\ \emph {et~al.}(2019)\citenamefont {Li},
  \citenamefont {Terzic}, \citenamefont {Baity}, \citenamefont {Popovi\'{c}},
  \citenamefont {Gu}, \citenamefont {Li}, \citenamefont {Tsvelik},\ and\
  \citenamefont {Tranquada}}]{li19a}%
  \BibitemOpen
  \bibfield  {author} {\bibinfo {author} {\bibfnamefont {Y.}~\bibnamefont
  {Li}}, \bibinfo {author} {\bibfnamefont {J.}~\bibnamefont {Terzic}}, \bibinfo
  {author} {\bibfnamefont {P.~G.}\ \bibnamefont {Baity}}, \bibinfo {author}
  {\bibfnamefont {D.}~\bibnamefont {Popovi\'{c}}}, \bibinfo {author}
  {\bibfnamefont {G.~D.}\ \bibnamefont {Gu}}, \bibinfo {author} {\bibfnamefont
  {Q.}~\bibnamefont {Li}}, \bibinfo {author} {\bibfnamefont {A.~M.}\
  \bibnamefont {Tsvelik}},\ and\ \bibinfo {author} {\bibfnamefont {J.~M.}\
  \bibnamefont {Tranquada}},\ }\bibfield  {title} {\bibinfo {title} {{Tuning
  from failed superconductor to failed insulator with magnetic field}},\ }\href
  {https://doi.org/10.1126/sciadv.aav7686} {\bibfield  {journal} {\bibinfo
  {journal} {Sci. Adv.}\ }\textbf {\bibinfo {volume} {5}},\ \bibinfo {pages}
  {{eaav7686}} (\bibinfo {year} {2019})}\BibitemShut {NoStop}%
\bibitem [{\citenamefont {Li}\ \emph {et~al.}(2018)\citenamefont {Li},
  \citenamefont {Zhong}, \citenamefont {Stone}, \citenamefont {Kolesnikov},
  \citenamefont {Gu}, \citenamefont {Zaliznyak},\ and\ \citenamefont
  {Tranquada}}]{li18}%
  \BibitemOpen
  \bibfield  {author} {\bibinfo {author} {\bibfnamefont {Y.}~\bibnamefont
  {Li}}, \bibinfo {author} {\bibfnamefont {R.}~\bibnamefont {Zhong}}, \bibinfo
  {author} {\bibfnamefont {M.~B.}\ \bibnamefont {Stone}}, \bibinfo {author}
  {\bibfnamefont {A.~I.}\ \bibnamefont {Kolesnikov}}, \bibinfo {author}
  {\bibfnamefont {G.~D.}\ \bibnamefont {Gu}}, \bibinfo {author} {\bibfnamefont
  {I.~A.}\ \bibnamefont {Zaliznyak}},\ and\ \bibinfo {author} {\bibfnamefont
  {J.~M.}\ \bibnamefont {Tranquada}},\ }\bibfield  {title} {\bibinfo {title}
  {{Low-energy antiferromagnetic spin fluctuations limit the coherent
  superconducting gap in cuprates}},\ }\href
  {https://doi.org/10.1103/PhysRevB.98.224508} {\bibfield  {journal} {\bibinfo
  {journal} {Phys. Rev. B}\ }\textbf {\bibinfo {volume} {98}},\ \bibinfo
  {pages} {224508} (\bibinfo {year} {2018})}\BibitemShut {NoStop}%
\bibitem [{\citenamefont {Edkins}\ \emph {et~al.}(2019)\citenamefont {Edkins},
  \citenamefont {Kostin}, \citenamefont {Fujita}, \citenamefont {Mackenzie},
  \citenamefont {Eisaki}, \citenamefont {Uchida}, \citenamefont {Sachdev},
  \citenamefont {Lawler}, \citenamefont {Kim}, \citenamefont
  {S{\'e}amus~Davis},\ and\ \citenamefont {Hamidian}}]{edki19}%
  \BibitemOpen
  \bibfield  {author} {\bibinfo {author} {\bibfnamefont {S.~D.}\ \bibnamefont
  {Edkins}}, \bibinfo {author} {\bibfnamefont {A.}~\bibnamefont {Kostin}},
  \bibinfo {author} {\bibfnamefont {K.}~\bibnamefont {Fujita}}, \bibinfo
  {author} {\bibfnamefont {A.~P.}\ \bibnamefont {Mackenzie}}, \bibinfo {author}
  {\bibfnamefont {H.}~\bibnamefont {Eisaki}}, \bibinfo {author} {\bibfnamefont
  {S.}~\bibnamefont {Uchida}}, \bibinfo {author} {\bibfnamefont
  {S.}~\bibnamefont {Sachdev}}, \bibinfo {author} {\bibfnamefont {M.~J.}\
  \bibnamefont {Lawler}}, \bibinfo {author} {\bibfnamefont {E.-A.}\
  \bibnamefont {Kim}}, \bibinfo {author} {\bibfnamefont {J.~C.}\ \bibnamefont
  {S{\'e}amus~Davis}},\ and\ \bibinfo {author} {\bibfnamefont {M.~H.}\
  \bibnamefont {Hamidian}},\ }\bibfield  {title} {\bibinfo {title} {{Magnetic
  field--induced pair density wave state in the cuprate vortex halo}},\ }\href
  {https://doi.org/10.1126/science.aat1773} {\bibfield  {journal} {\bibinfo
  {journal} {Science}\ }\textbf {\bibinfo {volume} {364}},\ \bibinfo {pages}
  {976} (\bibinfo {year} {2019})}\BibitemShut {NoStop}%
\bibitem [{\citenamefont {Du}\ \emph {et~al.}(2020)\citenamefont {Du},
  \citenamefont {Li}, \citenamefont {Joo}, \citenamefont {Donoway},
  \citenamefont {Lee}, \citenamefont {Davis}, \citenamefont {Gu}, \citenamefont
  {Johnson},\ and\ \citenamefont {Fujita}}]{du20}%
  \BibitemOpen
  \bibfield  {author} {\bibinfo {author} {\bibfnamefont {Z.}~\bibnamefont
  {Du}}, \bibinfo {author} {\bibfnamefont {H.}~\bibnamefont {Li}}, \bibinfo
  {author} {\bibfnamefont {S.~H.}\ \bibnamefont {Joo}}, \bibinfo {author}
  {\bibfnamefont {E.~P.}\ \bibnamefont {Donoway}}, \bibinfo {author}
  {\bibfnamefont {J.}~\bibnamefont {Lee}}, \bibinfo {author} {\bibfnamefont
  {J.~C.~S.}\ \bibnamefont {Davis}}, \bibinfo {author} {\bibfnamefont
  {G.}~\bibnamefont {Gu}}, \bibinfo {author} {\bibfnamefont {P.~D.}\
  \bibnamefont {Johnson}},\ and\ \bibinfo {author} {\bibfnamefont
  {K.}~\bibnamefont {Fujita}},\ }\bibfield  {title} {\bibinfo {title} {{Imaging
  the energy gap modulations of the cuprate pair-density-wave state}},\ }\href
  {https://doi.org/10.1038/s41586-020-2143-x} {\bibfield  {journal} {\bibinfo
  {journal} {Nature}\ }\textbf {\bibinfo {volume} {580}},\ \bibinfo {pages}
  {65} (\bibinfo {year} {2020})}\BibitemShut {NoStop}%
\bibitem [{\citenamefont {Merritt}\ \emph {et~al.}(2019)\citenamefont
  {Merritt}, \citenamefont {Reznik}, \citenamefont {Garlea}, \citenamefont
  {Gu},\ and\ \citenamefont {Tranquada}}]{merr19}%
  \BibitemOpen
  \bibfield  {author} {\bibinfo {author} {\bibfnamefont {A.~M.}\ \bibnamefont
  {Merritt}}, \bibinfo {author} {\bibfnamefont {D.}~\bibnamefont {Reznik}},
  \bibinfo {author} {\bibfnamefont {V.~O.}\ \bibnamefont {Garlea}}, \bibinfo
  {author} {\bibfnamefont {G.~D.}\ \bibnamefont {Gu}},\ and\ \bibinfo {author}
  {\bibfnamefont {J.~M.}\ \bibnamefont {Tranquada}},\ }\bibfield  {title}
  {\bibinfo {title} {{Nature and impact of stripe freezing in
  ${\mathrm{La}}_{1.67}{\mathrm{Sr}}_{0.33}{\mathrm{NiO}}_{4}$}},\ }\href
  {https://doi.org/10.1103/PhysRevB.100.195122} {\bibfield  {journal} {\bibinfo
   {journal} {Phys. Rev. B}\ }\textbf {\bibinfo {volume} {100}},\ \bibinfo
  {pages} {195122} (\bibinfo {year} {2019})}\BibitemShut {NoStop}%
\bibitem [{\citenamefont {Xu}\ \emph {et~al.}(2007)\citenamefont {Xu},
  \citenamefont {Tranquada}, \citenamefont {Perring}, \citenamefont {Gu},
  \citenamefont {Fujita},\ and\ \citenamefont {Yamada}}]{xu07}%
  \BibitemOpen
  \bibfield  {author} {\bibinfo {author} {\bibfnamefont {G.}~\bibnamefont
  {Xu}}, \bibinfo {author} {\bibfnamefont {J.~M.}\ \bibnamefont {Tranquada}},
  \bibinfo {author} {\bibfnamefont {T.~G.}\ \bibnamefont {Perring}}, \bibinfo
  {author} {\bibfnamefont {G.~D.}\ \bibnamefont {Gu}}, \bibinfo {author}
  {\bibfnamefont {M.}~\bibnamefont {Fujita}},\ and\ \bibinfo {author}
  {\bibfnamefont {K.}~\bibnamefont {Yamada}},\ }\bibfield  {title} {\bibinfo
  {title} {{High-energy magnetic excitations from dynamic stripes in
  ${\mathrm{La}}_{1.875}{\mathrm{Ba}}_{0.125}\mathrm{Cu}{\mathrm{O}}_{4}$}},\
  }\href {https://doi.org/10.1103/PhysRevB.76.014508} {\bibfield  {journal}
  {\bibinfo  {journal} {Phys. Rev. B}\ }\textbf {\bibinfo {volume} {76}},\
  \bibinfo {pages} {014508} (\bibinfo {year} {2007})}\BibitemShut {NoStop}%
\bibitem [{\citenamefont {Chen}\ \emph {et~al.}(1991)\citenamefont {Chen},
  \citenamefont {Sette}, \citenamefont {Ma}, \citenamefont {Hybertsen},
  \citenamefont {Stechel}, \citenamefont {Foulkes}, \citenamefont {Schulter},
  \citenamefont {Cheong}, \citenamefont {Cooper}, \citenamefont {Rupp},
  \citenamefont {Batlogg}, \citenamefont {Soo}, \citenamefont {Ming},
  \citenamefont {Krol},\ and\ \citenamefont {Kao}}]{chen91}%
  \BibitemOpen
  \bibfield  {author} {\bibinfo {author} {\bibfnamefont {C.~T.}\ \bibnamefont
  {Chen}}, \bibinfo {author} {\bibfnamefont {F.}~\bibnamefont {Sette}},
  \bibinfo {author} {\bibfnamefont {Y.}~\bibnamefont {Ma}}, \bibinfo {author}
  {\bibfnamefont {M.~S.}\ \bibnamefont {Hybertsen}}, \bibinfo {author}
  {\bibfnamefont {E.~B.}\ \bibnamefont {Stechel}}, \bibinfo {author}
  {\bibfnamefont {W.~M.~C.}\ \bibnamefont {Foulkes}}, \bibinfo {author}
  {\bibfnamefont {M.}~\bibnamefont {Schulter}}, \bibinfo {author}
  {\bibfnamefont {S.-W.}\ \bibnamefont {Cheong}}, \bibinfo {author}
  {\bibfnamefont {A.~S.}\ \bibnamefont {Cooper}}, \bibinfo {author}
  {\bibfnamefont {L.~W.}\ \bibnamefont {Rupp}}, \bibinfo {author}
  {\bibfnamefont {B.}~\bibnamefont {Batlogg}}, \bibinfo {author} {\bibfnamefont
  {Y.~L.}\ \bibnamefont {Soo}}, \bibinfo {author} {\bibfnamefont {Z.~H.}\
  \bibnamefont {Ming}}, \bibinfo {author} {\bibfnamefont {A.}~\bibnamefont
  {Krol}},\ and\ \bibinfo {author} {\bibfnamefont {Y.~H.}\ \bibnamefont
  {Kao}},\ }\bibfield  {title} {\bibinfo {title} {{Electronic states in
  ${\mathrm{La}}_{2\mathrm{\ensuremath{-}}\mathit{x}}$${\mathrm{Sr}}_{\mathit{x}}$${\mathrm{CuO}}_{4+\mathrm{\ensuremath{\delta}}}$
  probed by soft-x-ray absorption}},\ }\href
  {https://doi.org/10.1103/PhysRevLett.66.104} {\bibfield  {journal} {\bibinfo
  {journal} {Phys. Rev. Lett.}\ }\textbf {\bibinfo {volume} {66}},\ \bibinfo
  {pages} {104} (\bibinfo {year} {1991})}\BibitemShut {NoStop}%
\bibitem [{\citenamefont {Emery}\ and\ \citenamefont {Reiter}(1988)}]{emer88}%
  \BibitemOpen
  \bibfield  {author} {\bibinfo {author} {\bibfnamefont {V.~J.}\ \bibnamefont
  {Emery}}\ and\ \bibinfo {author} {\bibfnamefont {G.}~\bibnamefont {Reiter}},\
  }\bibfield  {title} {\bibinfo {title} {{Mechanism for high-temperature
  superconductivity}},\ }\href@noop {} {\bibfield  {journal} {\bibinfo
  {journal} {Phys. Rev. B}\ }\textbf {\bibinfo {volume} {38}},\ \bibinfo
  {pages} {4547} (\bibinfo {year} {1988})}\BibitemShut {NoStop}%
\bibitem [{\citenamefont {Matsuda}\ \emph {et~al.}(2002)\citenamefont
  {Matsuda}, \citenamefont {Fujita}, \citenamefont {Yamada}, \citenamefont
  {Birgeneau}, \citenamefont {Endoh},\ and\ \citenamefont {Shirane}}]{mats02}%
  \BibitemOpen
  \bibfield  {author} {\bibinfo {author} {\bibfnamefont {M.}~\bibnamefont
  {Matsuda}}, \bibinfo {author} {\bibfnamefont {M.}~\bibnamefont {Fujita}},
  \bibinfo {author} {\bibfnamefont {K.}~\bibnamefont {Yamada}}, \bibinfo
  {author} {\bibfnamefont {R.~J.}\ \bibnamefont {Birgeneau}}, \bibinfo {author}
  {\bibfnamefont {Y.}~\bibnamefont {Endoh}},\ and\ \bibinfo {author}
  {\bibfnamefont {G.}~\bibnamefont {Shirane}},\ }\bibfield  {title} {\bibinfo
  {title} {{Electronic phase separation in lightly doped
  ${\mathrm{La}}_{2\ensuremath{-}x}{\mathrm{Sr}}_{x}\mathrm{Cu}{\mathrm{O}}_{4}$}},\
  }\href {https://doi.org/10.1103/PhysRevB.65.134515} {\bibfield  {journal}
  {\bibinfo  {journal} {Phys. Rev. B}\ }\textbf {\bibinfo {volume} {65}},\
  \bibinfo {pages} {134515} (\bibinfo {year} {2002})}\BibitemShut {NoStop}%
\bibitem [{\citenamefont {Vajk}\ \emph {et~al.}(2002)\citenamefont {Vajk},
  \citenamefont {Mang}, \citenamefont {Greven}, \citenamefont {Gehring},\ and\
  \citenamefont {Lynn}}]{vajk02}%
  \BibitemOpen
  \bibfield  {author} {\bibinfo {author} {\bibfnamefont {O.~P.}\ \bibnamefont
  {Vajk}}, \bibinfo {author} {\bibfnamefont {P.~K.}\ \bibnamefont {Mang}},
  \bibinfo {author} {\bibfnamefont {M.}~\bibnamefont {Greven}}, \bibinfo
  {author} {\bibfnamefont {P.~M.}\ \bibnamefont {Gehring}},\ and\ \bibinfo
  {author} {\bibfnamefont {J.~W.}\ \bibnamefont {Lynn}},\ }\bibfield  {title}
  {\bibinfo {title} {{Quantum Impurities in the Two-Dimensional Spin One-Half
  Heisenberg Antiferromagnet}},\ }\href@noop {} {\bibfield  {journal} {\bibinfo
   {journal} {Science}\ }\textbf {\bibinfo {volume} {295}},\ \bibinfo {pages}
  {1691} (\bibinfo {year} {2002})}\BibitemShut {NoStop}%
\bibitem [{\citenamefont {Tranquada}\ \emph {et~al.}(1994)\citenamefont
  {Tranquada}, \citenamefont {Buttrey}, \citenamefont {Sachan},\ and\
  \citenamefont {Lorenzo}}]{tran94a}%
  \BibitemOpen
  \bibfield  {author} {\bibinfo {author} {\bibfnamefont {J.~M.}\ \bibnamefont
  {Tranquada}}, \bibinfo {author} {\bibfnamefont {D.~J.}\ \bibnamefont
  {Buttrey}}, \bibinfo {author} {\bibfnamefont {V.}~\bibnamefont {Sachan}},\
  and\ \bibinfo {author} {\bibfnamefont {J.~E.}\ \bibnamefont {Lorenzo}},\
  }\bibfield  {title} {\bibinfo {title} {{Simultaneous Ordering of Holes and
  Spins in ${\mathrm{La}}_{2}$Ni${\mathrm{O}}_{4.125}$}},\ }\href@noop {}
  {\bibfield  {journal} {\bibinfo  {journal} {Phys. Rev. Lett.}\ }\textbf
  {\bibinfo {volume} {73}},\ \bibinfo {pages} {1003} (\bibinfo {year}
  {1994})}\BibitemShut {NoStop}%
\bibitem [{\citenamefont {Yoshizawa}\ \emph {et~al.}(2000)\citenamefont
  {Yoshizawa}, \citenamefont {Kakeshita}, \citenamefont {Kajimoto},
  \citenamefont {Tanabe}, \citenamefont {Katsufuji},\ and\ \citenamefont
  {Tokura}}]{yosh00}%
  \BibitemOpen
  \bibfield  {author} {\bibinfo {author} {\bibfnamefont {H.}~\bibnamefont
  {Yoshizawa}}, \bibinfo {author} {\bibfnamefont {T.}~\bibnamefont
  {Kakeshita}}, \bibinfo {author} {\bibfnamefont {R.}~\bibnamefont {Kajimoto}},
  \bibinfo {author} {\bibfnamefont {T.}~\bibnamefont {Tanabe}}, \bibinfo
  {author} {\bibfnamefont {T.}~\bibnamefont {Katsufuji}},\ and\ \bibinfo
  {author} {\bibfnamefont {Y.}~\bibnamefont {Tokura}},\ }\bibfield  {title}
  {\bibinfo {title} {{Stripe order at low temperatures in
  ${\mathrm{La}}_{2\ensuremath{-}x}{\mathrm{Sr}}_{x}{\mathrm{NiO}}_{4}$ with
  $0.289\ensuremath{\lesssim}x\ensuremath{\lesssim}0.5$}},\ }\href
  {https://doi.org/10.1103/PhysRevB.61.R854} {\bibfield  {journal} {\bibinfo
  {journal} {Phys. Rev. B}\ }\textbf {\bibinfo {volume} {61}},\ \bibinfo
  {pages} {R854} (\bibinfo {year} {2000})}\BibitemShut {NoStop}%
\bibitem [{\citenamefont {Ulbrich}\ and\ \citenamefont
  {Braden}(2012)}]{ulbr12b}%
  \BibitemOpen
  \bibfield  {author} {\bibinfo {author} {\bibfnamefont {H.}~\bibnamefont
  {Ulbrich}}\ and\ \bibinfo {author} {\bibfnamefont {M.}~\bibnamefont
  {Braden}},\ }\bibfield  {title} {\bibinfo {title} {{Neutron scattering
  studies on stripe phases in non-cuprate materials}},\ }\href
  {https://doi.org/https://doi.org/10.1016/j.physc.2012.04.039} {\bibfield
  {journal} {\bibinfo  {journal} {Physica C: Superconductivity}\ }\textbf
  {\bibinfo {volume} {481}},\ \bibinfo {pages} {31} (\bibinfo {year}
  {2012})}\BibitemShut {NoStop}%
\bibitem [{\citenamefont {Lee}\ and\ \citenamefont {Cheong}(1997)}]{lee97}%
  \BibitemOpen
  \bibfield  {author} {\bibinfo {author} {\bibfnamefont {S.-H.}\ \bibnamefont
  {Lee}}\ and\ \bibinfo {author} {\bibfnamefont {S.-W.}\ \bibnamefont
  {Cheong}},\ }\bibfield  {title} {\bibinfo {title} {{Melting of
  Quasi-Two-Dimensional Charge Stripes in
  ${\mathrm{La}}_{5\mathrm{/}3}{\mathrm{Sr}}_{1\mathrm{/}3}{\mathrm{NiO}}_{4}$}},\
  }\href {https://doi.org/10.1103/PhysRevLett.79.2514} {\bibfield  {journal}
  {\bibinfo  {journal} {Phys. Rev. Lett.}\ }\textbf {\bibinfo {volume} {79}},\
  \bibinfo {pages} {2514} (\bibinfo {year} {1997})}\BibitemShut {NoStop}%
\bibitem [{\citenamefont {Boothroyd}\ \emph
  {et~al.}(2003{\natexlab{a}})\citenamefont {Boothroyd}, \citenamefont
  {Prabhakaran}, \citenamefont {Freeman}, \citenamefont {Lister}, \citenamefont
  {Enderle}, \citenamefont {Hiess},\ and\ \citenamefont {Kulda}}]{boot03a}%
  \BibitemOpen
  \bibfield  {author} {\bibinfo {author} {\bibfnamefont {A.~T.}\ \bibnamefont
  {Boothroyd}}, \bibinfo {author} {\bibfnamefont {D.}~\bibnamefont
  {Prabhakaran}}, \bibinfo {author} {\bibfnamefont {P.~G.}\ \bibnamefont
  {Freeman}}, \bibinfo {author} {\bibfnamefont {S.~J.~S.}\ \bibnamefont
  {Lister}}, \bibinfo {author} {\bibfnamefont {M.}~\bibnamefont {Enderle}},
  \bibinfo {author} {\bibfnamefont {A.}~\bibnamefont {Hiess}},\ and\ \bibinfo
  {author} {\bibfnamefont {J.}~\bibnamefont {Kulda}},\ }\bibfield  {title}
  {\bibinfo {title} {{Spin dynamics in stripe-ordered
  ${\mathrm{La}}_{5/3}{\mathrm{Sr}}_{1/3}{\mathrm{NiO}}_{4}$}},\ }\href
  {https://doi.org/10.1103/PhysRevB.67.100407} {\bibfield  {journal} {\bibinfo
  {journal} {Phys. Rev. B}\ }\textbf {\bibinfo {volume} {67}},\ \bibinfo
  {pages} {100407(R)} (\bibinfo {year} {2003}{\natexlab{a}})}\BibitemShut
  {NoStop}%
\bibitem [{\citenamefont {Woo}\ \emph {et~al.}(2005)\citenamefont {Woo},
  \citenamefont {Boothroyd}, \citenamefont {Nakajima}, \citenamefont {Perring},
  \citenamefont {Frost}, \citenamefont {Freeman}, \citenamefont {Prabhakaran},
  \citenamefont {Yamada},\ and\ \citenamefont {Tranquada}}]{woo05}%
  \BibitemOpen
  \bibfield  {author} {\bibinfo {author} {\bibfnamefont {H.}~\bibnamefont
  {Woo}}, \bibinfo {author} {\bibfnamefont {A.~T.}\ \bibnamefont {Boothroyd}},
  \bibinfo {author} {\bibfnamefont {K.}~\bibnamefont {Nakajima}}, \bibinfo
  {author} {\bibfnamefont {T.~G.}\ \bibnamefont {Perring}}, \bibinfo {author}
  {\bibfnamefont {C.~D.}\ \bibnamefont {Frost}}, \bibinfo {author}
  {\bibfnamefont {P.~G.}\ \bibnamefont {Freeman}}, \bibinfo {author}
  {\bibfnamefont {D.}~\bibnamefont {Prabhakaran}}, \bibinfo {author}
  {\bibfnamefont {K.}~\bibnamefont {Yamada}},\ and\ \bibinfo {author}
  {\bibfnamefont {J.~M.}\ \bibnamefont {Tranquada}},\ }\bibfield  {title}
  {\bibinfo {title} {{Mapping spin-wave dispersions in stripe-ordered
  ${\mathrm{La}}_{2\ensuremath{-}x}{\mathrm{Sr}}_{x}\mathrm{Ni}{\mathrm{O}}_{4}$
  ($x=0.275$, 0.333)}},\ }\href {https://doi.org/10.1103/PhysRevB.72.064437}
  {\bibfield  {journal} {\bibinfo  {journal} {Phys. Rev. B}\ }\textbf {\bibinfo
  {volume} {72}},\ \bibinfo {pages} {064437} (\bibinfo {year}
  {2005})}\BibitemShut {NoStop}%
\bibitem [{\citenamefont {Boothroyd}\ \emph
  {et~al.}(2003{\natexlab{b}})\citenamefont {Boothroyd}, \citenamefont
  {Freeman}, \citenamefont {Prabhakaran}, \citenamefont {Hiess}, \citenamefont
  {Enderle}, \citenamefont {Kulda},\ and\ \citenamefont {Altorfer}}]{boot03b}%
  \BibitemOpen
  \bibfield  {author} {\bibinfo {author} {\bibfnamefont {A.~T.}\ \bibnamefont
  {Boothroyd}}, \bibinfo {author} {\bibfnamefont {P.~G.}\ \bibnamefont
  {Freeman}}, \bibinfo {author} {\bibfnamefont {D.}~\bibnamefont
  {Prabhakaran}}, \bibinfo {author} {\bibfnamefont {A.}~\bibnamefont {Hiess}},
  \bibinfo {author} {\bibfnamefont {M.}~\bibnamefont {Enderle}}, \bibinfo
  {author} {\bibfnamefont {J.}~\bibnamefont {Kulda}},\ and\ \bibinfo {author}
  {\bibfnamefont {F.}~\bibnamefont {Altorfer}},\ }\bibfield  {title} {\bibinfo
  {title} {{Spin Correlations among the Charge Carriers in an Ordered Stripe
  Phase}},\ }\href {https://doi.org/10.1103/PhysRevLett.91.257201} {\bibfield
  {journal} {\bibinfo  {journal} {Phys. Rev. Lett.}\ }\textbf {\bibinfo
  {volume} {91}},\ \bibinfo {pages} {257201} (\bibinfo {year}
  {2003}{\natexlab{b}})}\BibitemShut {NoStop}%
\bibitem [{\citenamefont {Tranquada}\ \emph {et~al.}(2004)\citenamefont
  {Tranquada}, \citenamefont {Woo}, \citenamefont {Perring}, \citenamefont
  {Goka}, \citenamefont {Gu}, \citenamefont {Xu}, \citenamefont {Fujita},\ and\
  \citenamefont {Yamada}}]{tran04}%
  \BibitemOpen
  \bibfield  {author} {\bibinfo {author} {\bibfnamefont {J.~M.}\ \bibnamefont
  {Tranquada}}, \bibinfo {author} {\bibfnamefont {H.}~\bibnamefont {Woo}},
  \bibinfo {author} {\bibfnamefont {T.~G.}\ \bibnamefont {Perring}}, \bibinfo
  {author} {\bibfnamefont {H.}~\bibnamefont {Goka}}, \bibinfo {author}
  {\bibfnamefont {G.~D.}\ \bibnamefont {Gu}}, \bibinfo {author} {\bibfnamefont
  {G.}~\bibnamefont {Xu}}, \bibinfo {author} {\bibfnamefont {M.}~\bibnamefont
  {Fujita}},\ and\ \bibinfo {author} {\bibfnamefont {K.}~\bibnamefont
  {Yamada}},\ }\bibfield  {title} {\bibinfo {title} {Quantum magnetic
  excitations from stripes in copper oxide superconductors},\ }\href@noop {}
  {\bibfield  {journal} {\bibinfo  {journal} {Nature}\ }\textbf {\bibinfo
  {volume} {429}},\ \bibinfo {pages} {534} (\bibinfo {year}
  {2004})}\BibitemShut {NoStop}%
\bibitem [{\citenamefont {Sato}\ \emph {et~al.}(2020)\citenamefont {Sato},
  \citenamefont {Ikeuchi}, \citenamefont {Kajimoto}, \citenamefont {Wakimoto},
  \citenamefont {Arai},\ and\ \citenamefont {Fujita}}]{sato20}%
  \BibitemOpen
  \bibfield  {author} {\bibinfo {author} {\bibfnamefont {K.}~\bibnamefont
  {Sato}}, \bibinfo {author} {\bibfnamefont {K.}~\bibnamefont {Ikeuchi}},
  \bibinfo {author} {\bibfnamefont {R.}~\bibnamefont {Kajimoto}}, \bibinfo
  {author} {\bibfnamefont {S.}~\bibnamefont {Wakimoto}}, \bibinfo {author}
  {\bibfnamefont {M.}~\bibnamefont {Arai}},\ and\ \bibinfo {author}
  {\bibfnamefont {M.}~\bibnamefont {Fujita}},\ }\bibfield  {title} {\bibinfo
  {title} {{Coexistence of Two Components in Magnetic Excitations of
  La$_{2-x}$Sr$_x$CuO$_4$ ($x = 0.10$ and 0.16)}},\ }\href
  {https://doi.org/10.7566/JPSJ.89.114703} {\bibfield  {journal} {\bibinfo
  {journal} {J. Phys. Soc. Jpn.}\ }\textbf {\bibinfo {volume} {89}},\ \bibinfo
  {pages} {114703} (\bibinfo {year} {2020})}\BibitemShut {NoStop}%
\bibitem [{\citenamefont {Barnes}\ \emph {et~al.}(1993)\citenamefont {Barnes},
  \citenamefont {Dagotto}, \citenamefont {Riera},\ and\ \citenamefont
  {Swanson}}]{barn93}%
  \BibitemOpen
  \bibfield  {author} {\bibinfo {author} {\bibfnamefont {T.}~\bibnamefont
  {Barnes}}, \bibinfo {author} {\bibfnamefont {E.}~\bibnamefont {Dagotto}},
  \bibinfo {author} {\bibfnamefont {J.}~\bibnamefont {Riera}},\ and\ \bibinfo
  {author} {\bibfnamefont {E.~S.}\ \bibnamefont {Swanson}},\ }\bibfield
  {title} {\bibinfo {title} {{Excitation spectrum of Heisenberg spin
  ladders}},\ }\href {https://doi.org/10.1103/PhysRevB.47.3196} {\bibfield
  {journal} {\bibinfo  {journal} {Phys. Rev. B}\ }\textbf {\bibinfo {volume}
  {47}},\ \bibinfo {pages} {3196} (\bibinfo {year} {1993})}\BibitemShut
  {NoStop}%
\bibitem [{\citenamefont {Poilblanc}\ \emph {et~al.}(2003)\citenamefont
  {Poilblanc}, \citenamefont {Scalapino},\ and\ \citenamefont
  {Capponi}}]{poil03}%
  \BibitemOpen
  \bibfield  {author} {\bibinfo {author} {\bibfnamefont {D.}~\bibnamefont
  {Poilblanc}}, \bibinfo {author} {\bibfnamefont {D.~J.}\ \bibnamefont
  {Scalapino}},\ and\ \bibinfo {author} {\bibfnamefont {S.}~\bibnamefont
  {Capponi}},\ }\bibfield  {title} {\bibinfo {title} {{Superconducting Gap for
  a Two-Leg $t\mathrm{\text{\ensuremath{-}}}J$ Ladder}},\ }\href
  {https://doi.org/10.1103/PhysRevLett.91.137203} {\bibfield  {journal}
  {\bibinfo  {journal} {Phys. Rev. Lett.}\ }\textbf {\bibinfo {volume} {91}},\
  \bibinfo {pages} {137203} (\bibinfo {year} {2003})}\BibitemShut {NoStop}%
\bibitem [{\citenamefont {Anderson}(1987)}]{ande87}%
  \BibitemOpen
  \bibfield  {author} {\bibinfo {author} {\bibfnamefont {P.~W.}\ \bibnamefont
  {Anderson}},\ }\bibfield  {title} {\bibinfo {title} {{The Resonating Valence
  Bond State in La$_2$CuO$_4$ and Superconductivity}},\ }\href
  {https://doi.org/10.1126/science.235.4793.1196} {\bibfield  {journal}
  {\bibinfo  {journal} {Science}\ }\textbf {\bibinfo {volume} {235}},\ \bibinfo
  {pages} {1196} (\bibinfo {year} {1987})}\BibitemShut {NoStop}%
\bibitem [{\citenamefont {Kivelson}\ \emph {et~al.}(1987)\citenamefont
  {Kivelson}, \citenamefont {Rokhsar},\ and\ \citenamefont {Sethna}}]{kive87}%
  \BibitemOpen
  \bibfield  {author} {\bibinfo {author} {\bibfnamefont {S.~A.}\ \bibnamefont
  {Kivelson}}, \bibinfo {author} {\bibfnamefont {D.~S.}\ \bibnamefont
  {Rokhsar}},\ and\ \bibinfo {author} {\bibfnamefont {J.~P.}\ \bibnamefont
  {Sethna}},\ }\bibfield  {title} {\bibinfo {title} {{Topology of the
  resonating valence-bond state: Solitons and high-${T}_{c}$
  superconductivity}},\ }\href {https://doi.org/10.1103/PhysRevB.35.8865}
  {\bibfield  {journal} {\bibinfo  {journal} {Phys. Rev. B}\ }\textbf {\bibinfo
  {volume} {35}},\ \bibinfo {pages} {8865(R)} (\bibinfo {year}
  {1987})}\BibitemShut {NoStop}%
\bibitem [{\citenamefont {Reznik}\ \emph {et~al.}(2006)\citenamefont {Reznik},
  \citenamefont {Pintschovius}, \citenamefont {Ito}, \citenamefont {Iikubo},
  \citenamefont {Sato}, \citenamefont {Goka}, \citenamefont {Fujita},
  \citenamefont {Yamada}, \citenamefont {Gu},\ and\ \citenamefont
  {Tranquada}}]{rezn06}%
  \BibitemOpen
  \bibfield  {author} {\bibinfo {author} {\bibfnamefont {D.}~\bibnamefont
  {Reznik}}, \bibinfo {author} {\bibfnamefont {L.}~\bibnamefont
  {Pintschovius}}, \bibinfo {author} {\bibfnamefont {M.}~\bibnamefont {Ito}},
  \bibinfo {author} {\bibfnamefont {S.}~\bibnamefont {Iikubo}}, \bibinfo
  {author} {\bibfnamefont {M.}~\bibnamefont {Sato}}, \bibinfo {author}
  {\bibfnamefont {H.}~\bibnamefont {Goka}}, \bibinfo {author} {\bibfnamefont
  {M.}~\bibnamefont {Fujita}}, \bibinfo {author} {\bibfnamefont
  {K.}~\bibnamefont {Yamada}}, \bibinfo {author} {\bibfnamefont {G.~D.}\
  \bibnamefont {Gu}},\ and\ \bibinfo {author} {\bibfnamefont {J.~M.}\
  \bibnamefont {Tranquada}},\ }\bibfield  {title} {\bibinfo {title}
  {Electron--phonon coupling reflecting dynamic charge inhomogeneity in
  copper-oxide superconductors},\ }\href@noop {} {\bibfield  {journal}
  {\bibinfo  {journal} {Nature}\ }\textbf {\bibinfo {volume} {440}},\ \bibinfo
  {pages} {1170} (\bibinfo {year} {2006})}\BibitemShut {NoStop}%
\bibitem [{\citenamefont {Miao}\ \emph {et~al.}(2018)\citenamefont {Miao},
  \citenamefont {Ishikawa}, \citenamefont {Heid}, \citenamefont {Le~Tacon},
  \citenamefont {Fabbris}, \citenamefont {Meyers}, \citenamefont {Gu},
  \citenamefont {Baron},\ and\ \citenamefont {Dean}}]{miao18}%
  \BibitemOpen
  \bibfield  {author} {\bibinfo {author} {\bibfnamefont {H.}~\bibnamefont
  {Miao}}, \bibinfo {author} {\bibfnamefont {D.}~\bibnamefont {Ishikawa}},
  \bibinfo {author} {\bibfnamefont {R.}~\bibnamefont {Heid}}, \bibinfo {author}
  {\bibfnamefont {M.}~\bibnamefont {Le~Tacon}}, \bibinfo {author}
  {\bibfnamefont {G.}~\bibnamefont {Fabbris}}, \bibinfo {author} {\bibfnamefont
  {D.}~\bibnamefont {Meyers}}, \bibinfo {author} {\bibfnamefont {G.~D.}\
  \bibnamefont {Gu}}, \bibinfo {author} {\bibfnamefont {A.~Q.~R.}\ \bibnamefont
  {Baron}},\ and\ \bibinfo {author} {\bibfnamefont {M.~P.~M.}\ \bibnamefont
  {Dean}},\ }\bibfield  {title} {\bibinfo {title} {{Incommensurate Phonon
  Anomaly and the Nature of Charge Density Waves in Cuprates}},\ }\href
  {https://doi.org/10.1103/PhysRevX.8.011008} {\bibfield  {journal} {\bibinfo
  {journal} {Phys. Rev. X}\ }\textbf {\bibinfo {volume} {8}},\ \bibinfo {pages}
  {011008} (\bibinfo {year} {2018})}\BibitemShut {NoStop}%
\bibitem [{\citenamefont {Peng}\ \emph {et~al.}(2020)\citenamefont {Peng},
  \citenamefont {Husain}, \citenamefont {Mitrano}, \citenamefont {Sun},
  \citenamefont {Johnson}, \citenamefont {Zakrzewski}, \citenamefont
  {MacDougall}, \citenamefont {Barbour}, \citenamefont {Jarrige}, \citenamefont
  {Bisogni},\ and\ \citenamefont {Abbamonte}}]{peng20}%
  \BibitemOpen
  \bibfield  {author} {\bibinfo {author} {\bibfnamefont {Y.~Y.}\ \bibnamefont
  {Peng}}, \bibinfo {author} {\bibfnamefont {A.~A.}\ \bibnamefont {Husain}},
  \bibinfo {author} {\bibfnamefont {M.}~\bibnamefont {Mitrano}}, \bibinfo
  {author} {\bibfnamefont {S.~X.-L.}\ \bibnamefont {Sun}}, \bibinfo {author}
  {\bibfnamefont {T.~A.}\ \bibnamefont {Johnson}}, \bibinfo {author}
  {\bibfnamefont {A.~V.}\ \bibnamefont {Zakrzewski}}, \bibinfo {author}
  {\bibfnamefont {G.~J.}\ \bibnamefont {MacDougall}}, \bibinfo {author}
  {\bibfnamefont {A.}~\bibnamefont {Barbour}}, \bibinfo {author} {\bibfnamefont
  {I.}~\bibnamefont {Jarrige}}, \bibinfo {author} {\bibfnamefont
  {V.}~\bibnamefont {Bisogni}},\ and\ \bibinfo {author} {\bibfnamefont
  {P.}~\bibnamefont {Abbamonte}},\ }\bibfield  {title} {\bibinfo {title}
  {{Enhanced Electron-Phonon Coupling for Charge-Density-Wave Formation in
  ${\mathrm{La}}_{1.8\ensuremath{-}x}{\mathrm{Eu}}_{0.2}{\mathrm{Sr}}_{x}{\mathrm{CuO}}_{4+\ensuremath{\delta}}$}},\
  }\href {https://doi.org/10.1103/PhysRevLett.125.097002} {\bibfield  {journal}
  {\bibinfo  {journal} {Phys. Rev. Lett.}\ }\textbf {\bibinfo {volume} {125}},\
  \bibinfo {pages} {097002} (\bibinfo {year} {2020})}\BibitemShut {NoStop}%
\bibitem [{\citenamefont {Wang}\ \emph {et~al.}(2021)\citenamefont {Wang},
  \citenamefont {von Arx}, \citenamefont {Horio}, \citenamefont
  {Mukkattukavil}, \citenamefont {K{\"u}spert}, \citenamefont {Sassa},
  \citenamefont {Schmitt}, \citenamefont {Nag}, \citenamefont {Pyon},
  \citenamefont {Takayama}, \citenamefont {Takagi}, \citenamefont
  {Garcia-Fernandez}, \citenamefont {Zhou},\ and\ \citenamefont
  {Chang}}]{wang21a}%
  \BibitemOpen
  \bibfield  {author} {\bibinfo {author} {\bibfnamefont {Q.}~\bibnamefont
  {Wang}}, \bibinfo {author} {\bibfnamefont {K.}~\bibnamefont {von Arx}},
  \bibinfo {author} {\bibfnamefont {M.}~\bibnamefont {Horio}}, \bibinfo
  {author} {\bibfnamefont {D.~J.}\ \bibnamefont {Mukkattukavil}}, \bibinfo
  {author} {\bibfnamefont {J.}~\bibnamefont {K{\"u}spert}}, \bibinfo {author}
  {\bibfnamefont {Y.}~\bibnamefont {Sassa}}, \bibinfo {author} {\bibfnamefont
  {T.}~\bibnamefont {Schmitt}}, \bibinfo {author} {\bibfnamefont
  {A.}~\bibnamefont {Nag}}, \bibinfo {author} {\bibfnamefont {S.}~\bibnamefont
  {Pyon}}, \bibinfo {author} {\bibfnamefont {T.}~\bibnamefont {Takayama}},
  \bibinfo {author} {\bibfnamefont {H.}~\bibnamefont {Takagi}}, \bibinfo
  {author} {\bibfnamefont {M.}~\bibnamefont {Garcia-Fernandez}}, \bibinfo
  {author} {\bibfnamefont {K.-J.}\ \bibnamefont {Zhou}},\ and\ \bibinfo
  {author} {\bibfnamefont {J.}~\bibnamefont {Chang}},\ }\bibfield  {title}
  {\bibinfo {title} {{Charge order lock-in by electron-phonon coupling in
  La$_{1.675}$Eu$_{0.2}$Sr$_{0.125}$CuO$_4$}},\ }\href
  {https://doi.org/10.1126/sciadv.abg7394} {\bibfield  {journal} {\bibinfo
  {journal} {Sci. Adv.}\ }\textbf {\bibinfo {volume} {7}},\ \bibinfo {pages}
  {{eabg7394}} (\bibinfo {year} {2021})}\BibitemShut {NoStop}%
\bibitem [{\citenamefont {Kivelson}\ \emph {et~al.}(1998)\citenamefont
  {Kivelson}, \citenamefont {Fradkin},\ and\ \citenamefont {Emery}}]{kive98}%
  \BibitemOpen
  \bibfield  {author} {\bibinfo {author} {\bibfnamefont {S.~A.}\ \bibnamefont
  {Kivelson}}, \bibinfo {author} {\bibfnamefont {E.}~\bibnamefont {Fradkin}},\
  and\ \bibinfo {author} {\bibfnamefont {V.~J.}\ \bibnamefont {Emery}},\
  }\bibfield  {title} {\bibinfo {title} {{Electronic liquid-crystal phases of a
  doped Mott insulator}},\ }\href@noop {} {\bibfield  {journal} {\bibinfo
  {journal} {Nature}\ }\textbf {\bibinfo {volume} {393}},\ \bibinfo {pages}
  {550} (\bibinfo {year} {1998})}\BibitemShut {NoStop}%
\bibitem [{\citenamefont {Dolfi}\ \emph {et~al.}(2015)\citenamefont {Dolfi},
  \citenamefont {Bauer}, \citenamefont {Keller},\ and\ \citenamefont
  {Troyer}}]{dolf15}%
  \BibitemOpen
  \bibfield  {author} {\bibinfo {author} {\bibfnamefont {M.}~\bibnamefont
  {Dolfi}}, \bibinfo {author} {\bibfnamefont {B.}~\bibnamefont {Bauer}},
  \bibinfo {author} {\bibfnamefont {S.}~\bibnamefont {Keller}},\ and\ \bibinfo
  {author} {\bibfnamefont {M.}~\bibnamefont {Troyer}},\ }\bibfield  {title}
  {\bibinfo {title} {{Pair correlations in doped Hubbard ladders}},\ }\href
  {https://doi.org/10.1103/PhysRevB.92.195139} {\bibfield  {journal} {\bibinfo
  {journal} {Phys. Rev. B}\ }\textbf {\bibinfo {volume} {92}},\ \bibinfo
  {pages} {195139} (\bibinfo {year} {2015})}\BibitemShut {NoStop}%
\bibitem [{\citenamefont {Song}\ \emph {et~al.}(2021)\citenamefont {Song},
  \citenamefont {Mazumdar},\ and\ \citenamefont {Clay}}]{song21}%
  \BibitemOpen
  \bibfield  {author} {\bibinfo {author} {\bibfnamefont {J.-P.}\ \bibnamefont
  {Song}}, \bibinfo {author} {\bibfnamefont {S.}~\bibnamefont {Mazumdar}},\
  and\ \bibinfo {author} {\bibfnamefont {R.~T.}\ \bibnamefont {Clay}},\
  }\bibfield  {title} {\bibinfo {title} {{Absence of Luther-Emery
  superconducting phase in the three-band model for cuprate ladders}},\ }\href
  {https://doi.org/10.1103/PhysRevB.104.104504} {\bibfield  {journal} {\bibinfo
   {journal} {Phys. Rev. B}\ }\textbf {\bibinfo {volume} {104}},\ \bibinfo
  {pages} {104504} (\bibinfo {year} {2021})}\BibitemShut {NoStop}%
\bibitem [{\citenamefont {Tranquada}\ \emph {et~al.}(2008)\citenamefont
  {Tranquada}, \citenamefont {Gu}, \citenamefont {H{\"u}cker}, \citenamefont
  {Jie}, \citenamefont {Kang}, \citenamefont {Klingeler}, \citenamefont {Li},
  \citenamefont {Tristan}, \citenamefont {Wen}, \citenamefont {Xu},
  \citenamefont {Xu}, \citenamefont {Zhou},\ and\ \citenamefont
  {v.~Zimmermann}}]{tran08}%
  \BibitemOpen
  \bibfield  {author} {\bibinfo {author} {\bibfnamefont {J.~M.}\ \bibnamefont
  {Tranquada}}, \bibinfo {author} {\bibfnamefont {G.~D.}\ \bibnamefont {Gu}},
  \bibinfo {author} {\bibfnamefont {M.}~\bibnamefont {H{\"u}cker}}, \bibinfo
  {author} {\bibfnamefont {Q.}~\bibnamefont {Jie}}, \bibinfo {author}
  {\bibfnamefont {H.-J.}\ \bibnamefont {Kang}}, \bibinfo {author}
  {\bibfnamefont {R.}~\bibnamefont {Klingeler}}, \bibinfo {author}
  {\bibfnamefont {Q.}~\bibnamefont {Li}}, \bibinfo {author} {\bibfnamefont
  {N.}~\bibnamefont {Tristan}}, \bibinfo {author} {\bibfnamefont {J.~S.}\
  \bibnamefont {Wen}}, \bibinfo {author} {\bibfnamefont {G.~Y.}\ \bibnamefont
  {Xu}}, \bibinfo {author} {\bibfnamefont {Z.~J.}\ \bibnamefont {Xu}}, \bibinfo
  {author} {\bibfnamefont {J.}~\bibnamefont {Zhou}},\ and\ \bibinfo {author}
  {\bibfnamefont {M.}~\bibnamefont {v.~Zimmermann}},\ }\bibfield  {title}
  {\bibinfo {title} {{Evidence for unusual superconducting correlations
  coexisting with stripe order in La$_{1.875}$Ba$_{0.125}$CuO$_4$}},\
  }\href@noop {} {\bibfield  {journal} {\bibinfo  {journal} {Phys. Rev. B}\
  }\textbf {\bibinfo {volume} {78}},\ \bibinfo {eid} {174529} (\bibinfo {year}
  {2008})}\BibitemShut {NoStop}%
\bibitem [{\citenamefont {Ding}\ \emph {et~al.}(2008)\citenamefont {Ding},
  \citenamefont {Xiang}, \citenamefont {Zhang}, \citenamefont {Liu},\ and\
  \citenamefont {Li}}]{ding08}%
  \BibitemOpen
  \bibfield  {author} {\bibinfo {author} {\bibfnamefont {J.~F.}\ \bibnamefont
  {Ding}}, \bibinfo {author} {\bibfnamefont {X.~Q.}\ \bibnamefont {Xiang}},
  \bibinfo {author} {\bibfnamefont {Y.~Q.}\ \bibnamefont {Zhang}}, \bibinfo
  {author} {\bibfnamefont {H.}~\bibnamefont {Liu}},\ and\ \bibinfo {author}
  {\bibfnamefont {X.~G.}\ \bibnamefont {Li}},\ }\bibfield  {title} {\bibinfo
  {title} {{Two-dimensional superconductivity in stripe-ordered
  ${\text{La}}_{1.6\ensuremath{-}x}{\text{Nd}}_{0.4}{\text{Sr}}_{x}{\text{CuO}}_{4}$
  single crystals}},\ }\href {https://doi.org/10.1103/PhysRevB.77.214524}
  {\bibfield  {journal} {\bibinfo  {journal} {Phys. Rev. B}\ }\textbf {\bibinfo
  {volume} {77}},\ \bibinfo {pages} {214524} (\bibinfo {year}
  {2008})}\BibitemShut {NoStop}%
\bibitem [{\citenamefont {B\"uchner}\ \emph {et~al.}(1994)\citenamefont
  {B\"uchner}, \citenamefont {Breuer}, \citenamefont {Freimuth},\ and\
  \citenamefont {Kampf}}]{buch94a}%
  \BibitemOpen
  \bibfield  {author} {\bibinfo {author} {\bibfnamefont {B.}~\bibnamefont
  {B\"uchner}}, \bibinfo {author} {\bibfnamefont {M.}~\bibnamefont {Breuer}},
  \bibinfo {author} {\bibfnamefont {A.}~\bibnamefont {Freimuth}},\ and\
  \bibinfo {author} {\bibfnamefont {A.~P.}\ \bibnamefont {Kampf}},\ }\bibfield
  {title} {\bibinfo {title} {{Critical Buckling for the Disappearance of
  Superconductivity in Rare-Earth-Doped
  ${\mathrm{La}}_{2\ensuremath{-}x}{\mathrm{Sr}}_{x}\mathrm{Cu}{\mathrm{O}}_{4}$}},\
  }\href {https://doi.org/10.1103/PhysRevLett.73.1841} {\bibfield  {journal}
  {\bibinfo  {journal} {Phys. Rev. Lett.}\ }\textbf {\bibinfo {volume} {73}},\
  \bibinfo {pages} {1841} (\bibinfo {year} {1994})}\BibitemShut {NoStop}%
\bibitem [{\citenamefont {Axe}\ and\ \citenamefont {Crawford}(1994)}]{axe94}%
  \BibitemOpen
  \bibfield  {author} {\bibinfo {author} {\bibfnamefont {J.~D.}\ \bibnamefont
  {Axe}}\ and\ \bibinfo {author} {\bibfnamefont {M.~K.}\ \bibnamefont
  {Crawford}},\ }\bibfield  {title} {\bibinfo {title} {{Structural
  instabilities in lanthanum cuprate superconductors}},\ }\href
  {https://doi.org/10.1007/BF00754942} {\bibfield  {journal} {\bibinfo
  {journal} {J. Low Temp. Phys.}\ }\textbf {\bibinfo {volume} {95}},\ \bibinfo
  {pages} {271} (\bibinfo {year} {1994})}\BibitemShut {NoStop}%
\bibitem [{\citenamefont {Fulde}\ and\ \citenamefont {Ferrell}(1964)}]{fuld64}%
  \BibitemOpen
  \bibfield  {author} {\bibinfo {author} {\bibfnamefont {P.}~\bibnamefont
  {Fulde}}\ and\ \bibinfo {author} {\bibfnamefont {R.~A.}\ \bibnamefont
  {Ferrell}},\ }\bibfield  {title} {\bibinfo {title} {Superconductivity in a
  strong spin-exchange field},\ }\href@noop {} {\bibfield  {journal} {\bibinfo
  {journal} {Phys. Rev.}\ }\textbf {\bibinfo {volume} {135}},\ \bibinfo {pages}
  {A550} (\bibinfo {year} {1964})}\BibitemShut {NoStop}%
\bibitem [{\citenamefont {Larkin}\ and\ \citenamefont
  {Ovchinnikov}(1964)}]{lark64}%
  \BibitemOpen
  \bibfield  {author} {\bibinfo {author} {\bibfnamefont {A.~I.}\ \bibnamefont
  {Larkin}}\ and\ \bibinfo {author} {\bibfnamefont {Y.~N.}\ \bibnamefont
  {Ovchinnikov}},\ }\href@noop {} {\bibfield  {journal} {\bibinfo  {journal}
  {Zh. Eksp. Teor. Fiz.}\ }\textbf {\bibinfo {volume} {47}},\ \bibinfo {pages}
  {1136} (\bibinfo {year} {1964})}\BibitemShut {NoStop}%
\bibitem [{\citenamefont {Koutroulakis}\ \emph {et~al.}(2016)\citenamefont
  {Koutroulakis}, \citenamefont {K\"uhne}, \citenamefont {Schlueter},
  \citenamefont {Wosnitza},\ and\ \citenamefont {Brown}}]{kout16}%
  \BibitemOpen
  \bibfield  {author} {\bibinfo {author} {\bibfnamefont {G.}~\bibnamefont
  {Koutroulakis}}, \bibinfo {author} {\bibfnamefont {H.}~\bibnamefont
  {K\"uhne}}, \bibinfo {author} {\bibfnamefont {J.~A.}\ \bibnamefont
  {Schlueter}}, \bibinfo {author} {\bibfnamefont {J.}~\bibnamefont
  {Wosnitza}},\ and\ \bibinfo {author} {\bibfnamefont {S.~E.}\ \bibnamefont
  {Brown}},\ }\bibfield  {title} {\bibinfo {title} {{Microscopic Study of the
  Fulde-Ferrell-Larkin-Ovchinnikov State in an All-Organic Superconductor}},\
  }\href {https://doi.org/10.1103/PhysRevLett.116.067003} {\bibfield  {journal}
  {\bibinfo  {journal} {Phys. Rev. Lett.}\ }\textbf {\bibinfo {volume} {116}},\
  \bibinfo {pages} {067003} (\bibinfo {year} {2016})}\BibitemShut {NoStop}%
\bibitem [{\citenamefont {Sugiura}\ \emph {et~al.}(2019)\citenamefont
  {Sugiura}, \citenamefont {Isono}, \citenamefont {Terashima}, \citenamefont
  {Yasuzuka}, \citenamefont {Schlueter},\ and\ \citenamefont {Uji}}]{sugi19}%
  \BibitemOpen
  \bibfield  {author} {\bibinfo {author} {\bibfnamefont {S.}~\bibnamefont
  {Sugiura}}, \bibinfo {author} {\bibfnamefont {T.}~\bibnamefont {Isono}},
  \bibinfo {author} {\bibfnamefont {T.}~\bibnamefont {Terashima}}, \bibinfo
  {author} {\bibfnamefont {S.}~\bibnamefont {Yasuzuka}}, \bibinfo {author}
  {\bibfnamefont {J.~A.}\ \bibnamefont {Schlueter}},\ and\ \bibinfo {author}
  {\bibfnamefont {S.}~\bibnamefont {Uji}},\ }\bibfield  {title} {\bibinfo
  {title} {{Fulde--Ferrell--Larkin--Ovchinnikov and vortex phases in a layered
  organic superconductor}},\ }\href {https://doi.org/10.1038/s41535-019-0147-2}
  {\bibfield  {journal} {\bibinfo  {journal} {npj Quantum Mater.}\ }\textbf
  {\bibinfo {volume} {4}},\ \bibinfo {pages} {7} (\bibinfo {year}
  {2019})}\BibitemShut {NoStop}%
\bibitem [{\citenamefont {Castellani}\ \emph {et~al.}(1995)\citenamefont
  {Castellani}, \citenamefont {Di~Castro},\ and\ \citenamefont
  {Grilli}}]{cast95}%
  \BibitemOpen
  \bibfield  {author} {\bibinfo {author} {\bibfnamefont {C.}~\bibnamefont
  {Castellani}}, \bibinfo {author} {\bibfnamefont {C.}~\bibnamefont
  {Di~Castro}},\ and\ \bibinfo {author} {\bibfnamefont {M.}~\bibnamefont
  {Grilli}},\ }\bibfield  {title} {\bibinfo {title} {{Singular Quasiparticle
  Scattering in the Proximity of Charge Instabilities}},\ }\href@noop {}
  {\bibfield  {journal} {\bibinfo  {journal} {Phys. Rev. Lett.}\ }\textbf
  {\bibinfo {volume} {75}},\ \bibinfo {pages} {4650} (\bibinfo {year}
  {1995})}\BibitemShut {NoStop}%
\bibitem [{\citenamefont {Grilli}\ \emph {et~al.}(2009)\citenamefont {Grilli},
  \citenamefont {Seibold}, \citenamefont {Di~Ciolo},\ and\ \citenamefont
  {Lorenzana}}]{gril09}%
  \BibitemOpen
  \bibfield  {author} {\bibinfo {author} {\bibfnamefont {M.}~\bibnamefont
  {Grilli}}, \bibinfo {author} {\bibfnamefont {G.}~\bibnamefont {Seibold}},
  \bibinfo {author} {\bibfnamefont {A.}~\bibnamefont {Di~Ciolo}},\ and\
  \bibinfo {author} {\bibfnamefont {J.}~\bibnamefont {Lorenzana}},\ }\bibfield
  {title} {\bibinfo {title} {{Fermi surface dichotomy in systems with
  fluctuating order}},\ }\href {https://doi.org/10.1103/PhysRevB.79.125111}
  {\bibfield  {journal} {\bibinfo  {journal} {Phys. Rev. B}\ }\textbf {\bibinfo
  {volume} {79}},\ \bibinfo {pages} {125111} (\bibinfo {year}
  {2009})}\BibitemShut {NoStop}%
\bibitem [{\citenamefont {Caprara}\ \emph {et~al.}(2017)\citenamefont
  {Caprara}, \citenamefont {Di~Castro}, \citenamefont {Seibold},\ and\
  \citenamefont {Grilli}}]{capr17}%
  \BibitemOpen
  \bibfield  {author} {\bibinfo {author} {\bibfnamefont {S.}~\bibnamefont
  {Caprara}}, \bibinfo {author} {\bibfnamefont {C.}~\bibnamefont {Di~Castro}},
  \bibinfo {author} {\bibfnamefont {G.}~\bibnamefont {Seibold}},\ and\ \bibinfo
  {author} {\bibfnamefont {M.}~\bibnamefont {Grilli}},\ }\bibfield  {title}
  {\bibinfo {title} {{Dynamical charge density waves rule the phase diagram of
  cuprates}},\ }\href {https://doi.org/10.1103/PhysRevB.95.224511} {\bibfield
  {journal} {\bibinfo  {journal} {Phys. Rev. B}\ }\textbf {\bibinfo {volume}
  {95}},\ \bibinfo {pages} {224511} (\bibinfo {year} {2017})}\BibitemShut
  {NoStop}%
\bibitem [{\citenamefont {Wang}\ \emph {et~al.}(2016)\citenamefont {Wang},
  \citenamefont {Abanov}, \citenamefont {Altshuler}, \citenamefont
  {Yuzbashyan},\ and\ \citenamefont {Chubukov}}]{wang16}%
  \BibitemOpen
  \bibfield  {author} {\bibinfo {author} {\bibfnamefont {Y.}~\bibnamefont
  {Wang}}, \bibinfo {author} {\bibfnamefont {A.}~\bibnamefont {Abanov}},
  \bibinfo {author} {\bibfnamefont {B.~L.}\ \bibnamefont {Altshuler}}, \bibinfo
  {author} {\bibfnamefont {E.~A.}\ \bibnamefont {Yuzbashyan}},\ and\ \bibinfo
  {author} {\bibfnamefont {A.~V.}\ \bibnamefont {Chubukov}},\ }\bibfield
  {title} {\bibinfo {title} {{Superconductivity near a Quantum-Critical Point:
  The Special Role of the First Matsubara Frequency}},\ }\href
  {https://doi.org/10.1103/PhysRevLett.117.157001} {\bibfield  {journal}
  {\bibinfo  {journal} {Phys. Rev. Lett.}\ }\textbf {\bibinfo {volume} {117}},\
  \bibinfo {pages} {157001} (\bibinfo {year} {2016})}\BibitemShut {NoStop}%
\bibitem [{\citenamefont {Fradkin}\ \emph {et~al.}(2015)\citenamefont
  {Fradkin}, \citenamefont {Kivelson},\ and\ \citenamefont
  {Tranquada}}]{frad15}%
  \BibitemOpen
  \bibfield  {author} {\bibinfo {author} {\bibfnamefont {E.}~\bibnamefont
  {Fradkin}}, \bibinfo {author} {\bibfnamefont {S.~A.}\ \bibnamefont
  {Kivelson}},\ and\ \bibinfo {author} {\bibfnamefont {J.~M.}\ \bibnamefont
  {Tranquada}},\ }\bibfield  {title} {\bibinfo {title} {{{Colloquium} : Theory
  of intertwined orders in high temperature superconductors}},\ }\href
  {https://doi.org/10.1103/RevModPhys.87.457} {\bibfield  {journal} {\bibinfo
  {journal} {Rev. Mod. Phys.}\ }\textbf {\bibinfo {volume} {87}},\ \bibinfo
  {pages} {457} (\bibinfo {year} {2015})}\BibitemShut {NoStop}%
\bibitem [{\citenamefont {Carlson}\ \emph {et~al.}(2004)\citenamefont
  {Carlson}, \citenamefont {Emery}, \citenamefont {Kivelson},\ and\
  \citenamefont {Orgad}}]{carl03}%
  \BibitemOpen
  \bibfield  {author} {\bibinfo {author} {\bibfnamefont {E.~W.}\ \bibnamefont
  {Carlson}}, \bibinfo {author} {\bibfnamefont {V.~J.}\ \bibnamefont {Emery}},
  \bibinfo {author} {\bibfnamefont {S.~A.}\ \bibnamefont {Kivelson}},\ and\
  \bibinfo {author} {\bibfnamefont {D.}~\bibnamefont {Orgad}},\ }\bibfield
  {title} {\bibinfo {title} {Concepts in high temperature superconductivity},\
  }in\ \href@noop {} {\emph {\bibinfo {booktitle} {The Physics of
  Superconductors Vol II: Superconductivity in Nanostructures, High-$T_c$ and
  Novel Superconductors, Organic Superconducors}}},\ \bibinfo {editor} {edited
  by\ \bibinfo {editor} {\bibfnamefont {K.~H.}\ \bibnamefont {Bennemann}}\ and\
  \bibinfo {editor} {\bibfnamefont {J.~B.}\ \bibnamefont {Ketterson}}}\
  (\bibinfo  {publisher} {Springer-Verlag},\ \bibinfo {address} {Berlin},\
  \bibinfo {year} {2004})\ \Eprint
  {https://arxiv.org/abs/arXiv:cond-mat/0206217} {arXiv:cond-mat/0206217}
  \BibitemShut {NoStop}%
\bibitem [{\citenamefont {Tsai}\ \emph {et~al.}(2008)\citenamefont {Tsai},
  \citenamefont {Yao}, \citenamefont {L\"auchli},\ and\ \citenamefont
  {Kivelson}}]{tsai08}%
  \BibitemOpen
  \bibfield  {author} {\bibinfo {author} {\bibfnamefont {W.-F.}\ \bibnamefont
  {Tsai}}, \bibinfo {author} {\bibfnamefont {H.}~\bibnamefont {Yao}}, \bibinfo
  {author} {\bibfnamefont {A.}~\bibnamefont {L\"auchli}},\ and\ \bibinfo
  {author} {\bibfnamefont {S.~A.}\ \bibnamefont {Kivelson}},\ }\bibfield
  {title} {\bibinfo {title} {{Optimal inhomogeneity for superconductivity:
  Finite-size studies}},\ }\href {https://doi.org/10.1103/PhysRevB.77.214502}
  {\bibfield  {journal} {\bibinfo  {journal} {Phys. Rev. B}\ }\textbf {\bibinfo
  {volume} {77}},\ \bibinfo {pages} {214502} (\bibinfo {year}
  {2008})}\BibitemShut {NoStop}%
\bibitem [{\citenamefont {Jiang}\ and\ \citenamefont
  {Kivelson}(2021)}]{jian21b}%
  \BibitemOpen
  \bibfield  {author} {\bibinfo {author} {\bibfnamefont {H.-C.}\ \bibnamefont
  {Jiang}}\ and\ \bibinfo {author} {\bibfnamefont {S.~A.}\ \bibnamefont
  {Kivelson}},\ }\href@noop {} {\bibinfo {title} {{Stripe order enhanced
  superconductivity in the Hubbard model}}},\ \bibinfo {howpublished}
  {arXiv:2105.07048} (\bibinfo {year} {2021})\BibitemShut {NoStop}%
\bibitem [{\citenamefont {Yang}(2013)}]{yang13}%
  \BibitemOpen
  \bibfield  {author} {\bibinfo {author} {\bibfnamefont {K.}~\bibnamefont
  {Yang}},\ }\bibfield  {title} {\bibinfo {title} {{Detection of Striped
  Superconductors Using Magnetic Field Modulated Josephson Effect}},\ }\href
  {https://doi.org/10.1007/s10948-012-2075-2} {\bibfield  {journal} {\bibinfo
  {journal} {J. Supercond. Nov. Magn.}\ }\textbf {\bibinfo {volume} {26}},\
  \bibinfo {pages} {2741} (\bibinfo {year} {2013})}\BibitemShut {NoStop}%
\bibitem [{\citenamefont {Shi}\ \emph {et~al.}(2021)\citenamefont {Shi},
  \citenamefont {Baity}, \citenamefont {Terzic}, \citenamefont {Pokharel},
  \citenamefont {Sasagawa},\ and\ \citenamefont {Popovi{\'c}}}]{shi21}%
  \BibitemOpen
  \bibfield  {author} {\bibinfo {author} {\bibfnamefont {Z.}~\bibnamefont
  {Shi}}, \bibinfo {author} {\bibfnamefont {P.~G.}\ \bibnamefont {Baity}},
  \bibinfo {author} {\bibfnamefont {J.}~\bibnamefont {Terzic}}, \bibinfo
  {author} {\bibfnamefont {B.~K.}\ \bibnamefont {Pokharel}}, \bibinfo {author}
  {\bibfnamefont {T.}~\bibnamefont {Sasagawa}},\ and\ \bibinfo {author}
  {\bibfnamefont {D.}~\bibnamefont {Popovi{\'c}}},\ }\bibfield  {title}
  {\bibinfo {title} {{Magnetic field reveals vanishing Hall response in the
  normal state of stripe-ordered cuprates}},\ }\href
  {https://doi.org/10.1038/s41467-021-24000-3} {\bibfield  {journal} {\bibinfo
  {journal} {Nat. Commun.}\ }\textbf {\bibinfo {volume} {12}},\ \bibinfo
  {pages} {3724} (\bibinfo {year} {2021})}\BibitemShut {NoStop}%
\bibitem [{\citenamefont {Ren}\ and\ \citenamefont {Tsvelik}(2020)}]{ren20}%
  \BibitemOpen
  \bibfield  {author} {\bibinfo {author} {\bibfnamefont {T.}~\bibnamefont
  {Ren}}\ and\ \bibinfo {author} {\bibfnamefont {A.~M.}\ \bibnamefont
  {Tsvelik}},\ }\bibfield  {title} {\bibinfo {title} {{How magnetic field can
  transform a superconductor into a Bose metal}},\ }\href
  {https://doi.org/10.1088/1367-2630/abbc2b} {\bibfield  {journal} {\bibinfo
  {journal} {New J. Phys.}\ }\textbf {\bibinfo {volume} {22}},\ \bibinfo
  {pages} {103021} (\bibinfo {year} {2020})}\BibitemShut {NoStop}%
\bibitem [{\citenamefont {Zhou}\ \emph {et~al.}(2019)\citenamefont {Zhou},
  \citenamefont {Chen}, \citenamefont {Liu}, \citenamefont {Sochnikov},
  \citenamefont {Bollinger}, \citenamefont {Han}, \citenamefont {Zhu},
  \citenamefont {He}, \citenamefont {Bo\v{z}ovi\'{c}},\ and\ \citenamefont
  {Natelson}}]{zhou19}%
  \BibitemOpen
  \bibfield  {author} {\bibinfo {author} {\bibfnamefont {P.}~\bibnamefont
  {Zhou}}, \bibinfo {author} {\bibfnamefont {L.}~\bibnamefont {Chen}}, \bibinfo
  {author} {\bibfnamefont {Y.}~\bibnamefont {Liu}}, \bibinfo {author}
  {\bibfnamefont {I.}~\bibnamefont {Sochnikov}}, \bibinfo {author}
  {\bibfnamefont {A.~T.}\ \bibnamefont {Bollinger}}, \bibinfo {author}
  {\bibfnamefont {M.-G.}\ \bibnamefont {Han}}, \bibinfo {author} {\bibfnamefont
  {Y.}~\bibnamefont {Zhu}}, \bibinfo {author} {\bibfnamefont {X.}~\bibnamefont
  {He}}, \bibinfo {author} {\bibfnamefont {I.}~\bibnamefont
  {Bo\v{z}ovi\'{c}}},\ and\ \bibinfo {author} {\bibfnamefont {D.}~\bibnamefont
  {Natelson}},\ }\bibfield  {title} {\bibinfo {title} {{Electron pairing in the
  pseudogap state revealed by shot noise in copper oxide junctions}},\ }\href
  {https://doi.org/10.1038/s41586-019-1486-7} {\bibfield  {journal} {\bibinfo
  {journal} {Nature}\ }\textbf {\bibinfo {volume} {572}},\ \bibinfo {pages}
  {493} (\bibinfo {year} {2019})}\BibitemShut {NoStop}%
\bibitem [{\citenamefont {Enoki}\ \emph {et~al.}(2013)\citenamefont {Enoki},
  \citenamefont {Fujita}, \citenamefont {Nishizaki}, \citenamefont {Iikubo},
  \citenamefont {Singh}, \citenamefont {Chang}, \citenamefont {Tranquada},\
  and\ \citenamefont {Yamada}}]{enok13}%
  \BibitemOpen
  \bibfield  {author} {\bibinfo {author} {\bibfnamefont {M.}~\bibnamefont
  {Enoki}}, \bibinfo {author} {\bibfnamefont {M.}~\bibnamefont {Fujita}},
  \bibinfo {author} {\bibfnamefont {T.}~\bibnamefont {Nishizaki}}, \bibinfo
  {author} {\bibfnamefont {S.}~\bibnamefont {Iikubo}}, \bibinfo {author}
  {\bibfnamefont {D.~K.}\ \bibnamefont {Singh}}, \bibinfo {author}
  {\bibfnamefont {S.}~\bibnamefont {Chang}}, \bibinfo {author} {\bibfnamefont
  {J.~M.}\ \bibnamefont {Tranquada}},\ and\ \bibinfo {author} {\bibfnamefont
  {K.}~\bibnamefont {Yamada}},\ }\bibfield  {title} {\bibinfo {title}
  {{Spin-Stripe Density Varies Linearly With the Hole Content in Single-Layer
  ${\mathrm{Bi}}_{2+x}{\mathrm{Sr}}_{2\ensuremath{-}x}{\mathrm{CuO}}_{6+y}$
  Cuprate Superconductors}},\ }\href
  {https://doi.org/10.1103/PhysRevLett.110.017004} {\bibfield  {journal}
  {\bibinfo  {journal} {Phys. Rev. Lett.}\ }\textbf {\bibinfo {volume} {110}},\
  \bibinfo {pages} {017004} (\bibinfo {year} {2013})}\BibitemShut {NoStop}%
\bibitem [{\citenamefont {Valla}\ \emph {et~al.}(2006)\citenamefont {Valla},
  \citenamefont {Federov}, \citenamefont {Lee}, \citenamefont {Davis},\ and\
  \citenamefont {Gu}}]{vall06}%
  \BibitemOpen
  \bibfield  {author} {\bibinfo {author} {\bibfnamefont {T.}~\bibnamefont
  {Valla}}, \bibinfo {author} {\bibfnamefont {A.~V.}\ \bibnamefont {Federov}},
  \bibinfo {author} {\bibfnamefont {J.}~\bibnamefont {Lee}}, \bibinfo {author}
  {\bibfnamefont {J.~C.}\ \bibnamefont {Davis}},\ and\ \bibinfo {author}
  {\bibfnamefont {G.~D.}\ \bibnamefont {Gu}},\ }\bibfield  {title} {\bibinfo
  {title} {{The Ground State of the Pseudogap in Cuprate Superconductors}},\
  }\href@noop {} {\bibfield  {journal} {\bibinfo  {journal} {Science}\ }\textbf
  {\bibinfo {volume} {314}},\ \bibinfo {pages} {1914} (\bibinfo {year}
  {2006})}\BibitemShut {NoStop}%
\bibitem [{\citenamefont {He}\ \emph {et~al.}(2009)\citenamefont {He},
  \citenamefont {Tanaka}, \citenamefont {Mo}, \citenamefont {Sasagawa},
  \citenamefont {Fujita}, \citenamefont {Adachi}, \citenamefont {Mannella},
  \citenamefont {Yamada}, \citenamefont {Koike}, \citenamefont {Hussain},\ and\
  \citenamefont {Shen}}]{he09}%
  \BibitemOpen
  \bibfield  {author} {\bibinfo {author} {\bibfnamefont {R.-H.}\ \bibnamefont
  {He}}, \bibinfo {author} {\bibfnamefont {K.}~\bibnamefont {Tanaka}}, \bibinfo
  {author} {\bibfnamefont {S.-K.}\ \bibnamefont {Mo}}, \bibinfo {author}
  {\bibfnamefont {T.}~\bibnamefont {Sasagawa}}, \bibinfo {author}
  {\bibfnamefont {M.}~\bibnamefont {Fujita}}, \bibinfo {author} {\bibfnamefont
  {T.}~\bibnamefont {Adachi}}, \bibinfo {author} {\bibfnamefont
  {N.}~\bibnamefont {Mannella}}, \bibinfo {author} {\bibfnamefont
  {K.}~\bibnamefont {Yamada}}, \bibinfo {author} {\bibfnamefont
  {Y.}~\bibnamefont {Koike}}, \bibinfo {author} {\bibfnamefont
  {Z.}~\bibnamefont {Hussain}},\ and\ \bibinfo {author} {\bibfnamefont {Z.-X.}\
  \bibnamefont {Shen}},\ }\bibfield  {title} {\bibinfo {title} {{Energy gaps in
  the failed high-$T_c$ superconductor La$_{1.875}$Ba$_{0.125}$CuO$_4$}},\
  }\href@noop {} {\bibfield  {journal} {\bibinfo  {journal} {Nat. Phys.}\
  }\textbf {\bibinfo {volume} {5}},\ \bibinfo {pages} {119} (\bibinfo {year}
  {2009})}\BibitemShut {NoStop}%
\bibitem [{\citenamefont {Homes}\ \emph {et~al.}(2012)\citenamefont {Homes},
  \citenamefont {H\"ucker}, \citenamefont {Li}, \citenamefont {Xu},
  \citenamefont {Wen}, \citenamefont {Gu},\ and\ \citenamefont
  {Tranquada}}]{home12}%
  \BibitemOpen
  \bibfield  {author} {\bibinfo {author} {\bibfnamefont {C.~C.}\ \bibnamefont
  {Homes}}, \bibinfo {author} {\bibfnamefont {M.}~\bibnamefont {H\"ucker}},
  \bibinfo {author} {\bibfnamefont {Q.}~\bibnamefont {Li}}, \bibinfo {author}
  {\bibfnamefont {Z.~J.}\ \bibnamefont {Xu}}, \bibinfo {author} {\bibfnamefont
  {J.~S.}\ \bibnamefont {Wen}}, \bibinfo {author} {\bibfnamefont {G.~D.}\
  \bibnamefont {Gu}},\ and\ \bibinfo {author} {\bibfnamefont {J.~M.}\
  \bibnamefont {Tranquada}},\ }\bibfield  {title} {\bibinfo {title}
  {{Determination of the optical properties of La$_{2-x}$Ba$_{x}$CuO$_{4}$ for
  several dopings, including the anomalous $x=\frac{1}{8}$ phase}},\
  }\href@noop {} {\bibfield  {journal} {\bibinfo  {journal} {Phys. Rev. B}\
  }\textbf {\bibinfo {volume} {85}},\ \bibinfo {pages} {134510} (\bibinfo
  {year} {2012})}\BibitemShut {NoStop}%
\bibitem [{\citenamefont {Damascelli}\ \emph {et~al.}(2003)\citenamefont
  {Damascelli}, \citenamefont {Hussain},\ and\ \citenamefont {Shen}}]{dama03}%
  \BibitemOpen
  \bibfield  {author} {\bibinfo {author} {\bibfnamefont {A.}~\bibnamefont
  {Damascelli}}, \bibinfo {author} {\bibfnamefont {Z.}~\bibnamefont
  {Hussain}},\ and\ \bibinfo {author} {\bibfnamefont {Z.-X.}\ \bibnamefont
  {Shen}},\ }\bibfield  {title} {\bibinfo {title} {{Angle-resolved
  photoemission studies of the cuprate superconductors}},\ }\href
  {https://doi.org/10.1103/RevModPhys.75.473} {\bibfield  {journal} {\bibinfo
  {journal} {Rev. Mod. Phys.}\ }\textbf {\bibinfo {volume} {75}},\ \bibinfo
  {pages} {473} (\bibinfo {year} {2003})}\BibitemShut {NoStop}%
\bibitem [{\citenamefont {Basov}\ and\ \citenamefont {Timusk}(2005)}]{baso05}%
  \BibitemOpen
  \bibfield  {author} {\bibinfo {author} {\bibfnamefont {D.~N.}\ \bibnamefont
  {Basov}}\ and\ \bibinfo {author} {\bibfnamefont {T.}~\bibnamefont {Timusk}},\
  }\bibfield  {title} {\bibinfo {title} {{Electrodynamics of high-${T}_{c}$
  superconductors}},\ }\href@noop {} {\bibfield  {journal} {\bibinfo  {journal}
  {Rev. Mod. Phys.}\ }\textbf {\bibinfo {volume} {77}},\ \bibinfo {pages} {721}
  (\bibinfo {year} {2005})}\BibitemShut {NoStop}%
\bibitem [{\citenamefont {Chakravarty}\ \emph {et~al.}(1988)\citenamefont
  {Chakravarty}, \citenamefont {Halperin},\ and\ \citenamefont
  {Nelson}}]{chak88}%
  \BibitemOpen
  \bibfield  {author} {\bibinfo {author} {\bibfnamefont {S.}~\bibnamefont
  {Chakravarty}}, \bibinfo {author} {\bibfnamefont {B.~I.}\ \bibnamefont
  {Halperin}},\ and\ \bibinfo {author} {\bibfnamefont {D.~R.}\ \bibnamefont
  {Nelson}},\ }\bibfield  {title} {\bibinfo {title} {{Low-temperature behavior
  of two-dimensional quantum antiferromagnets}},\ }\href
  {https://doi.org/10.1103/PhysRevLett.60.1057} {\bibfield  {journal} {\bibinfo
   {journal} {Phys. Rev. Lett.}\ }\textbf {\bibinfo {volume} {60}},\ \bibinfo
  {pages} {1057} (\bibinfo {year} {1988})}\BibitemShut {NoStop}%
\bibitem [{\citenamefont {White}\ and\ \citenamefont
  {Scalapino}(2015)}]{whit15}%
  \BibitemOpen
  \bibfield  {author} {\bibinfo {author} {\bibfnamefont {S.~R.}\ \bibnamefont
  {White}}\ and\ \bibinfo {author} {\bibfnamefont {D.~J.}\ \bibnamefont
  {Scalapino}},\ }\bibfield  {title} {\bibinfo {title} {{Doping asymmetry and
  striping in a three-orbital ${\mathrm{CuO}}_{2}$ Hubbard model}},\ }\href
  {https://doi.org/10.1103/PhysRevB.92.205112} {\bibfield  {journal} {\bibinfo
  {journal} {Phys. Rev. B}\ }\textbf {\bibinfo {volume} {92}},\ \bibinfo
  {pages} {205112} (\bibinfo {year} {2015})}\BibitemShut {NoStop}%
\bibitem [{\citenamefont {Lee}\ \emph {et~al.}(2007)\citenamefont {Lee},
  \citenamefont {Vishik}, \citenamefont {Tanaka}, \citenamefont {Lu},
  \citenamefont {Sasagawa}, \citenamefont {Nagaosa}, \citenamefont {Devereaux},
  \citenamefont {Hussain},\ and\ \citenamefont {Shen}}]{lee07}%
  \BibitemOpen
  \bibfield  {author} {\bibinfo {author} {\bibfnamefont {W.~S.}\ \bibnamefont
  {Lee}}, \bibinfo {author} {\bibfnamefont {I.~M.}\ \bibnamefont {Vishik}},
  \bibinfo {author} {\bibfnamefont {K.}~\bibnamefont {Tanaka}}, \bibinfo
  {author} {\bibfnamefont {D.~H.}\ \bibnamefont {Lu}}, \bibinfo {author}
  {\bibfnamefont {T.}~\bibnamefont {Sasagawa}}, \bibinfo {author}
  {\bibfnamefont {N.}~\bibnamefont {Nagaosa}}, \bibinfo {author} {\bibfnamefont
  {T.~P.}\ \bibnamefont {Devereaux}}, \bibinfo {author} {\bibfnamefont
  {Z.}~\bibnamefont {Hussain}},\ and\ \bibinfo {author} {\bibfnamefont {Z.~X.}\
  \bibnamefont {Shen}},\ }\bibfield  {title} {\bibinfo {title} {{Abrupt onset
  of a second energy gap at the superconducting transition of underdoped
  Bi2212}},\ }\href {https://doi.org/10.1038/nature06219} {\bibfield  {journal}
  {\bibinfo  {journal} {Nature}\ }\textbf {\bibinfo {volume} {450}},\ \bibinfo
  {pages} {81} (\bibinfo {year} {2007})}\BibitemShut {NoStop}%
\bibitem [{\citenamefont {Yoshida}\ \emph {et~al.}(2016)\citenamefont
  {Yoshida}, \citenamefont {Ishii}, \citenamefont {Naka}, \citenamefont
  {Ishihara}, \citenamefont {Jarrige}, \citenamefont {Ikeuchi}, \citenamefont
  {Murakami}, \citenamefont {Kudo}, \citenamefont {Koike}, \citenamefont
  {Nagata}, \citenamefont {Fukada}, \citenamefont {Ikeda},\ and\ \citenamefont
  {Mizuki}}]{yosh16}%
  \BibitemOpen
  \bibfield  {author} {\bibinfo {author} {\bibfnamefont {M.}~\bibnamefont
  {Yoshida}}, \bibinfo {author} {\bibfnamefont {K.}~\bibnamefont {Ishii}},
  \bibinfo {author} {\bibfnamefont {M.}~\bibnamefont {Naka}}, \bibinfo {author}
  {\bibfnamefont {S.}~\bibnamefont {Ishihara}}, \bibinfo {author}
  {\bibfnamefont {I.}~\bibnamefont {Jarrige}}, \bibinfo {author} {\bibfnamefont
  {K.}~\bibnamefont {Ikeuchi}}, \bibinfo {author} {\bibfnamefont
  {Y.}~\bibnamefont {Murakami}}, \bibinfo {author} {\bibfnamefont
  {K.}~\bibnamefont {Kudo}}, \bibinfo {author} {\bibfnamefont {Y.}~\bibnamefont
  {Koike}}, \bibinfo {author} {\bibfnamefont {T.}~\bibnamefont {Nagata}},
  \bibinfo {author} {\bibfnamefont {Y.}~\bibnamefont {Fukada}}, \bibinfo
  {author} {\bibfnamefont {N.}~\bibnamefont {Ikeda}},\ and\ \bibinfo {author}
  {\bibfnamefont {J.}~\bibnamefont {Mizuki}},\ }\bibfield  {title} {\bibinfo
  {title} {{Observation of momentum-resolved charge fluctuations proximate to
  the charge-order phase using resonant inelastic x-ray scattering}},\ }\href
  {http://dx.doi.org/10.1038/srep23611} {\bibfield  {journal} {\bibinfo
  {journal} {Sci. Rep.}\ }\textbf {\bibinfo {volume} {6}},\ \bibinfo {pages}
  {23611 EP } (\bibinfo {year} {2016})}\BibitemShut {NoStop}%
\bibitem [{\citenamefont {Drozdov}\ \emph {et~al.}(2018)\citenamefont
  {Drozdov}, \citenamefont {Pletikosi{\'c}}, \citenamefont {Kim}, \citenamefont
  {Fujita}, \citenamefont {Gu}, \citenamefont {Davis}, \citenamefont {Johnson},
  \citenamefont {Bo{\v z}ovi{\'c}},\ and\ \citenamefont {Valla}}]{droz18}%
  \BibitemOpen
  \bibfield  {author} {\bibinfo {author} {\bibfnamefont {I.~K.}\ \bibnamefont
  {Drozdov}}, \bibinfo {author} {\bibfnamefont {I.}~\bibnamefont
  {Pletikosi{\'c}}}, \bibinfo {author} {\bibfnamefont {C.~K.}\ \bibnamefont
  {Kim}}, \bibinfo {author} {\bibfnamefont {K.}~\bibnamefont {Fujita}},
  \bibinfo {author} {\bibfnamefont {G.~D.}\ \bibnamefont {Gu}}, \bibinfo
  {author} {\bibfnamefont {J.~C.~S.}\ \bibnamefont {Davis}}, \bibinfo {author}
  {\bibfnamefont {P.~D.}\ \bibnamefont {Johnson}}, \bibinfo {author}
  {\bibfnamefont {I.}~\bibnamefont {Bo{\v z}ovi{\'c}}},\ and\ \bibinfo {author}
  {\bibfnamefont {T.}~\bibnamefont {Valla}},\ }\bibfield  {title} {\bibinfo
  {title} {{Phase diagram of Bi$_2$Sr$_2$CaCu$_2$O$_{8+\delta}$ revisited}},\
  }\href {https://doi.org/10.1038/s41467-018-07686-w} {\bibfield  {journal}
  {\bibinfo  {journal} {Nat. Commun.}\ }\textbf {\bibinfo {volume} {9}},\
  \bibinfo {pages} {5210} (\bibinfo {year} {2018})}\BibitemShut {NoStop}%
\bibitem [{\citenamefont {Munnikes}\ \emph {et~al.}(2011)\citenamefont
  {Munnikes}, \citenamefont {Muschler}, \citenamefont {Venturini},
  \citenamefont {Tassini}, \citenamefont {Prestel}, \citenamefont {Ono},
  \citenamefont {Ando}, \citenamefont {Peets}, \citenamefont {Hardy},
  \citenamefont {Liang}, \citenamefont {Bonn}, \citenamefont {Damascelli},
  \citenamefont {Eisaki}, \citenamefont {Greven}, \citenamefont {Erb},\ and\
  \citenamefont {Hackl}}]{munn11}%
  \BibitemOpen
  \bibfield  {author} {\bibinfo {author} {\bibfnamefont {N.}~\bibnamefont
  {Munnikes}}, \bibinfo {author} {\bibfnamefont {B.}~\bibnamefont {Muschler}},
  \bibinfo {author} {\bibfnamefont {F.}~\bibnamefont {Venturini}}, \bibinfo
  {author} {\bibfnamefont {L.}~\bibnamefont {Tassini}}, \bibinfo {author}
  {\bibfnamefont {W.}~\bibnamefont {Prestel}}, \bibinfo {author} {\bibfnamefont
  {S.}~\bibnamefont {Ono}}, \bibinfo {author} {\bibfnamefont {Y.}~\bibnamefont
  {Ando}}, \bibinfo {author} {\bibfnamefont {D.~C.}\ \bibnamefont {Peets}},
  \bibinfo {author} {\bibfnamefont {W.~N.}\ \bibnamefont {Hardy}}, \bibinfo
  {author} {\bibfnamefont {R.}~\bibnamefont {Liang}}, \bibinfo {author}
  {\bibfnamefont {D.~A.}\ \bibnamefont {Bonn}}, \bibinfo {author}
  {\bibfnamefont {A.}~\bibnamefont {Damascelli}}, \bibinfo {author}
  {\bibfnamefont {H.}~\bibnamefont {Eisaki}}, \bibinfo {author} {\bibfnamefont
  {M.}~\bibnamefont {Greven}}, \bibinfo {author} {\bibfnamefont
  {A.}~\bibnamefont {Erb}},\ and\ \bibinfo {author} {\bibfnamefont
  {R.}~\bibnamefont {Hackl}},\ }\bibfield  {title} {\bibinfo {title} {{Pair
  breaking versus symmetry breaking: Origin of the Raman modes in
  superconducting cuprates}},\ }\href
  {https://doi.org/10.1103/PhysRevB.84.144523} {\bibfield  {journal} {\bibinfo
  {journal} {Phys. Rev. B}\ }\textbf {\bibinfo {volume} {84}},\ \bibinfo
  {pages} {144523} (\bibinfo {year} {2011})}\BibitemShut {NoStop}%
\bibitem [{\citenamefont {Sacuto}\ \emph {et~al.}(2013)\citenamefont {Sacuto},
  \citenamefont {Gallais}, \citenamefont {Cazayous}, \citenamefont {M\'easson},
  \citenamefont {Gu},\ and\ \citenamefont {Colson}}]{sacu13}%
  \BibitemOpen
  \bibfield  {author} {\bibinfo {author} {\bibfnamefont {A.}~\bibnamefont
  {Sacuto}}, \bibinfo {author} {\bibfnamefont {Y.}~\bibnamefont {Gallais}},
  \bibinfo {author} {\bibfnamefont {M.}~\bibnamefont {Cazayous}}, \bibinfo
  {author} {\bibfnamefont {M.-A.}\ \bibnamefont {M\'easson}}, \bibinfo {author}
  {\bibfnamefont {G.~D.}\ \bibnamefont {Gu}},\ and\ \bibinfo {author}
  {\bibfnamefont {D.}~\bibnamefont {Colson}},\ }\bibfield  {title} {\bibinfo
  {title} {{New insights into the phase diagram of the copper oxide
  superconductors from electronic Raman scattering}},\ }\href
  {http://stacks.iop.org/0034-4885/76/i=2/a=022502} {\bibfield  {journal}
  {\bibinfo  {journal} {Rep. Prog. Phys.}\ }\textbf {\bibinfo {volume} {76}},\
  \bibinfo {pages} {022502} (\bibinfo {year} {2013})}\BibitemShut {NoStop}%
\bibitem [{\citenamefont {Guguchia}\ \emph {et~al.}(2020)\citenamefont
  {Guguchia}, \citenamefont {Das}, \citenamefont {Wang}, \citenamefont
  {Adachi}, \citenamefont {Kitajima}, \citenamefont {Elender}, \citenamefont
  {Br\"uckner}, \citenamefont {Ghosh}, \citenamefont {Grinenko}, \citenamefont
  {Shiroka}, \citenamefont {M\"uller}, \citenamefont {Mudry}, \citenamefont
  {Baines}, \citenamefont {Bartkowiak}, \citenamefont {Koike}, \citenamefont
  {Amato}, \citenamefont {Tranquada}, \citenamefont {Klauss}, \citenamefont
  {Hicks},\ and\ \citenamefont {Luetkens}}]{gugu20}%
  \BibitemOpen
  \bibfield  {author} {\bibinfo {author} {\bibfnamefont {Z.}~\bibnamefont
  {Guguchia}}, \bibinfo {author} {\bibfnamefont {D.}~\bibnamefont {Das}},
  \bibinfo {author} {\bibfnamefont {C.~N.}\ \bibnamefont {Wang}}, \bibinfo
  {author} {\bibfnamefont {T.}~\bibnamefont {Adachi}}, \bibinfo {author}
  {\bibfnamefont {N.}~\bibnamefont {Kitajima}}, \bibinfo {author}
  {\bibfnamefont {M.}~\bibnamefont {Elender}}, \bibinfo {author} {\bibfnamefont
  {F.}~\bibnamefont {Br\"uckner}}, \bibinfo {author} {\bibfnamefont
  {S.}~\bibnamefont {Ghosh}}, \bibinfo {author} {\bibfnamefont
  {V.}~\bibnamefont {Grinenko}}, \bibinfo {author} {\bibfnamefont
  {T.}~\bibnamefont {Shiroka}}, \bibinfo {author} {\bibfnamefont
  {M.}~\bibnamefont {M\"uller}}, \bibinfo {author} {\bibfnamefont
  {C.}~\bibnamefont {Mudry}}, \bibinfo {author} {\bibfnamefont
  {C.}~\bibnamefont {Baines}}, \bibinfo {author} {\bibfnamefont
  {M.}~\bibnamefont {Bartkowiak}}, \bibinfo {author} {\bibfnamefont
  {Y.}~\bibnamefont {Koike}}, \bibinfo {author} {\bibfnamefont
  {A.}~\bibnamefont {Amato}}, \bibinfo {author} {\bibfnamefont {J.~M.}\
  \bibnamefont {Tranquada}}, \bibinfo {author} {\bibfnamefont {H.-H.}\
  \bibnamefont {Klauss}}, \bibinfo {author} {\bibfnamefont {C.~W.}\
  \bibnamefont {Hicks}},\ and\ \bibinfo {author} {\bibfnamefont
  {H.}~\bibnamefont {Luetkens}},\ }\bibfield  {title} {\bibinfo {title} {{Using
  Uniaxial Stress to Probe the Relationship between Competing Superconducting
  States in a Cuprate with Spin-stripe Order}},\ }\href
  {https://doi.org/10.1103/PhysRevLett.125.097005} {\bibfield  {journal}
  {\bibinfo  {journal} {Phys. Rev. Lett.}\ }\textbf {\bibinfo {volume} {125}},\
  \bibinfo {pages} {097005} (\bibinfo {year} {2020})}\BibitemShut {NoStop}%
\bibitem [{\citenamefont {Leroux}\ \emph {et~al.}(2019)\citenamefont {Leroux},
  \citenamefont {Mishra}, \citenamefont {Ruff}, \citenamefont {Claus},
  \citenamefont {Smylie}, \citenamefont {Opagiste}, \citenamefont
  {Rodi{\`e}re}, \citenamefont {Kayani}, \citenamefont {Gu}, \citenamefont
  {Tranquada}, \citenamefont {Kwok}, \citenamefont {Islam},\ and\ \citenamefont
  {Welp}}]{lero19}%
  \BibitemOpen
  \bibfield  {author} {\bibinfo {author} {\bibfnamefont {M.}~\bibnamefont
  {Leroux}}, \bibinfo {author} {\bibfnamefont {V.}~\bibnamefont {Mishra}},
  \bibinfo {author} {\bibfnamefont {J.~P.~C.}\ \bibnamefont {Ruff}}, \bibinfo
  {author} {\bibfnamefont {H.}~\bibnamefont {Claus}}, \bibinfo {author}
  {\bibfnamefont {M.~P.}\ \bibnamefont {Smylie}}, \bibinfo {author}
  {\bibfnamefont {C.}~\bibnamefont {Opagiste}}, \bibinfo {author}
  {\bibfnamefont {P.}~\bibnamefont {Rodi{\`e}re}}, \bibinfo {author}
  {\bibfnamefont {A.}~\bibnamefont {Kayani}}, \bibinfo {author} {\bibfnamefont
  {G.~D.}\ \bibnamefont {Gu}}, \bibinfo {author} {\bibfnamefont {J.~M.}\
  \bibnamefont {Tranquada}}, \bibinfo {author} {\bibfnamefont {W.-K.}\
  \bibnamefont {Kwok}}, \bibinfo {author} {\bibfnamefont {Z.}~\bibnamefont
  {Islam}},\ and\ \bibinfo {author} {\bibfnamefont {U.}~\bibnamefont {Welp}},\
  }\bibfield  {title} {\bibinfo {title} {{Disorder raises the critical
  temperature of a cuprate superconductor}},\ }\href
  {https://doi.org/10.1073/pnas.1817134116} {\bibfield  {journal} {\bibinfo
  {journal} {Proc. Natl. Acad. Sci. USA}\ }\textbf {\bibinfo {volume} {116}},\
  \bibinfo {pages} {10691} (\bibinfo {year} {2019})}\BibitemShut {NoStop}%
\bibitem [{\citenamefont {Lee}(2014)}]{lee14}%
  \BibitemOpen
  \bibfield  {author} {\bibinfo {author} {\bibfnamefont {P.~A.}\ \bibnamefont
  {Lee}},\ }\bibfield  {title} {\bibinfo {title} {{Amperean Pairing and the
  Pseudogap Phase of Cuprate Superconductors}},\ }\href@noop {} {\bibfield
  {journal} {\bibinfo  {journal} {Phys. Rev. X}\ }\textbf {\bibinfo {volume}
  {4}},\ \bibinfo {pages} {031017} (\bibinfo {year} {2014})}\BibitemShut
  {NoStop}%
\bibitem [{\citenamefont {Miao}\ \emph {et~al.}(2017)\citenamefont {Miao},
  \citenamefont {Lorenzana}, \citenamefont {Seibold}, \citenamefont {Peng},
  \citenamefont {Amorese}, \citenamefont {Yakhou-Harris}, \citenamefont
  {Kummer}, \citenamefont {Brookes}, \citenamefont {Konik}, \citenamefont
  {Thampy}, \citenamefont {Gu}, \citenamefont {Ghiringhelli}, \citenamefont
  {Braicovich},\ and\ \citenamefont {Dean}}]{miao17}%
  \BibitemOpen
  \bibfield  {author} {\bibinfo {author} {\bibfnamefont {H.}~\bibnamefont
  {Miao}}, \bibinfo {author} {\bibfnamefont {J.}~\bibnamefont {Lorenzana}},
  \bibinfo {author} {\bibfnamefont {G.}~\bibnamefont {Seibold}}, \bibinfo
  {author} {\bibfnamefont {Y.~Y.}\ \bibnamefont {Peng}}, \bibinfo {author}
  {\bibfnamefont {A.}~\bibnamefont {Amorese}}, \bibinfo {author} {\bibfnamefont
  {F.}~\bibnamefont {Yakhou-Harris}}, \bibinfo {author} {\bibfnamefont
  {K.}~\bibnamefont {Kummer}}, \bibinfo {author} {\bibfnamefont {N.~B.}\
  \bibnamefont {Brookes}}, \bibinfo {author} {\bibfnamefont {R.~M.}\
  \bibnamefont {Konik}}, \bibinfo {author} {\bibfnamefont {V.}~\bibnamefont
  {Thampy}}, \bibinfo {author} {\bibfnamefont {G.~D.}\ \bibnamefont {Gu}},
  \bibinfo {author} {\bibfnamefont {G.}~\bibnamefont {Ghiringhelli}}, \bibinfo
  {author} {\bibfnamefont {L.}~\bibnamefont {Braicovich}},\ and\ \bibinfo
  {author} {\bibfnamefont {M.~P.~M.}\ \bibnamefont {Dean}},\ }\bibfield
  {title} {\bibinfo {title} {{High-temperature charge density wave correlations
  in La$_{1.875}$Ba$_{0.125}$CuO$_4$ without spin-charge locking}},\ }\href
  {https://doi.org/10.1073/pnas.1708549114} {\bibfield  {journal} {\bibinfo
  {journal} {Proc. Natl. Acad. Sci. USA}\ }\textbf {\bibinfo {volume} {114}},\
  \bibinfo {pages} {12430} (\bibinfo {year} {2017})},\ \Eprint
  {https://arxiv.org/abs/https://www.pnas.org/content/114/47/12430.full.pdf}
  {https://www.pnas.org/content/114/47/12430.full.pdf} \BibitemShut {NoStop}%
\bibitem [{\citenamefont {H\"ucker}\ \emph {et~al.}(2014)\citenamefont
  {H\"ucker}, \citenamefont {Christensen}, \citenamefont {Holmes},
  \citenamefont {Blackburn}, \citenamefont {Forgan}, \citenamefont {Liang},
  \citenamefont {Bonn}, \citenamefont {Hardy}, \citenamefont {Gutowski},
  \citenamefont {Zimmermann}, \citenamefont {Hayden},\ and\ \citenamefont
  {Chang}}]{huck14}%
  \BibitemOpen
  \bibfield  {author} {\bibinfo {author} {\bibfnamefont {M.}~\bibnamefont
  {H\"ucker}}, \bibinfo {author} {\bibfnamefont {N.~B.}\ \bibnamefont
  {Christensen}}, \bibinfo {author} {\bibfnamefont {A.~T.}\ \bibnamefont
  {Holmes}}, \bibinfo {author} {\bibfnamefont {E.}~\bibnamefont {Blackburn}},
  \bibinfo {author} {\bibfnamefont {E.~M.}\ \bibnamefont {Forgan}}, \bibinfo
  {author} {\bibfnamefont {R.}~\bibnamefont {Liang}}, \bibinfo {author}
  {\bibfnamefont {D.~A.}\ \bibnamefont {Bonn}}, \bibinfo {author}
  {\bibfnamefont {W.~N.}\ \bibnamefont {Hardy}}, \bibinfo {author}
  {\bibfnamefont {O.}~\bibnamefont {Gutowski}}, \bibinfo {author}
  {\bibfnamefont {M.~v.}\ \bibnamefont {Zimmermann}}, \bibinfo {author}
  {\bibfnamefont {S.~M.}\ \bibnamefont {Hayden}},\ and\ \bibinfo {author}
  {\bibfnamefont {J.}~\bibnamefont {Chang}},\ }\bibfield  {title} {\bibinfo
  {title} {{Competing charge, spin, and superconducting orders in underdoped
  ${\mathrm{YBa}}_{2}{\mathrm{Cu}}_{3}{\mathrm{O}}_{y}$}},\ }\href
  {https://doi.org/10.1103/PhysRevB.90.054514} {\bibfield  {journal} {\bibinfo
  {journal} {Phys. Rev. B}\ }\textbf {\bibinfo {volume} {90}},\ \bibinfo
  {pages} {054514} (\bibinfo {year} {2014})}\BibitemShut {NoStop}%
\bibitem [{\citenamefont {Birgeneau}\ \emph {et~al.}(2006)\citenamefont
  {Birgeneau}, \citenamefont {Stock}, \citenamefont {Tranquada},\ and\
  \citenamefont {Yamada}}]{birg06}%
  \BibitemOpen
  \bibfield  {author} {\bibinfo {author} {\bibfnamefont {R.~J.}\ \bibnamefont
  {Birgeneau}}, \bibinfo {author} {\bibfnamefont {C.}~\bibnamefont {Stock}},
  \bibinfo {author} {\bibfnamefont {J.~M.}\ \bibnamefont {Tranquada}},\ and\
  \bibinfo {author} {\bibfnamefont {K.}~\bibnamefont {Yamada}},\ }\bibfield
  {title} {\bibinfo {title} {{Magnetic Neutron Scattering in Hole-Doped Cuprate
  Superconductors}},\ }\href {https://doi.org/10.1143/JPSJ.75.111003}
  {\bibfield  {journal} {\bibinfo  {journal} {J. Phys. Soc. Jpn.}\ }\textbf
  {\bibinfo {volume} {75}},\ \bibinfo {pages} {111003} (\bibinfo {year}
  {2006})}\BibitemShut {NoStop}%
\bibitem [{\citenamefont {Blanco-Canosa}\ \emph {et~al.}(2014)\citenamefont
  {Blanco-Canosa}, \citenamefont {Frano}, \citenamefont {Schierle},
  \citenamefont {Porras}, \citenamefont {Loew}, \citenamefont {Minola},
  \citenamefont {Bluschke}, \citenamefont {Weschke}, \citenamefont {Keimer},\
  and\ \citenamefont {Le~Tacon}}]{blan14}%
  \BibitemOpen
  \bibfield  {author} {\bibinfo {author} {\bibfnamefont {S.}~\bibnamefont
  {Blanco-Canosa}}, \bibinfo {author} {\bibfnamefont {A.}~\bibnamefont
  {Frano}}, \bibinfo {author} {\bibfnamefont {E.}~\bibnamefont {Schierle}},
  \bibinfo {author} {\bibfnamefont {J.}~\bibnamefont {Porras}}, \bibinfo
  {author} {\bibfnamefont {T.}~\bibnamefont {Loew}}, \bibinfo {author}
  {\bibfnamefont {M.}~\bibnamefont {Minola}}, \bibinfo {author} {\bibfnamefont
  {M.}~\bibnamefont {Bluschke}}, \bibinfo {author} {\bibfnamefont
  {E.}~\bibnamefont {Weschke}}, \bibinfo {author} {\bibfnamefont
  {B.}~\bibnamefont {Keimer}},\ and\ \bibinfo {author} {\bibfnamefont
  {M.}~\bibnamefont {Le~Tacon}},\ }\bibfield  {title} {\bibinfo {title}
  {{Resonant x-ray scattering study of charge-density wave correlations in
  ${\mathrm{YBa}}_{2}{\mathrm{Cu}}_{3}{\mathrm{O}}_{6+x}$}},\ }\href
  {https://doi.org/10.1103/PhysRevB.90.054513} {\bibfield  {journal} {\bibinfo
  {journal} {Phys. Rev. B}\ }\textbf {\bibinfo {volume} {90}},\ \bibinfo
  {pages} {054513} (\bibinfo {year} {2014})}\BibitemShut {NoStop}%
\bibitem [{\citenamefont {Lee}\ \emph {et~al.}(2021)\citenamefont {Lee},
  \citenamefont {Huang}, \citenamefont {Johnson}, \citenamefont {Guo},
  \citenamefont {Husain}, \citenamefont {Mitrano}, \citenamefont {Lu},
  \citenamefont {Zakrzewski}, \citenamefont {de~la Pe\~na}, \citenamefont
  {Peng}, \citenamefont {Lee}, \citenamefont {Jang}, \citenamefont {Lee},
  \citenamefont {Joe}, \citenamefont {Dorisese}, \citenamefont {Szypryt},
  \citenamefont {Swetz}, \citenamefont {Aczel}, \citenamefont {Macdougall},
  \citenamefont {Kivelson}, \citenamefont {Fradkin},\ and\ \citenamefont
  {Abbamonte}}]{lee21c}%
  \BibitemOpen
  \bibfield  {author} {\bibinfo {author} {\bibfnamefont {S.}~\bibnamefont
  {Lee}}, \bibinfo {author} {\bibfnamefont {E.~W.}\ \bibnamefont {Huang}},
  \bibinfo {author} {\bibfnamefont {T.~A.}\ \bibnamefont {Johnson}}, \bibinfo
  {author} {\bibfnamefont {X.}~\bibnamefont {Guo}}, \bibinfo {author}
  {\bibfnamefont {A.~A.}\ \bibnamefont {Husain}}, \bibinfo {author}
  {\bibfnamefont {M.}~\bibnamefont {Mitrano}}, \bibinfo {author} {\bibfnamefont
  {K.}~\bibnamefont {Lu}}, \bibinfo {author} {\bibfnamefont {A.~V.}\
  \bibnamefont {Zakrzewski}}, \bibinfo {author} {\bibfnamefont
  {G.}~\bibnamefont {de~la Pe\~na}}, \bibinfo {author} {\bibfnamefont
  {Y.}~\bibnamefont {Peng}}, \bibinfo {author} {\bibfnamefont {S.-J.}\
  \bibnamefont {Lee}}, \bibinfo {author} {\bibfnamefont {H.}~\bibnamefont
  {Jang}}, \bibinfo {author} {\bibfnamefont {J.-S.}\ \bibnamefont {Lee}},
  \bibinfo {author} {\bibfnamefont {Y.~I.}\ \bibnamefont {Joe}}, \bibinfo
  {author} {\bibfnamefont {W.~B.}\ \bibnamefont {Dorisese}}, \bibinfo {author}
  {\bibfnamefont {P.}~\bibnamefont {Szypryt}}, \bibinfo {author} {\bibfnamefont
  {D.~S.}\ \bibnamefont {Swetz}}, \bibinfo {author} {\bibfnamefont {A.~A.}\
  \bibnamefont {Aczel}}, \bibinfo {author} {\bibfnamefont {G.~J.}\ \bibnamefont
  {Macdougall}}, \bibinfo {author} {\bibfnamefont {S.~A.}\ \bibnamefont
  {Kivelson}}, \bibinfo {author} {\bibfnamefont {E.}~\bibnamefont {Fradkin}},\
  and\ \bibinfo {author} {\bibfnamefont {P.}~\bibnamefont {Abbamonte}},\
  }\href@noop {} {\bibinfo {title} {{Generic character of charge and spin
  density waves in superconducting cuprates}}} (\bibinfo {year} {2021}),\
  \Eprint {https://arxiv.org/abs/2110.13991} {arXiv:2110.13991} \BibitemShut
  {NoStop}%
\bibitem [{\citenamefont {Hoffman}\ \emph {et~al.}(2002)\citenamefont
  {Hoffman}, \citenamefont {Hudson}, \citenamefont {Lang}, \citenamefont
  {Madhavan}, \citenamefont {Eisaki}, \citenamefont {Uchida},\ and\
  \citenamefont {Davis}}]{hoff02}%
  \BibitemOpen
  \bibfield  {author} {\bibinfo {author} {\bibfnamefont {J.~E.}\ \bibnamefont
  {Hoffman}}, \bibinfo {author} {\bibfnamefont {E.~W.}\ \bibnamefont {Hudson}},
  \bibinfo {author} {\bibfnamefont {K.~M.}\ \bibnamefont {Lang}}, \bibinfo
  {author} {\bibfnamefont {V.}~\bibnamefont {Madhavan}}, \bibinfo {author}
  {\bibfnamefont {H.}~\bibnamefont {Eisaki}}, \bibinfo {author} {\bibfnamefont
  {S.}~\bibnamefont {Uchida}},\ and\ \bibinfo {author} {\bibfnamefont {J.~C.}\
  \bibnamefont {Davis}},\ }\bibfield  {title} {\bibinfo {title} {{A Four Unit
  Cell Periodic Pattern of Quasi-Particle States Surrounding Vortex Cores in
  Bi$_2$Sr$_2$CaCu$_2$O$_{8+\delta}$ }},\ }\href@noop {} {\bibfield  {journal}
  {\bibinfo  {journal} {Science}\ }\textbf {\bibinfo {volume} {295}},\ \bibinfo
  {pages} {466} (\bibinfo {year} {2002})}\BibitemShut {NoStop}%
\bibitem [{\citenamefont {Xu}\ \emph {et~al.}(2009)\citenamefont {Xu},
  \citenamefont {Gu}, \citenamefont {Hucker}, \citenamefont {Fauque},
  \citenamefont {Perring}, \citenamefont {Regnault},\ and\ \citenamefont
  {Tranquada}}]{xu09}%
  \BibitemOpen
  \bibfield  {author} {\bibinfo {author} {\bibfnamefont {G.}~\bibnamefont
  {Xu}}, \bibinfo {author} {\bibfnamefont {G.~D.}\ \bibnamefont {Gu}}, \bibinfo
  {author} {\bibfnamefont {M.}~\bibnamefont {Hucker}}, \bibinfo {author}
  {\bibfnamefont {B.}~\bibnamefont {Fauque}}, \bibinfo {author} {\bibfnamefont
  {T.~G.}\ \bibnamefont {Perring}}, \bibinfo {author} {\bibfnamefont {L.~P.}\
  \bibnamefont {Regnault}},\ and\ \bibinfo {author} {\bibfnamefont {J.~M.}\
  \bibnamefont {Tranquada}},\ }\bibfield  {title} {\bibinfo {title} {{Testing
  the itinerancy of spin dynamics in superconducting
  Bi$_2$Sr$_2$CaCu$_2$O$_{8+\delta}$}},\ }\href
  {https://doi.org/10.1038/nphys1360} {\bibfield  {journal} {\bibinfo
  {journal} {Nat. Phys.}\ }\textbf {\bibinfo {volume} {5}},\ \bibinfo {pages}
  {642} (\bibinfo {year} {2009})}\BibitemShut {NoStop}%
\bibitem [{\citenamefont {Nachumi}\ \emph {et~al.}(1996)\citenamefont
  {Nachumi}, \citenamefont {Keren}, \citenamefont {Kojima}, \citenamefont
  {Larkin}, \citenamefont {Luke}, \citenamefont {Merrin}, \citenamefont
  {Tchernysh\"ov}, \citenamefont {Uemura}, \citenamefont {Ichikawa},
  \citenamefont {Goto},\ and\ \citenamefont {Uchida}}]{nach96}%
  \BibitemOpen
  \bibfield  {author} {\bibinfo {author} {\bibfnamefont {B.}~\bibnamefont
  {Nachumi}}, \bibinfo {author} {\bibfnamefont {A.}~\bibnamefont {Keren}},
  \bibinfo {author} {\bibfnamefont {K.}~\bibnamefont {Kojima}}, \bibinfo
  {author} {\bibfnamefont {M.}~\bibnamefont {Larkin}}, \bibinfo {author}
  {\bibfnamefont {G.~M.}\ \bibnamefont {Luke}}, \bibinfo {author}
  {\bibfnamefont {J.}~\bibnamefont {Merrin}}, \bibinfo {author} {\bibfnamefont
  {O.}~\bibnamefont {Tchernysh\"ov}}, \bibinfo {author} {\bibfnamefont {Y.~J.}\
  \bibnamefont {Uemura}}, \bibinfo {author} {\bibfnamefont {N.}~\bibnamefont
  {Ichikawa}}, \bibinfo {author} {\bibfnamefont {M.}~\bibnamefont {Goto}},\
  and\ \bibinfo {author} {\bibfnamefont {S.}~\bibnamefont {Uchida}},\
  }\bibfield  {title} {\bibinfo {title} {{Muon Spin Relaxation Studies of
  Zn-Substitution Effects in High- ${T}_{\mathit{c}}$ Cuprate
  Superconductors}},\ }\href {https://doi.org/10.1103/PhysRevLett.77.5421}
  {\bibfield  {journal} {\bibinfo  {journal} {Phys. Rev. Lett.}\ }\textbf
  {\bibinfo {volume} {77}},\ \bibinfo {pages} {5421} (\bibinfo {year}
  {1996})}\BibitemShut {NoStop}%
\bibitem [{\citenamefont {Tranquada}\ \emph {et~al.}(2021)\citenamefont
  {Tranquada}, \citenamefont {Dean},\ and\ \citenamefont {Li}}]{tran21b}%
  \BibitemOpen
  \bibfield  {author} {\bibinfo {author} {\bibfnamefont {J.~M.}\ \bibnamefont
  {Tranquada}}, \bibinfo {author} {\bibfnamefont {M.~P.~M.}\ \bibnamefont
  {Dean}},\ and\ \bibinfo {author} {\bibfnamefont {Q.}~\bibnamefont {Li}},\
  }\bibfield  {title} {\bibinfo {title} {{Superconductivity from Charge Order
  in Cuprates}},\ }\href {https://doi.org/10.7566/JPSJ.90.111002} {\bibfield
  {journal} {\bibinfo  {journal} {J. Phys. Soc. Jpn.}\ }\textbf {\bibinfo
  {volume} {90}},\ \bibinfo {pages} {111002} (\bibinfo {year}
  {2021})}\BibitemShut {NoStop}%
\bibitem [{\citenamefont {Wen}\ \emph {et~al.}(2012)\citenamefont {Wen},
  \citenamefont {Jie}, \citenamefont {Li}, \citenamefont {H\"ucker},
  \citenamefont {v.~Zimmermann}, \citenamefont {Han}, \citenamefont {Xu},
  \citenamefont {Singh}, \citenamefont {Konik}, \citenamefont {Zhang},
  \citenamefont {Gu},\ and\ \citenamefont {Tranquada}}]{wen12b}%
  \BibitemOpen
  \bibfield  {author} {\bibinfo {author} {\bibfnamefont {J.}~\bibnamefont
  {Wen}}, \bibinfo {author} {\bibfnamefont {Q.}~\bibnamefont {Jie}}, \bibinfo
  {author} {\bibfnamefont {Q.}~\bibnamefont {Li}}, \bibinfo {author}
  {\bibfnamefont {M.}~\bibnamefont {H\"ucker}}, \bibinfo {author}
  {\bibfnamefont {M.}~\bibnamefont {v.~Zimmermann}}, \bibinfo {author}
  {\bibfnamefont {S.~J.}\ \bibnamefont {Han}}, \bibinfo {author} {\bibfnamefont
  {Z.}~\bibnamefont {Xu}}, \bibinfo {author} {\bibfnamefont {D.~K.}\
  \bibnamefont {Singh}}, \bibinfo {author} {\bibfnamefont {R.~M.}\ \bibnamefont
  {Konik}}, \bibinfo {author} {\bibfnamefont {L.}~\bibnamefont {Zhang}},
  \bibinfo {author} {\bibfnamefont {G.}~\bibnamefont {Gu}},\ and\ \bibinfo
  {author} {\bibfnamefont {J.~M.}\ \bibnamefont {Tranquada}},\ }\bibfield
  {title} {\bibinfo {title} {{Uniaxial linear resistivity of superconducting
  La${}_{1.905}$Ba${}_{0.095}$CuO${}_{4}$ induced by an external magnetic
  field}},\ }\href {https://doi.org/10.1103/PhysRevB.85.134513} {\bibfield
  {journal} {\bibinfo  {journal} {Phys. Rev. B}\ }\textbf {\bibinfo {volume}
  {85}},\ \bibinfo {pages} {134513} (\bibinfo {year} {2012})}\BibitemShut
  {NoStop}%
\bibitem [{\citenamefont {Stegen}\ \emph {et~al.}(2013)\citenamefont {Stegen},
  \citenamefont {Han}, \citenamefont {Wu}, \citenamefont {Pramanik},
  \citenamefont {H\"ucker}, \citenamefont {Gu}, \citenamefont {Li},
  \citenamefont {Park}, \citenamefont {Boebinger},\ and\ \citenamefont
  {Tranquada}}]{steg13}%
  \BibitemOpen
  \bibfield  {author} {\bibinfo {author} {\bibfnamefont {Z.}~\bibnamefont
  {Stegen}}, \bibinfo {author} {\bibfnamefont {S.~J.}\ \bibnamefont {Han}},
  \bibinfo {author} {\bibfnamefont {J.}~\bibnamefont {Wu}}, \bibinfo {author}
  {\bibfnamefont {A.~K.}\ \bibnamefont {Pramanik}}, \bibinfo {author}
  {\bibfnamefont {M.}~\bibnamefont {H\"ucker}}, \bibinfo {author}
  {\bibfnamefont {G.}~\bibnamefont {Gu}}, \bibinfo {author} {\bibfnamefont
  {Q.}~\bibnamefont {Li}}, \bibinfo {author} {\bibfnamefont {J.~H.}\
  \bibnamefont {Park}}, \bibinfo {author} {\bibfnamefont {G.~S.}\ \bibnamefont
  {Boebinger}},\ and\ \bibinfo {author} {\bibfnamefont {J.~M.}\ \bibnamefont
  {Tranquada}},\ }\bibfield  {title} {\bibinfo {title} {{Evolution of
  superconducting correlations within magnetic-field-decoupled
  La${}_{2\ensuremath{-}x}$Ba${}_{x}$CuO${}_{4}$ ($x=0.095$)}},\ }\href
  {https://doi.org/10.1103/PhysRevB.87.064509} {\bibfield  {journal} {\bibinfo
  {journal} {Phys. Rev. B}\ }\textbf {\bibinfo {volume} {87}},\ \bibinfo
  {pages} {064509} (\bibinfo {year} {2013})}\BibitemShut {NoStop}%
\bibitem [{\citenamefont {Lozano}\ \emph
  {et~al.}(2021{\natexlab{b}})\citenamefont {Lozano}, \citenamefont {Gu},
  \citenamefont {Tranquada},\ and\ \citenamefont {Li}}]{loza21a}%
  \BibitemOpen
  \bibfield  {author} {\bibinfo {author} {\bibfnamefont {P.~M.}\ \bibnamefont
  {Lozano}}, \bibinfo {author} {\bibfnamefont {G.~D.}\ \bibnamefont {Gu}},
  \bibinfo {author} {\bibfnamefont {J.~M.}\ \bibnamefont {Tranquada}},\ and\
  \bibinfo {author} {\bibfnamefont {Q.}~\bibnamefont {Li}},\ }\bibfield
  {title} {\bibinfo {title} {{Experimental evidence that zinc impurities pin
  pair-density-wave order in
  ${\mathrm{La}}_{2\text{\ensuremath{-}}x}{\mathrm{Ba}}_{x}{\mathrm{CuO}}_{4}$}},\
  }\href {https://doi.org/10.1103/PhysRevB.103.L020502} {\bibfield  {journal}
  {\bibinfo  {journal} {Phys. Rev. B}\ }\textbf {\bibinfo {volume} {103}},\
  \bibinfo {pages} {L020502} (\bibinfo {year}
  {2021}{\natexlab{b}})}\BibitemShut {NoStop}%
\bibitem [{\citenamefont {Huang}\ \emph {et~al.}(2021)\citenamefont {Huang},
  \citenamefont {Lee}, \citenamefont {Ikeda}, \citenamefont {Taniguchi},
  \citenamefont {Takahama}, \citenamefont {Kao}, \citenamefont {Fujita},\ and\
  \citenamefont {Lee}}]{huan21}%
  \BibitemOpen
  \bibfield  {author} {\bibinfo {author} {\bibfnamefont {H.}~\bibnamefont
  {Huang}}, \bibinfo {author} {\bibfnamefont {S.-J.}\ \bibnamefont {Lee}},
  \bibinfo {author} {\bibfnamefont {Y.}~\bibnamefont {Ikeda}}, \bibinfo
  {author} {\bibfnamefont {T.}~\bibnamefont {Taniguchi}}, \bibinfo {author}
  {\bibfnamefont {M.}~\bibnamefont {Takahama}}, \bibinfo {author}
  {\bibfnamefont {C.-C.}\ \bibnamefont {Kao}}, \bibinfo {author} {\bibfnamefont
  {M.}~\bibnamefont {Fujita}},\ and\ \bibinfo {author} {\bibfnamefont {J.-S.}\
  \bibnamefont {Lee}},\ }\bibfield  {title} {\bibinfo {title} {{Two-Dimensional
  Superconducting Fluctuations Associated with Charge-Density-Wave Stripes in
  ${\mathrm{La}}_{1.87}{\mathrm{Sr}}_{0.13}{\mathrm{Cu}}_{0.99}{\mathrm{Fe}}_{0.01}{\mathrm{O}}_{4}$}},\
  }\href {https://doi.org/10.1103/PhysRevLett.126.167001} {\bibfield  {journal}
  {\bibinfo  {journal} {Phys. Rev. Lett.}\ }\textbf {\bibinfo {volume} {126}},\
  \bibinfo {pages} {167001} (\bibinfo {year} {2021})}\BibitemShut {NoStop}%
\bibitem [{\citenamefont {Hirota}(2001)}]{hiro01}%
  \BibitemOpen
  \bibfield  {author} {\bibinfo {author} {\bibfnamefont {K.}~\bibnamefont
  {Hirota}},\ }\bibfield  {title} {\bibinfo {title} {{Neutron scattering
  studies of Zn-doped La$_{2-x}$Sr$_x$CuO$_4$}},\ }\href
  {https://doi.org/https://doi.org/10.1016/S0921-4534(01)00195-2} {\bibfield
  {journal} {\bibinfo  {journal} {Physica C}\ }\textbf {\bibinfo {volume}
  {357--360}},\ \bibinfo {pages} {61} (\bibinfo {year} {2001})}\BibitemShut
  {NoStop}%
\bibitem [{\citenamefont {Guguchia}\ \emph {et~al.}(2017)\citenamefont
  {Guguchia}, \citenamefont {Roessli}, \citenamefont {Khasanov}, \citenamefont
  {Amato}, \citenamefont {Pomjakushina}, \citenamefont {Conder}, \citenamefont
  {Uemura}, \citenamefont {Tranquada}, \citenamefont {Keller},\ and\
  \citenamefont {Shengelaya}}]{gugu17}%
  \BibitemOpen
  \bibfield  {author} {\bibinfo {author} {\bibfnamefont {Z.}~\bibnamefont
  {Guguchia}}, \bibinfo {author} {\bibfnamefont {B.}~\bibnamefont {Roessli}},
  \bibinfo {author} {\bibfnamefont {R.}~\bibnamefont {Khasanov}}, \bibinfo
  {author} {\bibfnamefont {A.}~\bibnamefont {Amato}}, \bibinfo {author}
  {\bibfnamefont {E.}~\bibnamefont {Pomjakushina}}, \bibinfo {author}
  {\bibfnamefont {K.}~\bibnamefont {Conder}}, \bibinfo {author} {\bibfnamefont
  {Y.~J.}\ \bibnamefont {Uemura}}, \bibinfo {author} {\bibfnamefont {J.~M.}\
  \bibnamefont {Tranquada}}, \bibinfo {author} {\bibfnamefont {H.}~\bibnamefont
  {Keller}},\ and\ \bibinfo {author} {\bibfnamefont {A.}~\bibnamefont
  {Shengelaya}},\ }\bibfield  {title} {\bibinfo {title} {{Complementary
  Response of Static Spin-Stripe Order and Superconductivity to Nonmagnetic
  Impurities in Cuprates}},\ }\href
  {https://doi.org/10.1103/PhysRevLett.119.087002} {\bibfield  {journal}
  {\bibinfo  {journal} {Phys. Rev. Lett.}\ }\textbf {\bibinfo {volume} {119}},\
  \bibinfo {pages} {087002} (\bibinfo {year} {2017})}\BibitemShut {NoStop}%
\bibitem [{\citenamefont {Ganin}\ \emph {et~al.}(2010)\citenamefont {Ganin},
  \citenamefont {Takabayashi}, \citenamefont {Jegli{\v c}}, \citenamefont
  {Ar{\v c}on}, \citenamefont {Poto{\v c}nik}, \citenamefont {Baker},
  \citenamefont {Ohishi}, \citenamefont {McDonald}, \citenamefont {Tzirakis},
  \citenamefont {McLennan}, \citenamefont {Darling}, \citenamefont {Takata},
  \citenamefont {Rosseinsky},\ and\ \citenamefont {Prassides}}]{gani10}%
  \BibitemOpen
  \bibfield  {author} {\bibinfo {author} {\bibfnamefont {A.~Y.}\ \bibnamefont
  {Ganin}}, \bibinfo {author} {\bibfnamefont {Y.}~\bibnamefont {Takabayashi}},
  \bibinfo {author} {\bibfnamefont {P.}~\bibnamefont {Jegli{\v c}}}, \bibinfo
  {author} {\bibfnamefont {D.}~\bibnamefont {Ar{\v c}on}}, \bibinfo {author}
  {\bibfnamefont {A.}~\bibnamefont {Poto{\v c}nik}}, \bibinfo {author}
  {\bibfnamefont {P.~J.}\ \bibnamefont {Baker}}, \bibinfo {author}
  {\bibfnamefont {Y.}~\bibnamefont {Ohishi}}, \bibinfo {author} {\bibfnamefont
  {M.~T.}\ \bibnamefont {McDonald}}, \bibinfo {author} {\bibfnamefont {M.~D.}\
  \bibnamefont {Tzirakis}}, \bibinfo {author} {\bibfnamefont {A.}~\bibnamefont
  {McLennan}}, \bibinfo {author} {\bibfnamefont {G.~R.}\ \bibnamefont
  {Darling}}, \bibinfo {author} {\bibfnamefont {M.}~\bibnamefont {Takata}},
  \bibinfo {author} {\bibfnamefont {M.~J.}\ \bibnamefont {Rosseinsky}},\ and\
  \bibinfo {author} {\bibfnamefont {K.}~\bibnamefont {Prassides}},\ }\bibfield
  {title} {\bibinfo {title} {{Polymorphism control of superconductivity and
  magnetism in Cs$_3$C$_{60}$ close to the Mott transition}},\ }\href
  {https://doi.org/10.1038/nature09120} {\bibfield  {journal} {\bibinfo
  {journal} {Nature}\ }\textbf {\bibinfo {volume} {466}},\ \bibinfo {pages}
  {221} (\bibinfo {year} {2010})}\BibitemShut {NoStop}%
\bibitem [{\citenamefont {Chakravarty}\ and\ \citenamefont
  {Kivelson}(2001)}]{chak01c}%
  \BibitemOpen
  \bibfield  {author} {\bibinfo {author} {\bibfnamefont {S.}~\bibnamefont
  {Chakravarty}}\ and\ \bibinfo {author} {\bibfnamefont {S.~A.}\ \bibnamefont
  {Kivelson}},\ }\bibfield  {title} {\bibinfo {title} {{Electronic mechanism of
  superconductivity in the cuprates, ${\mathrm{C}}_{60},$ and polyacenes}},\
  }\href {https://doi.org/10.1103/PhysRevB.64.064511} {\bibfield  {journal}
  {\bibinfo  {journal} {Phys. Rev. B}\ }\textbf {\bibinfo {volume} {64}},\
  \bibinfo {pages} {064511} (\bibinfo {year} {2001})}\BibitemShut {NoStop}%
\bibitem [{\citenamefont {Ardavan}\ \emph {et~al.}(2012)\citenamefont
  {Ardavan}, \citenamefont {Brown}, \citenamefont {Kagoshima}, \citenamefont
  {Kanoda}, \citenamefont {Kuroki}, \citenamefont {Mori}, \citenamefont
  {Ogata}, \citenamefont {Uji},\ and\ \citenamefont {Wosnitza}}]{arda12}%
  \BibitemOpen
  \bibfield  {author} {\bibinfo {author} {\bibfnamefont {A.}~\bibnamefont
  {Ardavan}}, \bibinfo {author} {\bibfnamefont {S.}~\bibnamefont {Brown}},
  \bibinfo {author} {\bibfnamefont {S.}~\bibnamefont {Kagoshima}}, \bibinfo
  {author} {\bibfnamefont {K.}~\bibnamefont {Kanoda}}, \bibinfo {author}
  {\bibfnamefont {K.}~\bibnamefont {Kuroki}}, \bibinfo {author} {\bibfnamefont
  {H.}~\bibnamefont {Mori}}, \bibinfo {author} {\bibfnamefont {M.}~\bibnamefont
  {Ogata}}, \bibinfo {author} {\bibfnamefont {S.}~\bibnamefont {Uji}},\ and\
  \bibinfo {author} {\bibfnamefont {J.}~\bibnamefont {Wosnitza}},\ }\bibfield
  {title} {\bibinfo {title} {{Recent Topics of Organic Superconductors}},\
  }\href {https://doi.org/10.1143/JPSJ.81.011004} {\bibfield  {journal}
  {\bibinfo  {journal} {J. Phys. Soc. Jpn.}\ }\textbf {\bibinfo {volume}
  {81}},\ \bibinfo {pages} {011004} (\bibinfo {year} {2012})}\BibitemShut
  {NoStop}%
\bibitem [{\citenamefont {Kang}\ \emph {et~al.}(2010)\citenamefont {Kang},
  \citenamefont {Salameh}, \citenamefont {Auban-Senzier}, \citenamefont
  {J\'erome}, \citenamefont {Pasquier},\ and\ \citenamefont
  {Brazovskii}}]{kang10}%
  \BibitemOpen
  \bibfield  {author} {\bibinfo {author} {\bibfnamefont {N.}~\bibnamefont
  {Kang}}, \bibinfo {author} {\bibfnamefont {B.}~\bibnamefont {Salameh}},
  \bibinfo {author} {\bibfnamefont {P.}~\bibnamefont {Auban-Senzier}}, \bibinfo
  {author} {\bibfnamefont {D.}~\bibnamefont {J\'erome}}, \bibinfo {author}
  {\bibfnamefont {C.~R.}\ \bibnamefont {Pasquier}},\ and\ \bibinfo {author}
  {\bibfnamefont {S.}~\bibnamefont {Brazovskii}},\ }\bibfield  {title}
  {\bibinfo {title} {Domain walls at the spin-density-wave endpoint of the
  organic superconductor ${(\text{TMTSF})}_{2}{\text{pf}}_{6}$ under
  pressure},\ }\href {https://doi.org/10.1103/PhysRevB.81.100509} {\bibfield
  {journal} {\bibinfo  {journal} {Phys. Rev. B}\ }\textbf {\bibinfo {volume}
  {81}},\ \bibinfo {pages} {100509} (\bibinfo {year} {2010})}\BibitemShut
  {NoStop}%
\bibitem [{\citenamefont {Kawasugi}\ \emph {et~al.}(2019)\citenamefont
  {Kawasugi}, \citenamefont {Seki}, \citenamefont {Tajima}, \citenamefont {Pu},
  \citenamefont {Takenobu}, \citenamefont {Yunoki}, \citenamefont {Yamamoto},\
  and\ \citenamefont {Kato}}]{kawa19}%
  \BibitemOpen
  \bibfield  {author} {\bibinfo {author} {\bibfnamefont {Y.}~\bibnamefont
  {Kawasugi}}, \bibinfo {author} {\bibfnamefont {K.}~\bibnamefont {Seki}},
  \bibinfo {author} {\bibfnamefont {S.}~\bibnamefont {Tajima}}, \bibinfo
  {author} {\bibfnamefont {J.}~\bibnamefont {Pu}}, \bibinfo {author}
  {\bibfnamefont {T.}~\bibnamefont {Takenobu}}, \bibinfo {author}
  {\bibfnamefont {S.}~\bibnamefont {Yunoki}}, \bibinfo {author} {\bibfnamefont
  {H.~M.}\ \bibnamefont {Yamamoto}},\ and\ \bibinfo {author} {\bibfnamefont
  {R.}~\bibnamefont {Kato}},\ }\bibfield  {title} {\bibinfo {title}
  {{Two-dimensional ground-state mapping of a Mott-Hubbard system in a flexible
  field-effect device}},\ }\href {https://doi.org/10.1126/sciadv.aav7282}
  {\bibfield  {journal} {\bibinfo  {journal} {Sci. Adv.}\ }\textbf {\bibinfo
  {volume} {5}},\ \bibinfo {pages} {eaav7282} (\bibinfo {year}
  {2019})}\BibitemShut {NoStop}%
\bibitem [{\citenamefont {Motoyama}\ \emph {et~al.}(2007)\citenamefont
  {Motoyama}, \citenamefont {Yu}, \citenamefont {Vishik}, \citenamefont {Vajk},
  \citenamefont {Mang},\ and\ \citenamefont {Greven}}]{moto07}%
  \BibitemOpen
  \bibfield  {author} {\bibinfo {author} {\bibfnamefont {E.~M.}\ \bibnamefont
  {Motoyama}}, \bibinfo {author} {\bibfnamefont {G.}~\bibnamefont {Yu}},
  \bibinfo {author} {\bibfnamefont {I.~M.}\ \bibnamefont {Vishik}}, \bibinfo
  {author} {\bibfnamefont {O.~P.}\ \bibnamefont {Vajk}}, \bibinfo {author}
  {\bibfnamefont {P.~K.}\ \bibnamefont {Mang}},\ and\ \bibinfo {author}
  {\bibfnamefont {M.}~\bibnamefont {Greven}},\ }\bibfield  {title} {\bibinfo
  {title} {{Spin correlations in the electron-doped high-transition-temperature
  superconductor Nd$_{2-x}$Ce$_x$CuO$_{4\pm\delta}$}},\ }\href@noop {}
  {\bibfield  {journal} {\bibinfo  {journal} {Nature}\ }\textbf {\bibinfo
  {volume} {445}},\ \bibinfo {pages} {186} (\bibinfo {year}
  {2007})}\BibitemShut {NoStop}%
\bibitem [{\citenamefont {Yamada}\ \emph {et~al.}(2003)\citenamefont {Yamada},
  \citenamefont {Kurahashi}, \citenamefont {Uefuji}, \citenamefont {Fujita},
  \citenamefont {Park}, \citenamefont {Lee},\ and\ \citenamefont
  {Endoh}}]{yama03}%
  \BibitemOpen
  \bibfield  {author} {\bibinfo {author} {\bibfnamefont {K.}~\bibnamefont
  {Yamada}}, \bibinfo {author} {\bibfnamefont {K.}~\bibnamefont {Kurahashi}},
  \bibinfo {author} {\bibfnamefont {T.}~\bibnamefont {Uefuji}}, \bibinfo
  {author} {\bibfnamefont {M.}~\bibnamefont {Fujita}}, \bibinfo {author}
  {\bibfnamefont {S.}~\bibnamefont {Park}}, \bibinfo {author} {\bibfnamefont
  {S.-H.}\ \bibnamefont {Lee}},\ and\ \bibinfo {author} {\bibfnamefont
  {Y.}~\bibnamefont {Endoh}},\ }\bibfield  {title} {\bibinfo {title}
  {{Commensurate Spin Dynamics in the Superconducting State of an
  Electron-Doped Cuprate Superconductor}},\ }\href
  {https://doi.org/10.1103/PhysRevLett.90.137004} {\bibfield  {journal}
  {\bibinfo  {journal} {Phys. Rev. Lett.}\ }\textbf {\bibinfo {volume} {90}},\
  \bibinfo {pages} {137004} (\bibinfo {year} {2003})}\BibitemShut {NoStop}%
\bibitem [{\citenamefont {Zhao}\ \emph {et~al.}(2007)\citenamefont {Zhao},
  \citenamefont {Gaulin}, \citenamefont {Castellan}, \citenamefont {Ruff},
  \citenamefont {Dunsiger}, \citenamefont {Gu},\ and\ \citenamefont
  {Dabkowska}}]{zhao07}%
  \BibitemOpen
  \bibfield  {author} {\bibinfo {author} {\bibfnamefont {Y.}~\bibnamefont
  {Zhao}}, \bibinfo {author} {\bibfnamefont {B.~D.}\ \bibnamefont {Gaulin}},
  \bibinfo {author} {\bibfnamefont {J.~P.}\ \bibnamefont {Castellan}}, \bibinfo
  {author} {\bibfnamefont {J.~P.~C.}\ \bibnamefont {Ruff}}, \bibinfo {author}
  {\bibfnamefont {S.~R.}\ \bibnamefont {Dunsiger}}, \bibinfo {author}
  {\bibfnamefont {G.~D.}\ \bibnamefont {Gu}},\ and\ \bibinfo {author}
  {\bibfnamefont {H.~A.}\ \bibnamefont {Dabkowska}},\ }\bibfield  {title}
  {\bibinfo {title} {{High-resolution x-ray scattering studies of structural
  phase transitions in underdoped
  ${\mathrm{La}}_{2\ensuremath{-}x}{\mathrm{Ba}}_{x}\mathrm{Cu}{\mathrm{O}}_{4}$}},\
  }\href {https://doi.org/10.1103/PhysRevB.76.184121} {\bibfield  {journal}
  {\bibinfo  {journal} {Phys. Rev. B}\ }\textbf {\bibinfo {volume} {76}},\
  \bibinfo {pages} {184121} (\bibinfo {year} {2007})}\BibitemShut {NoStop}%
\bibitem [{\citenamefont {Yu}\ \emph {et~al.}(2010)\citenamefont {Yu},
  \citenamefont {Li}, \citenamefont {Motoyama}, \citenamefont {Zhao},
  \citenamefont {Bari\v{s}i\'{c}}, \citenamefont {Cho}, \citenamefont
  {Bourges}, \citenamefont {Hradil}, \citenamefont {Mole},\ and\ \citenamefont
  {Greven}}]{yu10}%
  \BibitemOpen
  \bibfield  {author} {\bibinfo {author} {\bibfnamefont {G.}~\bibnamefont
  {Yu}}, \bibinfo {author} {\bibfnamefont {Y.}~\bibnamefont {Li}}, \bibinfo
  {author} {\bibfnamefont {E.~M.}\ \bibnamefont {Motoyama}}, \bibinfo {author}
  {\bibfnamefont {X.}~\bibnamefont {Zhao}}, \bibinfo {author} {\bibfnamefont
  {N.}~\bibnamefont {Bari\v{s}i\'{c}}}, \bibinfo {author} {\bibfnamefont
  {Y.}~\bibnamefont {Cho}}, \bibinfo {author} {\bibfnamefont {P.}~\bibnamefont
  {Bourges}}, \bibinfo {author} {\bibfnamefont {K.}~\bibnamefont {Hradil}},
  \bibinfo {author} {\bibfnamefont {R.~A.}\ \bibnamefont {Mole}},\ and\
  \bibinfo {author} {\bibfnamefont {M.}~\bibnamefont {Greven}},\ }\bibfield
  {title} {\bibinfo {title} {{Magnetic resonance in the model high-temperature
  superconductor HgBa$_{2}$CuO$_{4+\delta{}}$}},\ }\href@noop {} {\bibfield
  {journal} {\bibinfo  {journal} {Phys. Rev. B}\ }\textbf {\bibinfo {volume}
  {81}},\ \bibinfo {pages} {064518} (\bibinfo {year} {2010})}\BibitemShut
  {NoStop}%
\end{thebibliography}%

\end{document}